\documentclass[fleqn,usenatbib]{mnras}

\usepackage[T1]{fontenc}

\DeclareRobustCommand{\VAN}[3]{#2}
\let\VANthebibliography\thebibliography
\def\thebibliography{\DeclareRobustCommand{\VAN}[3]{##3}\VANthebibliography}

\usepackage{graphicx}	
\usepackage{amsmath}	
\usepackage{amssymb}	
\usepackage{graphics}
\usepackage{multirow} 

\usepackage{newtxtext,newtxmath}

\title[Identifying Pop III galaxies with JWST]{On the observability and identification of Population III galaxies with JWST}

\author[J. A. A. Trussler et al.]{James A. A. Trussler,$^{1}$\thanks{E-mail: james.trussler@manchester.ac.uk}
Christopher J. Conselice,$^{1}$
Nathan J. Adams,$^{1}$
Roberto Maiolino,$^{2,3,4}$
\newauthor
Kimihiko Nakajima,$^{5}$
Erik Zackrisson,$^{6}$
Duncan Austin,$^{1}$
Leonardo Ferreira$^{7}$
and Tom Harvey$^{1}$
\\
$^{1}$Jodrell Bank Centre for Astrophysics, University of Manchester, Oxford Road, Manchester M13 9PL, UK\\
$^{2}$Cavendish Laboratory, University of Cambridge, 19 J.J.\@ Thomson Avenue, Cambridge CB3 0HE, UK\\
$^{3}$Kavli Institute for Cosmology, University of Cambridge, Madingley Road, Cambridge CB3 0HA, UK\\
$^{4}$Department of Physics and Astronomy, University College London, Gower Street, London WC1E 6BT, UK\\
$^{5}$National Astronomical Observatory of Japan, 2-21-1 Osawa, Mitaka, Tokyo 181-8588, Japan\\
$^{6}$Observational Astrophysics, Department of Physics and Astronomy, Uppsala University, Box 516, SE-751 20 Uppsala, Sweden\\
$^{7}$Department of Physics \& Astronomy, University of Victoria, Finnerty Road, Victoria, British Columbia, V8P 1A1, Canada
}

\date{Accepted XXX. Received YYY; in original form ZZZ}

\pubyear{2023}

\begin{document}
\label{firstpage}
\pagerange{\pageref{firstpage}--\pageref{lastpage}}
\maketitle

\begin{abstract}
We utilise theoretical models of Population III stellar+nebular spectra to investigate the prospects of observing and accurately identifying Population III galaxies with \emph{JWST} using both deep imaging and spectroscopy.  We investigate a series of different colour cuts, finding that a combination of NIRCam and MIRI photometry through the F444W$-$F560W, F560W$-$F770W colours offers the most robust identifier of potential $z=8$ Pop III candidates. We calculate that NIRCam will have to reach ${\sim}28.5$--$30.0$~AB~mag depths (1--20~h), and MIRI F560W must reach ${\sim}27.5$--$29.0$~AB~mag depths (10--100~h) to achieve $5\sigma$ continuum detections of $M_* = 10^6~\mathrm{M}_\odot$ Pop III galaxies at $z=8$. We also discuss the prospects of identifying Pop III candidates through slitless and NIRSpec spectroscopic surveys that target Ly$\alpha$, H$\beta$ and/or \ion{He}{II} $\lambda$1640. We find small differences in the H$\beta$ rest-frame equivalent width (EW) between Pop III and non-Pop III galaxies, rendering this diagnostic likely impractical. Instead, we find that the detection of high EW \ion{He}{II} $\lambda 1640$ emission will serve as the definitive Pop III identifier, requiring (ultra-)deep integrations (5--150~h) with NIRSpec/G140M for $M_*=10^6~\mathrm{M}_\odot$ Pop III galaxies at $z=8$. However, MIRI F770W detections of Pop III galaxies will require substantial gravitational lensing ($\mu=10$) and/or fortuitous imaging of exceptionally massive ($M_* = 10^7~\mathrm{M}_\odot$) Pop III galaxies. Thus, NIRCam medium-band imaging surveys that can search for high EW \ion{He}{II} $\lambda 1640$ emitters in photometry may perhaps be a viable alternative for finding Pop III candidates.
\end{abstract}

\begin{keywords}
galaxies: formation -- galaxies: evolution -- galaxies: abundances -- galaxies: high-redshift -- stars: Population III
\end{keywords}



\section{Introduction}

The recently launched and now fully operational \emph{James Webb Space Telescope} (\emph{JWST}) is set to usher in a golden era of astronomy, being the central catalyst for a great new age of discovery. Indeed, our perception of the Universe will forever change, as we begin to see it with a clarity and depth that greatly surpasses anything that has come before. Among its many grand discoveries set to come, perhaps the greatest of all is the possibility of observing the light from the very first stars in the Universe \citep[e.g.\@][]{Bromm2004, Bromm2011}. These chemically pristine, so-called `Population III' stars, formed out of the primordial hydrogen and helium (and trace amounts of lithium), and were the first embers to ignite, producers of the starlight that ended the cosmic dark ages \citep[e.g.\@][]{Miralda-Escude2003} and paved the way for cosmic dawn \citep[e.g.\@][]{Laporte2021}. In studying these first stars, the foundation upon which all of cosmic history stands, we will gain unprecedented insights into both primordial star formation \citep[e.g.\@][]{Johnson2013}, as well as the characteristics of the `Population III galaxies' within which the first stars reside \citep[e.g.\@][]{Bromm2011}. Indeed, by analysing Pop III galaxies, we will be witnesses to the synthesis of the very first non-primordial elements and the subsequent chemical enrichment of the Universe \citep[e.g.\@][]{Ferrara2000, Madau2001, Bromm2003}. 

Forming in so-called `atomic cooling haloes', where hydrogen gas is sufficiently hot to collisionally ionise, and thus cool through recombination, Pop III galaxies are expected, from virial arguments, to have halo masses of $M_\mathrm{h} \sim 10^{8}~\mathrm{M}_\odot$ \citep[see e.g.\@][]{Johnson2008, Bromm2011}. Accounting for the cosmic baryon fraction and approximate estimates for the efficiency with which primordial baryonic gas is converted into stars, these first galaxies are believed to have a typical stellar mass of $M_* \sim 10^{5}~\mathrm{M}_\odot$ \citep{Bromm2011}. Thus the combination of their low stellar masses, together with their great cosmological distances from us, makes such Pop III galaxies exceptionally faint, and therefore beyond the detection capabilities of the telescopes that came before \emph{JWST}.

\emph{JWST}, with its exceptional sensitivity and resolution, is thus the ideal facility with which to search for these elusive Pop III galaxies \citep[e.g.\@][]{Gardner2006a, Bromm2011, Nakajima2022, Katz2023}. Its extensive suite of scientific instruments and diverse observing modes, spanning imaging to both slit and slitless spectroscopy, will all play an essential role in identifying Pop III candidates and further characterising their properties. The signatures of Pop III stars are encoded within their rest-UV-optical-NIR light. \emph{JWST}, spanning the NIR-to-MIR wavelengths to which these fingerprints of primordial star formation are redshifted to, therefore has the capabilities to capture the full wealth of information stored within this ancient starlight.

The photometric and spectroscopic signatures of Pop III stars and galaxies has been extensively studied. For example, \citet{Schaerer2002} presented realistic models for massive Pop III stars, including nebular continuum emission, finding that nebular line and continuum emission strongly affects the broad band photometric properties of Pop III objects. Indeed, strong (i.e.\@ high EW) emission in the \ion{He}{II} $\lambda 1640$ recombination line was found \citep[e.g.\@][]{Schaerer2002, Schaerer2003, Raiter2010b} to be a clear signature of chemically-pristine Pop III stars, reflecting their exceptional capability in generating hard ionising photons capable of doubly ionising helium. This in turn is due to the likely very massive nature of Pop III stars \citep[$\gtrapprox 10~\mathrm{M_\odot}$, see e.g.\@][]{Bromm1999, Tan2004}, being the net result of inefficient cooling and subsequent star formation in metal-free, primordial gas. 

Later works, such as \citet{Zackrisson2011}, built upon these earlier models to generate forecasts for the prospects of \emph{JWST} at observing and identifying Pop III galaxies. They concluded that ultra-deep exposures would be needed to detect ${\sim}10^5~\mathrm{M}_\odot$ Pop III galaxies at $z=10$, with colour--colour selections combining \emph{JWST}/NIRCam and \emph{JWST}/MIRI photometry enabling a clean selection of Pop III galaxies at $z \approx 7$--$8$. Indeed, fortuitous gravitational lensing of Pop III galaxies will greatly relax the otherwise demanding integration times needed \citep{Zackrisson2012}. At the same time, wide-field surveys with e.g.\@ \emph{Euclid} and the \emph{Roman Space Telescope} will likely play a crucial role in our photometric search for Pop III galaxies \citep{Vikaeus2022}. 

Follow-up spectroscopy, aiming to target bright emission lines such as H$\beta$ and \ion{He}{II} $\lambda 1640$, will be essential to verify the pristine nature of potential Pop III candidates. The application of spectroscopic diagnostics, such as the \ion{He}{II} $\lambda 4686$/H$\beta$ line ratio, will enable us to distinguish between different types of chemically pristine and extremely metal-poor systems \citep{Nakajima2022}. In this regard, \emph{JWST}, but also the next-generation of extremely large telescopes, will play a pivotal role in definitively identifying true Pop III galaxies \citep{Grisdale2021, Nakajima2022}.

The aim of this paper is to build on previous works, by providing a comprehensive overview of the capabilities of \emph{JWST} in observing and identifying Pop III galaxies. We wish to consider in detail the role of all four of \emph{JWST}'s scientific instruments, across all of their observing modes, from imaging to slitless and slit spectroscopy. Indeed, we intend this work to encompass all aspects of the observational Pop III search, from the initial identification of Pop III candidates using colour- and/or emission-line selections, to the removal of interlopers and stronger Pop III constraints derived from follow-up \emph{JWST} spectroscopy. We present a series of photometric and spectroscopic diagnostics that have been designed to be as viable as possible, targeting photometric and spectral features that can be detected in relatively short integration times (compared to alternatives). Simultaneously, we aim for these diagnostics to still be effective and practical, offering valuable constraints once observational errors on e.g.\@ colours or line fluxes, model uncertainties and potentially unknown source parameters (like the ionisation parameter $U$, which is the ratio between the number density of ionising photons and hydrogen atoms) are taken into account.

This paper is structured as follows. In Section~\ref{sec:models}, we discuss the models for Pop III (and non-Pop III) spectra that will be used in our analysis. In Section~\ref{sec:selection}, we discuss the prospects of observing and identifying Pop III $z\sim8$ galaxies with \emph{JWST}. We discuss their expected apparent magnitudes and line fluxes, as well as the integration times needed to achieve a $5\sigma$ detection with imaging and slitless spectroscopy, as well as the visibility timescales for Pop III galaxies. We introduce and discuss several colour selections for identifying Pop III candidates, as well as the prospects of identifying Pop III candidates from slitless spectroscopic emission-line surveys with \emph{JWST}. We undertake a similar analysis of $z\sim10$ Pop III galaxies in Appendix~\ref{app:z10}. In Section~\ref{sec:spectroscopy} we discuss the additional Pop III constraints that can be derived from follow-up spectroscopy on Pop III candidates, as well as the integration times needed to achieve $5\sigma$ line detections with NIRSpec. In Section~\ref{sec:caveats}, we discuss our assumptions regarding the stellar mass of Pop III galaxies, their redshifts and the general likelihood of encountering Pop III galaxies with \emph{JWST}. Finally, in Section~\ref{sec:conclusions} we summarise our main findings and conclude.

We assume a nominal stellar mass of $M_* = 10^6~\mathrm{M}_\odot$ in our analysis, rather than the typical Pop III stellar mass of $M_* = 10^5~\mathrm{M}_\odot$ expected from virial arguments. This nominal stellar mass was chosen as it likely reflects the least massive Pop III galaxies that will be detectable within feasible integration times with \emph{JWST}. We adopt a \citet{Planck2020} cosmology.

\section{Population III models} \label{sec:models}

We make use of both the \citet{Zackrisson2011} and \citet{Nakajima2022} models for Pop III (as well as non-Pop III) spectra in our analysis. This approach was adopted for two reasons. Firstly, to investigate the robustness of our colour selections and spectroscopic diagnostics at identifying and confirming Pop III candidates. Secondly, to draw upon the synergy between these two different but complementary models. With the \citet{Zackrisson2011} models we will determine the expected apparent magnitudes and line fluxes for $z=8$ Pop III sources, as well as investigate the time-dependence of these quantities. With the \citet{Nakajima2022} models we explore the dependence of our emission line diagnostics on the ionisation parameter $U$ and push our analysis of non-Pop III galaxies down to even lower (but still non-pristine) metallicities. We refer the reader to the respective papers for the full details on these models. In this section we briefly outline the features of these models that are most relevant for our analysis.

The \citet{Zackrisson2011} models provide rest UV-to-FIR spectra for both Pop III and non-Pop III populations. Most importantly, the spectra contain both stellar and nebular emission, with the pure stellar templates provided as inputs to the {\footnotesize CLOUDY} \citep{Ferland1998} photoionisation code \citep[following the procedure in][]{Zackrisson2001}, which in turn computes the associated nebular continuum and line emission. Models with ISM covering fractions of $f_\mathrm{cov} = 0, 0.5, 1$ are provided. We use the $f_\mathrm{cov} = 1$ spectra in our analysis, though we refer to the $f_\mathrm{cov} = 0.5$ results when relevant.

For the Pop III galaxies, three different IMFs are available: Pop III.1, Pop III.2 and Pop III with a Kroupa IMF. Roughly speaking, the characteristic mass of stars $M_{*, \mathrm{IMF}}$ formed in the Pop III.1, Pop III.2 and Pop III Kroupa IMFs are $\sim$$100~\mathrm{M}_\odot$, $\sim$$10~\mathrm{M}_\odot$ and $\sim$$1~\mathrm{M}_\odot$, respectively. In the case of Pop III.1, the \citet{Schaerer2002} stellar SSP with a power-law IMF of slope $\alpha=2.35$ across the mass range 50--500~$\mathrm{M}_\odot$ is used. For Pop III.2 galaxies, the \citet{Raiter2010b} model is adopted, which has a log-normal IMF with characteristic mass $M_\mathrm{c} = 10$~$\mathrm{M}_\odot$ and dispersion $\sigma = 1$~$\mathrm{M}_\odot$, with wings extending from 1 to 500~$\mathrm{M}_\odot$. For the Pop III Kroupa galaxies (which we include for completeness, and as a reference against the non-Pop III models which also have a Kroupa IMF) the \citet{Kroupa2001} IMF is adopted. \citet{Zackrisson2011} model spectra for non-Pop III (i.e.\@ Pop II and Pop I) galaxies are also analysed. These assume the \citet{Kroupa2001} IMF and are provided at $Z = [0.02, 0.2, 0.4, 1]~\mathrm{Z}_\odot$. 

The \citet{Nakajima2022} models also provide rest UV-to-FIR spectra for both Pop III and non-Pop III populations, including both the stellar and nebular \citep[generated using {\footnotesize CLOUDY,}][]{Ferland1998, Ferland2013} emission. 

For their Pop III galaxies, three metal-free stellar populations from \citet{Schaerer2003} and \citet{Raiter2010b} are used. A \citet{Salpeter1955} IMF is adopted, with three different mass ranges considered: 1--100~$\mathrm{M}_\odot$, 1--500~$\mathrm{M}_\odot$ and 50--500~$\mathrm{M}_\odot$. Three gas-phase metallicities are also available: $Z_\mathrm{gas} = 0,\ \mathrm{Z}_\odot / 1400,\ \mathrm{Z}_\odot / 140$. The non-zero metallicities are considered to explore the scenario of Pop III galaxies forming from pockets of pristine gas surviving in slightly enriched environments. We use the \citet{Nakajima2022} Pop III spectra with electron density $n_\mathrm{e} = 10^3$~cm$^{-3}$, across the ionisation parameter range $\log U = [-2.0, -1.5, -1.0, -0.5]$. 

Metal-poor Pop II models are also provided, spanning the metallicities $Z = [\mathrm{Z}_\odot / 1400, \mathrm{Z}_\odot / 140, \mathrm{Z}_\odot / 14]$, and across the ionisation parameter range  $\log U = [-2.0, -1.5, -1.0, -0.5]$. BPASS SEDs with a \citet{Kroupa2001} IMF with two possible upper mass limits are used, namely 100~$\mathrm{M}_\odot$ and 300~$\mathrm{M}_\odot$. Metal-enriched Pop I models are available at the following metallicities: $Z = [0.1, 0.2, 0.5, 1]~\mathrm{Z}_\odot$, and at the following ionisation parameter values:  $\log U = [-3.0, -2.5, -2.0, -1.5, -1.0, -0.5]$. 

Model spectra for active galactic nuclei (AGN) and direct collapse black holes (DCBH) are also used, though these are not central to our analysis. We only comment that while both AGN and DCBH can have bright \ion{He}{II} $\lambda 1640$ emission (similar to Pop III galaxies), they have very distinct red colours and so would not be confused with Pop III systems, see Fig.\@~\ref{fig:nakajima_colours1} but also \citet{Inayoshi2022}.

We use the \citet{Inoue2014} prescription for IGM attenuation in our analysis. This essentially removes all the flux blueward of the central wavelength of Ly$\alpha$. Though we note that this prescription does not result in any attenuation/scattering redward of the Ly$\alpha$ peak. When discussing our colour selections in Section~\ref{subsec:colour_selection}, the \citet{Inoue2014} IGM attenuation has been applied to the spectrum. However, when discussing spectroscopic diagnostics involving the Ly$\alpha$ line, as well as expected Ly$\alpha$ fluxes, we instead use the intrinsic (i.e.\@ unattenuated) Ly$\alpha$ line.

Finally, we assume the \citet{Calzetti2000} dust attenuation law in this work. Although we do not explicitly apply this attenuation law to any of the model spectra, we do show the imprint that dust reddening would have on galaxy colours within the colour--colour plane. 
\section{The observability and identification of Pop III candidates} \label{sec:selection}

In this section we discuss the prospects of observing and identifying Pop III candidates with \emph{JWST}. Our analysis will focus on Pop III candidates at $z\sim8$. We mirror this analysis for $z\sim10$ candidates in Appendix~\ref{app:z10}. In Section~\ref{subsec:model_spectrum} we introduce the identifying, distinct features of a Pop III spectrum and the expected observability windows after a starburst. In Section~\ref{subsec:colour_selection}, we discuss our colour--colour selections, outlining the principles behind these selections, the redshift windows of applicability and potential high-$z$ and Galactic contaminants. Finally, in Section~\ref{subsec:line_selection}, we discuss the feasibility of identifying Pop III candidates in \emph{JWST} slitless spectroscopic emission-line surveys.

\subsection{Pop III spectrum and photometry} \label{subsec:model_spectrum}

\begin{figure*}
\centering
\includegraphics[width=1\linewidth]{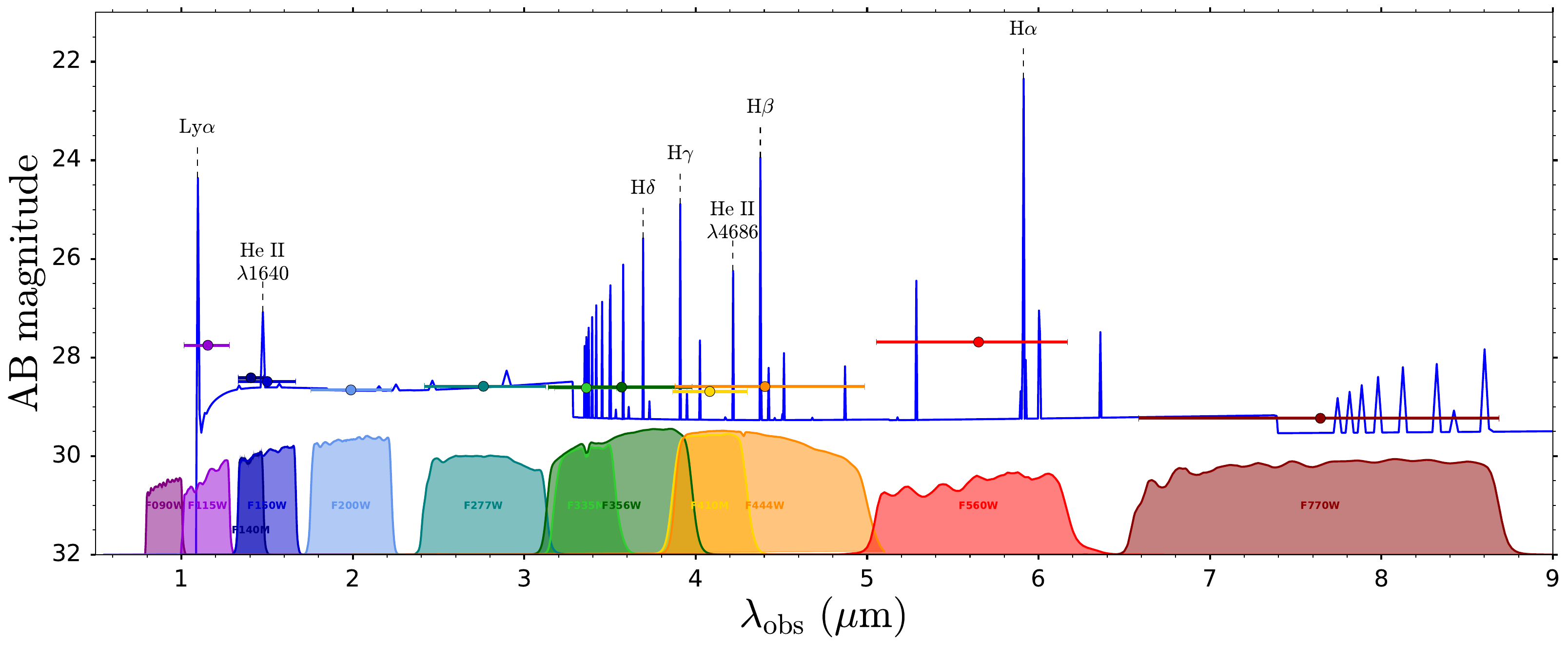}
\caption{The spectrum of a $z=8$ Pop III.1 galaxy (i.e.\@ with the characteristic mass of the individual Pop III stars $M_{*, \mathrm{IMF}}$ being $\sim$$100~\mathrm{M}_\odot$) at the nominal stellar mass $\log (M_*/\mathrm{M}_\odot) = 6$, observed immediately after ($0.01$~Myr) an instantaneous starburst. The filter throughputs and bandpass-averaged flux densities within the \emph{JWST} NIRCam + MIRI bands that are used throughout our $z\sim8$ analysis are also shown.}
\label{fig:popIII_spectrum}
\end{figure*}

We show an example of the spectrum of a $\log (M_*/\mathrm{M}_\odot) = 6$ Pop III galaxy at $z=8$, together with its associated \emph{JWST} NIRCam + MIRI photometry in Fig.\@~\ref{fig:popIII_spectrum}. Here we adopt the Pop III.1 model from \citet{Zackrisson2011}, which corresponds to the most top-heavy Pop III IMF in their models. We show the expected spectrum immediately after ($0.01$~Myr) an instantaneous starburst. The \emph{JWST} filters shown in Fig.\@~\ref{fig:popIII_spectrum} will be considered throughout our analysis in this paper. This set of NIRCam and MIRI filters has been chosen for two reasons. Firstly, because these filters are likely to be the standard filter set used by \emph{JWST}, having been widely adopted in Cycle 1 observations. Secondly, because we found that these filters proved to capture the most salient spectral features needed for Pop III identification. 

In the following section we briefly outline the key spectral features that distinguish Pop III galaxies from non-Pop III galaxies. This overview therefore forms the basis behind our subsequent colour-- and emission-line--selections in Sections \ref{subsec:colour_selection} and \ref{subsec:line_selection}, as well as our recommendations for spectroscopic follow-up observations in Section~\ref{sec:spectroscopy}. 

\begin{table*}
\begin{center}
\resizebox{\linewidth}{!}{
\begin{tabular}{ |c|c|c|c|c|c|c|c|c|c|c|c|c|c|c| } 
\hline
IMF & F090W & F115W & F140M & F150W & F200W & F277W & F335M & F356W & F410M & F444W & F560W & F770W & $\Delta m =0.25$ & $\Delta m =1.0$  \\
\hline
Pop III.1 & 38.15 & 27.75 & 28.41 & 28.49 & 28.66 & 28.58 & 28.62 & 28.60 & 28.69 & 28.59 & 27.69 & 29.23 & \multirow{2}{*}{2.2} & \multirow{2}{*}{2.8}\\
($M_{*, \mathrm{IMF}}\sim100~\mathrm{M}_\odot$) & N/A & 0.40 & 1.73 & 0.94 & 0.81 & 0.75 & 1.61 & 0.66 & 2.45 & 1.34 & 10.57 & 691.86 & & \\
\hline
Pop III.2 & 39.24 & 28.90 & 29.60 & 29.65 & 29.79 & 29.73 & 29.75 & 29.74 & 29.88 & 29.76 & 28.84 & 30.41 & \multirow{2}{*}{4.0} & \multirow{2}{*}{4.0} \\
($M_{*, \mathrm{IMF}}\sim10~\mathrm{M}_\odot$) & N/A  & 3.29 & 16.36 & 7.97 & 7.27 & 6.16 & 13.11 & 5.52 & 22.37 & 11.74 & 89.54 & 5970.61 & & \\
\hline
Pop III Kroupa & 41.13 & 30.86 & 31.55 & 31.58 & 31.72 & 31.68 & 31.71 & 31.71 & 31.87 & 31.75 & 30.84 & 32.40 & \multirow{2}{*}{4.0} & \multirow{2}{*}{5.6} \\
($M_{*, \mathrm{IMF}}\sim1~\mathrm{M}_\odot$) & N/A & 121.78 & 593.99 & 284.17 & 254.43 & 227.81 & 475.96 & 203.97 & 858.17 & 450.37 & 3499.60 & 237694.36 & & \\
\hline
\end{tabular}}
\caption{The expected apparent magnitudes in \emph{JWST} NIRCam+MIRI bands (top row) and integration times needed (in hours) to achieve a $5\sigma$ detection (bottom row) for the three \citet{Zackrisson2011} Pop III models at $z=8$ with $\log (M_*/\mathrm{M}_\odot) = 6$ observed immediately after ($0.01$~Myr) an instantaneous starburst. The timescales (in Myr) over which the average magnitude for Pop III galaxies in the listed \emph{JWST} bands drops by 0.25 and 1.0~mag are also provided, which should serve as an indication of the visibility window over which Pop III galaxies can be detected. Roughly speaking, the characteristic mass of stars $M_{*, \mathrm{IMF}}$ formed in the Pop III.1, Pop III.2 and Pop III Kroupa IMFs are $\sim$$100~\mathrm{M}_\odot$, $\sim$$10~\mathrm{M}_\odot$ and $\sim$$1~\mathrm{M}_\odot$, respectively. The integration times were estimated using the JWST ETC, assuming a point source, and adopting a 0.32~arcsec diameter circular extraction aperture for the NIRCam bands, and a 0.49~arcsec and 0.55~arcsec aperture for the MIRI F560W and F770W bands, respectively.}
\label{tab:imaging_observations}
\end{center}
\end{table*}
\subsubsection{Distinguishing Pop III spectral features} \label{subsec:popIII_features}

Owing to their metal-free nature, as well as their potentially more top-heavy IMF, Pop III stars (and galaxies) are expected to produce more ionising photons per unit stellar mass than non-Pop III stars \citep[e.g.\@][]{Schaerer2002, Schaerer2003, Raiter2010b, Zackrisson2011}. Hence we expect the recombination lines of hydrogen and helium to have a greater luminosity per unit stellar mass for Pop III galaxies. Furthermore, the ionising radiation produced will also be harder than for non-Pop III stars \citep[e.g.\@][]{Schaerer2002, Schaerer2003}, Hence, we also expect the recombination lines of doubly-ionised helium (\ion{He}{II}) to have much greater luminosity per unit stellar mass for Pop III galaxies. 

The great source of (hard) ionising photons therefore heats more ISM to a greater temperature than for non-Pop III stars, resulting in a much brighter nebular continuum contribution to the total (stellar+nebular) Pop III spectrum \citep[e.g.\@][]{Zackrisson2001, Schaerer2002, Zackrisson2011}. Although the brightest stars in a Pop III population are bluer than their non-Pop III counterparts, the greater contribution from the relatively cooler and redder nebular continuum results in the total Pop III spectrum being redder than for non-Pop III stars, which manifests itself in e.g.\@ a less negative 1500~\AA \ UV slope $\beta$ \citep[e.g.\@][]{Raiter2010b, Zackrisson2011, Dunlop2013}.

Given their higher ionising photon production rate (and hence recombination rate), one might expect the Balmer and Paschen jumps to be larger (in terms of an AB magnitude difference blueward and redward of the jump) for Pop III galaxies. However, we found that these jumps are in fact weaker for Pop III galaxies. Given the higher ISM temperatures \citep[e.g.\@][]{Schaerer2003}, the recombining electrons in Pop III systems have a greater spread of energies, with a smaller fraction of particles therefore recombining with energy $E=0$. Since it is precisely the recombination rate of the $E=0$ particles that drive the jump strength, the Balmer and Paschen jumps are weaker for Pop III galaxies.

Finally, since Pop III galaxies are by definition metal-free, their spectra do not contain any metal emission- or absorption-lines. Indeed, some of these metal emission-lines, such as the [\ion{O}{III}] $\lambda\lambda 4959,5007$ and [\ion{S}{III}] $\lambda\lambda 9069, 9531$ doublets are relatively bright, and therefore have very high equivalent widths. As we shall see in Section~\ref{subsec:colour_selection}, the absence of such bright lines in the Pop III spectrum can therefore lead to distinct colours which can be used to select Pop III candidates from photometry \citep[see also][]{Inoue2011, Zackrisson2011}.

\subsubsection{The observability of Pop III galaxies} \label{subsubsec:observability}

\begin{figure*}
\centering
\includegraphics[width=\linewidth]{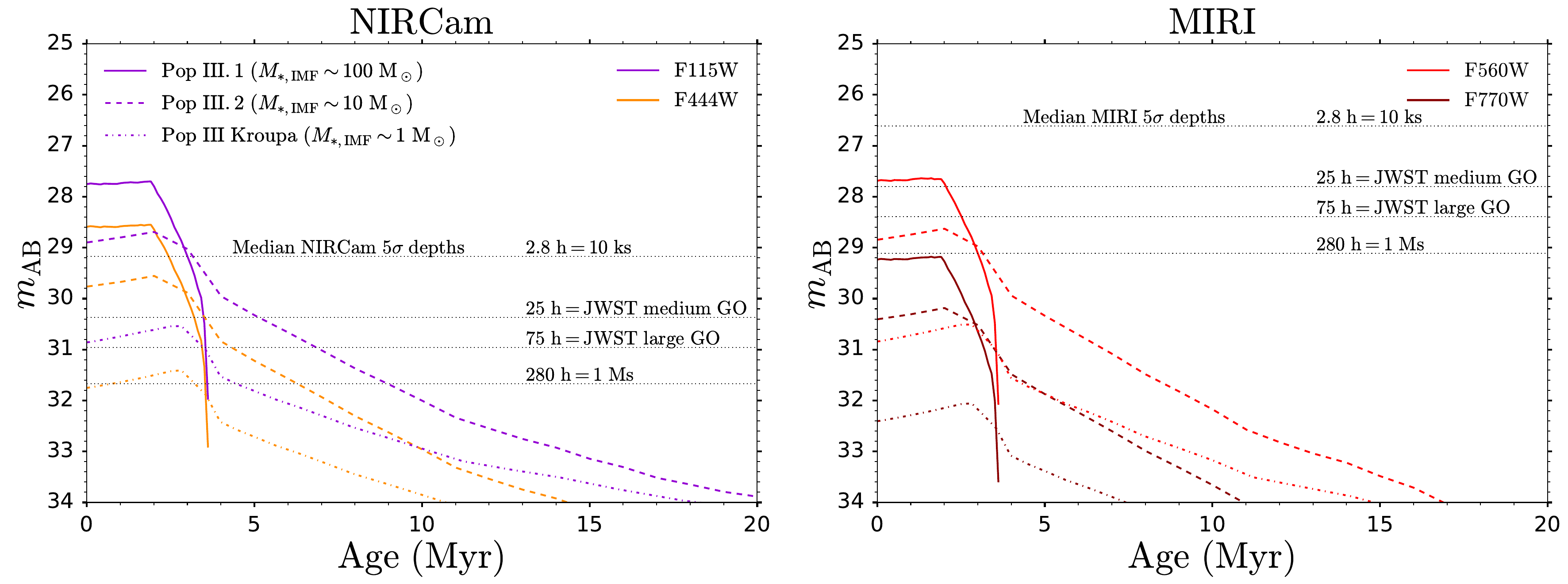}
\caption{The apparent magnitudes of $z=8$, $M_* = 10^6~\mathrm{M}_\odot$ Pop III galaxies with three different IMFs (Pop III.1: solid; Pop III.2: dashed; Pop III Kroupa: dash-dotted) as a function of time elapsed after an instantaneous starburst. Left panel: The bandpass-averaged flux densities in the NIRCam F115W (purple) and F444W (orange) filters. Right panel: The bandpass-averaged flux densities in the MIRI F560W (red) and F770W (dark red) filters. Shown also are the expected $5\sigma$ depths achieved (horizontal dotted lines) with NIRCam (left) and MIRI (right) in integration times of 2.8~h = 10~ks, 25~h = \emph{JWST} medium GO program, 75~h = \emph{JWST} large GO program and 280~h = 1~Ms. Note that the NIRCam depths reflect the median depth of the F115W, F150W, F200W, F277W, F335M, F356W, F410M and F444W filters. The MIRI depths reflect the median depth of the F560W and F770W filters. The elevated bandpass-averaged flux density in the F115W filter with respect to F444W (and the other NIRCam filters, which all have comparable flux densities, see Fig.\@~\ref{fig:popIII_spectrum} and Table~\ref{tab:imaging_observations}) is due to the high EW Ly$\alpha$ line which resides in this filter at $z=8$. Roughly speaking, the characteristic mass of stars $M_{*, \mathrm{IMF}}$ formed in the Pop III.1, Pop III.2 and Pop III Kroupa IMFs are $\sim$$100~\mathrm{M}_\odot$, $\sim$$10~\mathrm{M}_\odot$ and $\sim$$1~\mathrm{M}_\odot$, respectively.}
\label{fig:magnitudes_vs_time}
\end{figure*}

In Table~\ref{tab:imaging_observations} we show the expected AB magnitudes for Pop III galaxies at $z=8$ with $\log (M_*/\mathrm{M}_\odot) = 6$ immediately after ($0.01$~Myr) an instantaneous starburst. We also show the expected integration times (in hours) needed to achieve a $5\sigma$ detection within the various NIRCam + MIRI bands. For this estimation, we use the JWST Exposure Time Calculator \citep[ETC,][]{Pontoppidan2016}, assuming a point source, and adopting a 0.32~arcsec diameter circular extraction aperture for the NIRCam bands, and larger apertures for the MIRI F560W (0.49~arcsec) and MIRI 770W (0.55~arcsec) bands, obtained by scaling the 0.32~arcsec diameter by the ratio between the FWHM of the respective MIRI filter and the NIRCam/F444W filter. Furthermore, in Fig.\@~\ref{fig:magnitudes_vs_time}, we show how the apparent magnitudes of Pop III galaxies vary with time elapsed after an instantaneous starburst, showing both the bandpass-averaged flux densities in the F115W (purple) and F444W (orange) NIRCam filters (left panel), as well as the bandpass-averaged flux densities in the F560W (red) and F770W (dark red) MIRI filters (right). These are compared against the expected, median $5\sigma$ depths achieved (horizontal dotted lines) with NIRCam (left) and MIRI (right) in integration times of 2.8~h = 10~ks, 25~h = \emph{JWST} medium GO program, 75~h = \emph{JWST} large GO program and 280~h = 1~Ms.

Firstly, regarding detection with NIRCam, we see that Pop III galaxies with the more top-heavy III.1 and III.2 IMFs should be readily detectable at $5\sigma$ in medium-to-deep NIRCam surveys. On the other hand, Pop III galaxies with a Kroupa IMF will require extremely deep integrations ($\geq$ 100~h) to detect. 

Secondly, detections with MIRI imaging will prove to be challenging. Detections in the F560W band will require long integrations (10--100~h), even for the Pop III.1 and Pop III.2 IMFs. $5 \sigma$ detections in F560W for the Pop III Kroupa IMF, and in the F770W band for all three IMFs will likely be unachievable, given the exceptionally long integration times required (${\sim} 1000$~h).

However, the tabulated magnitudes and associated integration times assume an unlensed Pop III galaxy at the nominal stellar mass of $\log (M_*/\mathrm{M}_\odot) = 6$. If there is a flux boost $b=\mu \mathcal{M}$ for the Pop III galaxy, either because of a magnification factor $\mu$ from gravitational lensing, or because the galaxy is a multiple $\mathcal{M} = M_* / 10^6~\mathrm{M}_\odot$ of the nominal stellar mass then the required integration times will decrease. Assuming background-limited observations, the signal-to-noise S/N scales as $\mathrm{S/N} \propto f_\mathrm{cont} \sqrt{t}$, where $f_\mathrm{cont}$ is the continuum flux density in the appropriate \emph{JWST} imaging band. Hence with the flux boost $b$ the new continuum level is $f_\mathrm{cont}' = bf_\mathrm{cont}$, and the required integration times become $t' = t/b^2$. Thus with a flux boost factor $b=10$, a $5\sigma$ detection of F770W for Pop III.1 and Pop III.2 starts to become a real possibility, as the integration times needed get pushed down to ${\sim} 7$~h and ${\sim} 60$~h, respectively. Thus deep MIRI imaging on gravitationally lensed fields (with $\mu = 10$) and/or fortuitous imaging of massive Pop III galaxies (with $\log (M_*/\mathrm{M}_\odot) = 7$) will be necessary to get valuable flux constraints from the F770W filter. 

In Table~\ref{tab:imaging_observations} we also show the timescales over which the average magnitude for Pop III galaxies in the listed \emph{JWST} bands drops by 0.25 and 1.0 mag (following an instantaneous starburst), as an indication of the visibility window over which Pop III galaxies can be detected. Depending on the Pop III IMF, we see that the magnitudes remain roughly stable ($\Delta m = 0.25$) for 2.2--4.0~Myr, with their brightness having faded more strongly ($\Delta m = 1.0$) after 2.8--5.6~Myr. As we shall see in the next section on Pop III colour selection, Pop III galaxies maintain unique colours in the colour–colour plane over much longer timescales than the aforementioned visibility windows. Hence the limiting factor determining the timescale over which Pop III galaxies can be identified is their visibility window (${\sim}3$~Myr), rather than the time window over which they exhibit unique colours (${\sim}$10--25~Myr).

\begin{figure*}
\centering
\includegraphics[width=.65\linewidth]{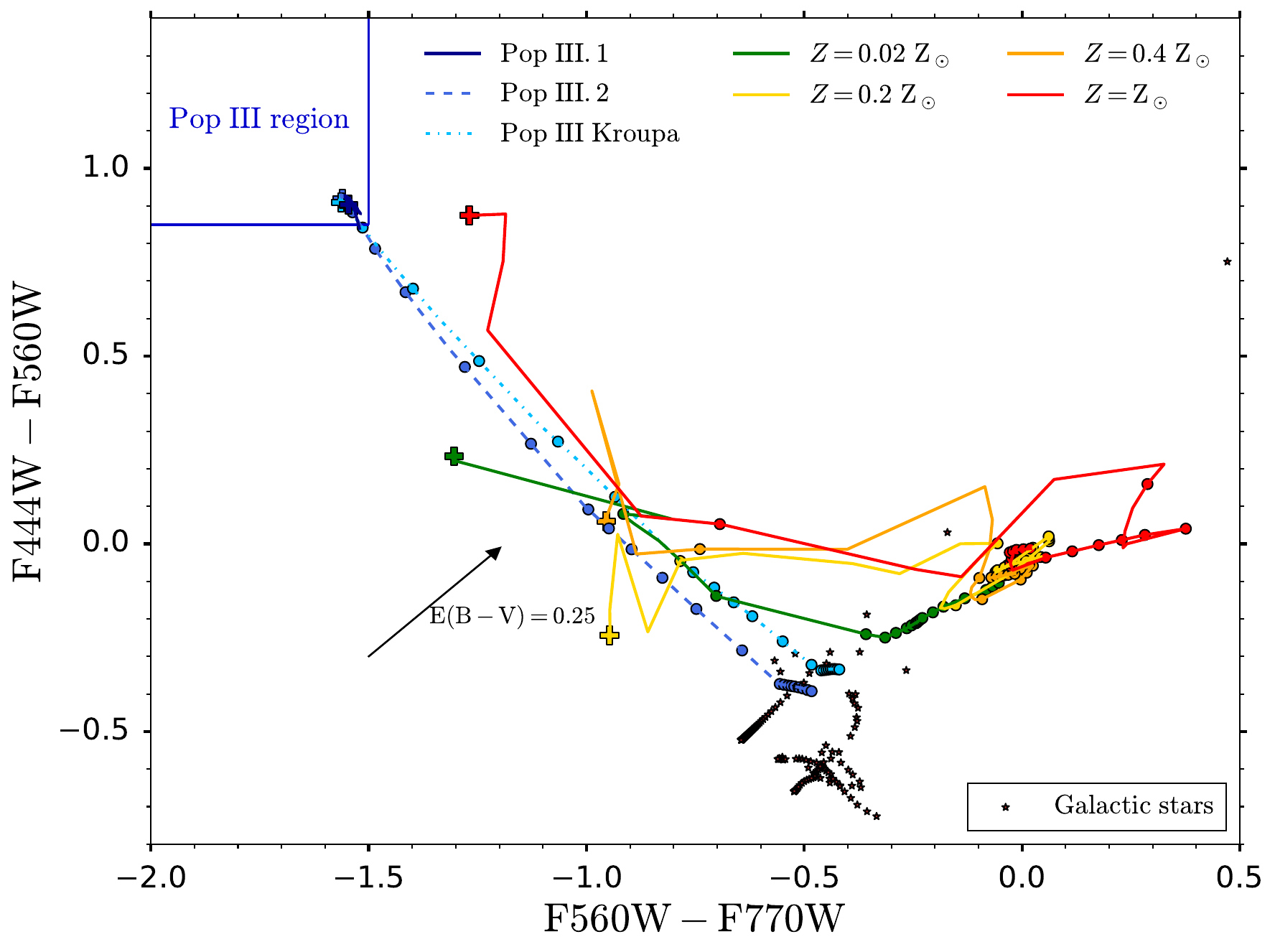}
\caption{The F444W$-$F560W, F560W$-$F770W colour--colour plane for selecting $z\sim8$ Pop III galaxies. The basis behind this selection is the marginally higher H$\alpha$ EW for Pop III galaxies in the F560W filter, together with the lack of [\ion{O}{III}] $\lambda\lambda 4959, 5007$ and [\ion{S}{III}] $\lambda\lambda 9069, 9531$ emission in the F444W and F770W filters, respectively. We show the colours for $z=8$ Pop III galaxies (different shades of blue) for three different IMFs: Pop III.1 (dark blue, solid), Pop III.2 (blue, dashed) and Pop III Kroupa (light blue, dash-dotted). Also shown are the colours for the $z=8$ non-Pop III galaxies with $Z = 0.02$~$\mathrm{Z}_\odot$ (green), $Z = 0.2$~$\mathrm{Z}_\odot$ (yellow), $Z = 0.4$~$\mathrm{Z}_\odot$ (orange), $Z = \mathrm{Z}_\odot$ (red). Plus symbols represent the galaxy colours immediately after an instantaneous starburst, while the circles show the colours after subsequent 5~Myr intervals up to 100~Myr. Note that the tracks for Pop III.1 galaxies (which overlap with Pop III.2 and Pop III Kroupa) are very short (and thus difficult to see in this Figure and other colour--colour diagrams), as these dim rapidly after a starburst (see Fig.\@~\ref{fig:magnitudes_vs_time}) and thus have no tabulated values beyond 3.6~Myr. The expected colours of Galactic stars \citep[for both theoretical SEDs for very low-mass stars and brown dwarfs from][as well as simple blackbody spectra]{Chabrier2000} are indicated by the brown stars. The shift in colours due to dust reddening, assuming the \citet{Calzetti2000} attenuation law, is also given, with the length of the dust vector representing the colour shift associated with $\mathrm{E(B}-\mathrm{V}) = 0.25$. Pop III galaxies clearly occupy a unique region within this colour--colour plane, the boundaries of which are given by the solid blue lines (see text for more details on the Pop III region), and thus potential Pop III candidates can be identified through a colour selection. This colour--colour selection is applicable over the redshift range $7.55 < z < 8.10$. If contamination from $Z = \mathrm{Z}_\odot$ galaxies is not deemed a concern, this selection can be applied over the wider redshift range $7.00 < z < 8.35$.}
\label{fig:zackrisson_colours1}
\end{figure*}

\subsection{Pop III galaxy colour selection} \label{subsec:colour_selection}

In this section we introduce colour selections that can be applied to identify $z\sim8$ Pop III candidates based off of their unique positions within colour--colour planes.  In Section~\ref{subsubsec:colour1} we discuss what we believe to be the optimal Pop III colour selection, which is based off of both NIRCam and MIRI photometry. In Section~\ref{subsubsec:colour_extra}, we discuss an alternative NIRCam+MIRI colour selection that can be more readily applied. In Sections~\ref{subsubsec:colour2} and \ref{subsubsec:colour3} we discuss colour selections that only require NIRCam photometry. In Section~\ref{subsec:continuum_slope}, we briefly examine how effective measurements of the continuum slope, as inferred from photometry, will be at identifying Pop III candidates. Finally, in Section~\ref{subsec:He_II_selection}, we discuss the prospects for identifying Pop III galaxies through the imprint their high EW \ion{He}{II} $\lambda$1640 emission leaves on NIRCam medium-band photometry.

\subsubsection{[\ion{O}{III}]$-$H$\alpha$, H$\alpha -$[\ion{S}{III}] colour selection} \label{subsubsec:colour1}

\subsubsection*{Colour selection}

We show the F444W$-$F560W, F560W$-$F770W colour--colour plane in Fig.\@~\ref{fig:zackrisson_colours1}. We note that this colour selection was also advocated for in \citet{Zackrisson2011}. $z = 8$ Pop III galaxies are shown in blue, with the Pop III.1, Pop III.2 and Pop III Kroupa IMFs given by the dark blue solid, blue dashed and light blue dash-dotted curves, respectively. The $z = 8$ non-Pop III galaxies are shown in non-blue colours, with the $Z = [0.02, 0.2, 0.4, 1]$~$\mathrm{Z}_\odot$ metallicities, given by the green, yellow, orange and red points, respectively. The plus symbols represent the galaxy colours immediately after an instantaneous starburst, while the circles show the colours after subsequent 5~Myr intervals. We see that the Pop III galaxies occupy a unique region within the colour--colour plane immediately after the instantaneous starburst. Indeed, these unique colours are maintained for 10--25~Myr (thus much longer than the ${\sim}3$~Myr visibility windows following an instantaneous starburst discussed in Section~\ref{subsubsec:observability}). Furthermore, the non-Pop III galaxies do not overlap within this region, even long after the initial starburst. The basis behind this colour selection is as follows.

At $z=8$, the H$\alpha$ line falls in the MIRI F560W filter. As briefly outlined in Section~\ref{subsec:popIII_features}, Pop III galaxies have higher Balmer recombination line luminosities per unit stellar mass. However, what is also the case is that the total (stellar+nebular) continuum is also higher per unit stellar mass. The net result, which will be discussed more in Section~\ref{sec:spectroscopy}, is that the H$\alpha$ equivalent width is only marginally higher (${\sim}0.1$~dex) for Pop III galaxies. This therefore results in only a marginally higher (${\sim}0.15$~mag) ``magnitude excess'' in the F560W filter (akin to the more familiar IRAC excess), compared to for non-Pop III galaxies.

Additionally, at $z=8$, the [\ion{S}{III}] $\lambda\lambda 9069,\ 9531$ doublets are redshifted into the MIRI F770W band. Although these lines are not very bright in terms of absolute brightness, they are bright relative to the rest-frame NIR continuum, having equivalent widths comparable to the more well-known [\ion{O}{III}] $\lambda\lambda 4959,\ 5007$ doublets in the rest-frame optical. Therefore the [\ion{S}{III}] doublets have a comparable effect on broadband photometry to the [\ion{O}{III}] doublets, generating a substantial magnitude excess in the F770W filter. $z=8$ Pop III galaxies, with their marginally higher H$\alpha$ EWs in F560W and lack of [\ion{S}{III}] in F770W, thus have bluer F560W$-$F770W colours compared to non-Pop III galaxies.

Finally, at $z=8$, the [\ion{O}{III}] $\lambda\lambda 4959,\ 5007$ doublet is redshifted into the NIRCam F444W band. This doublet drives a substantial magnitude excess in the F444W filter for non-Pop III galaxies. As a result, Pop III galaxies, with their lack of [\ion{O}{III}] in F444W, and their marginally higher H$\alpha$ EWs in F560W thus have redder F444W$-$F560W colours compared to non-Pop III galaxies.

\subsubsection*{Contamination}

It is therefore the presence (or absence) of bright emission lines that is driving our colour selection in Fig.\@~\ref{fig:zackrisson_colours1}. The reason why both $Z = 0.02 ~\mathrm{Z}_\odot$ and $Z = \mathrm{Z}_\odot$ galaxies have the most similar colours to Pop III is because their [\ion{O}{III}] and [\ion{S}{III}] emission lines have low EW. In the former case this is because of the lack of metals, while in the latter it is due to the lack of ionisation and/or heating of the ISM. Thus, whilst Pop III galaxies occupy a unique region within this colour--colour plane, they can start to become confused with $Z = 0.02~\mathrm{Z}_\odot$ and $Z = \mathrm{Z}_\odot$ galaxies once observational errors on the measured colours are taken into account. Indeed, with only a $5\sigma$ detection in each filter comprising the colour pair, the colour uncertainty is $\sigma_C = 0.28$, and there can be some contamination of the Pop III region within the colour--colour plane. This contamination is largely eliminated with $10\sigma$ detections within each filter, but given the challenge of detecting Pop III galaxies with MIRI (see Section~\ref{subsubsec:observability}), this seems impractical to achieve. 

Our colour selections in this work, with the exception of F444W$-$F560W, have been chosen such that Pop III galaxies exhibit \emph{bluer} colours than non-Pop III galaxies. This approach has been adopted for two reasons. 

Firstly, it ensures that non-Pop III galaxies, which can be reddened by dust, become even further separated from Pop III galaxies in the colour--colour plane. The shift in colour--colour caused by such reddening is shown by the reddening vector in each of our colour--colour diagrams. 

Secondly, it ensures that the typical \emph{red} contaminants that one encounters in high-$z$ searches, such as Galactic brown dwarfs, Balmer break galaxies and dusty galaxies are not a concern. Instead, one must instead worry about potentially \emph{blue} contaminants. At the long wavelengths typically probed by our colour selections, we capture the light redward of the blackbody peak in all but the coolest Galactic stars. Hence these stars appear blue in e.g.\@ the F444W$-$F560W and F560W$-$F770W colours. As can be seen from Fig.\@~\ref{fig:zackrisson_colours1}, there is no risk of confusing Pop III galaxies with Galactic stars, which are shown by the small brown star symbols and correspond to theoretical SEDs for very low-mass stars and brown dwarfs from \citet{Chabrier2000}, as well as simple blackbody spectra spanning the temperature range $500 < T$ (K) $ < 50000$. 

The other potential contaminants would be blue, lower-$z$ galaxies. In order to replicate the very blue F560W$-$F770W colours seen, these systems would have to have high EW emission lines that lie beyond H$\alpha$ in the rest-frame. We found that the Paschen recombination lines (in the \citealt{Zackrisson2011} models) simply do not have sufficient EW to achieve this, hence low-$z$ contaminants are not a concern for this colour--colour selection. Indeed, assuming an otherwise flat $f_\nu$ SED, the magnitude excess $\Delta m$ driven by emission lines in e.g.\@ the F560W filter (yielding a blue F560W$-$F770W colour) is related to the total rest-frame equivalent width $\mathrm{EW}_\mathrm{tot, rest}$ of all the emission lines that reside in the F560W filter, via $\Delta m = -2.5 \log_{10} (1 +\mathrm{EW}_\mathrm{tot, rest}(1+z)/\Delta \lambda)$, where $\Delta \lambda$ is the bandpass width of the filter. Hence a given observed blue F560W$-$F770W colour demands an increasingly higher rest-frame EW for contaminants at lower redshifts, which makes such low-$z$ contaminants (such as $z\sim2$ Pa$\alpha$ emitters, which would require $3\times$ the rest-frame EW of $z\sim8$ H$\alpha$ emitters) unlikely in this case.

Regarding potential contaminants at \emph{even higher} redshift, the only possibility would be $z\sim10$ galaxies with bright [\ion{O}{III}] emission in the F560W filter. However, as can be seen in Appendix~\ref{app:z10}, although such galaxies can mimic the red F444W$-$F560W colours seen in $z \sim 8$ Pop III galaxies, they have relatively flat F560W$-$F770W colours (due to strong H$\alpha$ emission in the F770W filter) compared to the blue colours seen in $z \sim 8$ Pop III galaxies. Hence there is no risk of confusing $z\sim8$ Pop III galaxies with $z\sim10$ [\ion{O}{III}] emitters in the F444W$-$F560W, F560W$-$F770W colour--colour plane.

\subsubsection*{Redshift range of applicability}

We now comment on the redshift range over which these colour selections can be applied. In principle, this redshift range is set by the redshift interval over which the emission lines that drive the colour selection fall within the intended filters. In practice however, the usable redshift range is narrower, as other (bright) emission lines get redshifted into and out of the adopted filters. This causes the locus of points occupied by Pop III and non-Pop III galaxies to drift in the colour--colour plane, causing overlap between Pop III and non-Pop III, and thus potential contamination. 

We therefore find that the F444W$-$F560W, F560W$-$F770W colour selection can safely be applied over the redshift range $7.55 < z < 8.10$. Beyond this redshift range $Z = \mathrm{Z}_\odot$ galaxies begin to occupy the Pop III region within the colour--colour plane. However, if such high metallicity galaxies are not deemed a concern at $z\sim8$, as one might not expect to find such enriched galaxies at these high redshifts \citep[see e.g.\@][]{Maiolino2008}, then our colour selection can be applied over the broader redshift range $7.00 < z < 8.35$. A reliable assessment of the photometric redshift of Pop III candidates from their Ly$\alpha$ breaks will therefore likely be important to rule out potential high-$z$ non-Pop III contaminants.

\subsubsection*{Pop III region}

The solid blue lines in Fig.\@~\ref{fig:zackrisson_colours1} denote our boundaries for what we define to be the ``Pop III region'' of the F444W$-$F560W, F560W$-$F770W colour--colour plane. This is the range of colours that are uniquely exhibited by Pop III galaxies within the redshift range of applicability for this colour selection. When defining the extent of the Pop III region, we only consider the colours displayed by Pop III galaxies up to 5~Myr after the initial starburst, as beyond this timescale Pop III galaxies (and their associated colours) are likely too faint to be observed with \emph{JWST} (see Section~\ref{subsubsec:observability}). This is why the Pop III region does not encompass the full range of Pop III tracks in the F444W$-$F560W, F560W$-$F770W colour--colour plane. In defining the Pop III region, we have aimed to be as restrictive as possible, by keeping the Pop III region as small as possible. Our priority is to minimise the risk of contamination, which comes at the cost of completeness, as we will likely miss some real Pop III galaxies in the \emph{JWST} data (as these get scattered out of the Pop III region by observational error or model uncertainties). We leave the boundaries of the Pop III region open-ended if we deem there to be little risk of contamination on the open side.

The boundaries of the Pop III region in the F444W$-$F560W, F560W$-$F770W colour--colour plane are given by: 
\begin{equation}
    \begin{aligned}
    &\mathrm{F444W}-\mathrm{F560W} > 0.85\\
    &\mathrm{F560W}-\mathrm{F770W} < -1.5 
\end{aligned}
\end{equation}

If contamination by $Z=\mathrm{Z}_\odot$ galaxies is not deemed a concern, these boundaries can be extended to:
\begin{equation}
    \begin{aligned}
    &\mathrm{F444W}-\mathrm{F560W} > 0.75\\
    &\mathrm{F560W}-\mathrm{F770W} < -1.25 
\end{aligned}
\end{equation}

We wish to stress that the identification of potential Pop III candidates through an application of the Pop III region criterion should be done with caution. Within the redshift range of applicability, the only galaxies (within the \citealt{Zackrisson2011} models) that can exhibit colours within the Pop III region of the colour--colour plane are Pop III galaxies (by definition, and ignoring observational error). However, as discussed earlier, it is possible for \emph{non-}Pop III galaxies \emph{outside} of this redshift range (and Galactic stars) to have colours that lie \emph{within} the Pop III region. Whenever applicable, we discuss these potential contaminants in the text and provide recommendations on how to distinguish them from actual Pop III galaxies. Additionally, the Pop III regions defined in this work were established by only considering the (limited) set of metallicities and galaxy properties probed by the \citet{Zackrisson2011} models. Hence it is possible that the Pop III regions we have defined suffer from more contamination than what is explored here. Additionally, owing to model uncertainties, different Pop III (and non-Pop III) models likely predict different locations for the Pop III region in the colour--colour plane.

\begin{figure*}
\centering
\includegraphics[width=.65\linewidth]{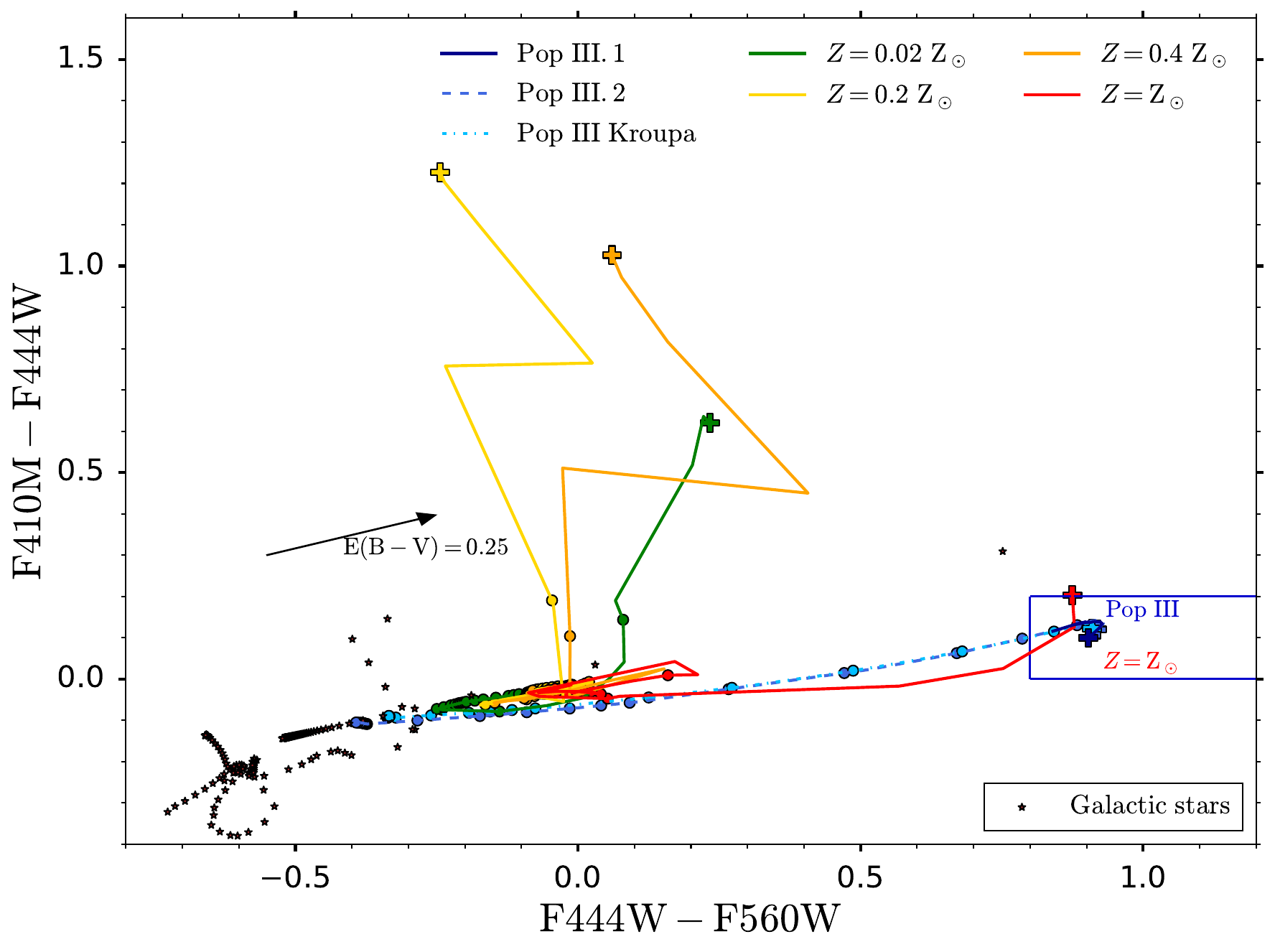}
\caption{Similar to Fig.\@~\ref{fig:zackrisson_colours1}, but now showing the F410M$-$F444W, F444W$-$F560W colour--colour plane. The basis behind this selection is similar to Fig.\@~\ref{fig:zackrisson_colours1}, though the MIRI F770W filter (which was probing [\ion{S}{III}]) has been substituted with the much more sensitive NIRCam F410M filter (which is now probing the continuum level around [\ion{O}{III}] at $z\sim8$, although F356W can also be adopted). Hence this colour selection is more practical to apply, as the integration times needed to detect Pop III galaxies in all of the adopted filters (most notably F560W) are much shorter. Therefore, this colour selection does not require as substantial of a flux boost from gravitational lensing and/or elevated Pop III stellar masses to be applied. The drawback of this colour selection is the potential confusion between Pop III galaxies and $Z=\mathrm{Z}_\odot$ galaxies, which have comparable colours. This redshift selection can be applied over the redshift interval $7.90 < z < 8.30$, with contamination by  $Z=\mathrm{Z}_\odot$ galaxies being a potential concern over this entire redshift range. The alternative colour selection, which uses F356W rather than F410M (i.e.\@ F356W$-$F444W, F444W$-$F560W) can be applied over the wider redshift range $7.20 < z < 8.30$, but still suffers contamination by $Z = \mathrm{Z}_\odot$ galaxies.}
\label{fig:zackrisson_colours_extra}
\end{figure*}

\subsubsection*{Caveats}

Finally, we comment on the robustness of these colour selections. In Appendix~\ref{app:nakajima_colours} we show the same colour--colour plane but applied to the \citet{Nakajima2022} models. Qualitatively, we obtain similar results to what was obtained using the \citet{Zackrisson2011} models. There are minor quantitative differences regarding the positions of Pop III galaxies in the colour--colour plane, owing to differences in the spectral shapes and the EWs of the relevant emission lines between the \citet{Zackrisson2011} and \citet{Nakajima2022} models. As the \citet{Nakajima2022} models extend below $Z = 0.02 ~\mathrm{Z}_\odot$, we find that our adopted colour selections likely also pick up $Z \leq \mathrm{Z}_\odot /140$ galaxies (even in the absence of any observational error). However, as we will discuss in Section~\ref{sec:spectroscopy}, deep follow-up spectroscopy will enable us to definitively distinguish between Pop III and very metal-poor galaxies. 

We note that our colour selections assume a covering fraction $f_\mathrm{cov} = 1$. Since these colour selections are primarily driven by the presence (or absence) of bright emission lines in the adopted filters, the resulting colours are sensitive to the strength of the emission lines and thus the covering fraction. For example, if we instead assume a covering fraction $f_\mathrm{cov} = 0.5$, then the emission line fluxes and equivalent widths will be smaller, resulting in a shift of the Pop III and non-Pop III regions in the colour--colour plane. Since the offset between Pop III and non-Pop III galaxies is due to their difference in emission line strength, this will also result in a reduced separation between Pop III and non-Pop III galaxies in the colour--colour plane. Therefore, while our colour selections are in principle effective at identifying Pop III candidates with $f_\mathrm{cov} = 1$, they may miss Pop III galaxies with lower covering fractions, as these could potentially overlap with the non-Pop III regions in our $f_\mathrm{cov} = 1$ colour--colour planes. 

\subsubsection{[\ion{O}{III}], [\ion{O}{III}]$-$H$\alpha$ colour selection} \label{subsubsec:colour_extra}

Due to the weaker sensitivity of the MIRI F770W filter, together with the lack of bright H/He emission lines in this filter at $z\sim8$ (as H$\alpha$ instead resides in the MIRI F560W filter at $z\sim8$) and lower continuum level, it will likely be very challenging to actually detect Pop III galaxies in the MIRI F770W filter (see Table~\ref{tab:imaging_observations} and Fig.\@~\ref{fig:magnitudes_vs_time}). Thus, we introduce an alternative colour selection in Fig.\@~\ref{fig:zackrisson_colours_extra}, that is still rather robust (barring any contamination from $Z=\mathrm{Z}_\odot$ galaxies), but much more practical to apply in practice. We substitute the MIRI F770W filter for the much more sensitive NIRCam F410M filter (though the F356W filter can also be used). Thus this colour selection does not require as considerable of a flux boost from gravitational lensing or elevated Pop III masses to be applied. Here we use the F410M$-$F444W, F444W$-$F560W filter pairs for our colour--colour selection. The basis behind this Pop III colour selection is as follows. 

As outlined earlier, [\ion{O}{III}] $\lambda 5007$ resides in the F444W filter at $z=8$. The F410M filter is chosen because it contains no bright metal emission lines and just measures the continuum level. Hence at this redshift the F410M$-$F444W colour is sensitive to the equivalent width of [\ion{O}{III}]. This colour is therefore red for non-Pop III galaxies and relatively flat for Pop III galaxies as there is no oxygen. As before, H$\alpha$ resides in the F560W filter, with the F444W$-$F560W colour being relatively red for $z\sim8$ Pop III galaxies.

We note that the F356W filter could be adopted instead of F410M. The benefit of selecting the wide-band filter is that it is more sensitive and can be used over a wider redshift interval. Indeed, [\ion{O}{III}] $\lambda 5007$ resides in the F410M filter at $6.8 < z < 7.6$, which drives \emph{blue} (rather than red) F410M$-$F444W colours for non-Pop III galaxies in that redshift range. Accounting for the effects of dust reddening and observational error on the measured colours, as well as uncertainties on the measured photometric redshifts, such galaxies could be confused with $z\sim8$ Pop III galaxies, which is why the F410M filter can only be used over a narrower redshift interval. The drawback of using the alternative F356W filter is that it also contains the [\ion{O}{II}] $\lambda\lambda 3726,\ 3729$ lines at $z\sim8$, which therefore diminishes the impact of [\ion{O}{III}] in driving redder colours in non-Pop III galaxies, thus reducing the separation between Pop III and non-Pop III galaxies in the colour--colour plane.

The F410M$-$F444W, F444W$-$F560W colour selection can be applied over the redshift interval $7.90 < z < 8.30$. Owing to their similar F410M$-$F444W and F444W$-$F560W colours to Pop III galaxies, contamination from $Z = \mathrm{Z}_\odot$ galaxies may be a concern over this entire redshift range. The alternative F356W$-$F444W, F444W$-$F560W colour selection can be applied over the wider redshift range $7.20 < z < 8.30$ (but still suffers contamination by $Z = \mathrm{Z}_\odot$  galaxies). 

Furthermore, [\ion{O}{III}] emitters at $9.15 < z < 11.40$ are also likely contaminants, due to their strong [\ion{O}{III}] emission in the F560W filter, together with their relatively flat F410M$-$F444W colours, thus mimicking the colours of $z\sim8$ Pop III galaxies. Given that these [\ion{O}{III}] emitters are at higher redshift, they should in principle (barring any strong Ly$\alpha$ emission, see the next section) exhibit a stronger Ly$\alpha$ break in the F115W filter, being partial or even full F115W dropouts (F115W$-$F150W $\geq 1.5$), as opposed to the $z\sim8$ Pop III galaxies which should only exhibit minor IGM attenuation (F115W$-$F150W $\leq 0.5$) in the F115W filter. 

The boundaries of the Pop III, $Z=\mathrm{Z}_\odot$ region in the F410M$-$F444W, F444W$-$F560W colour--colour plane are given by: 
\begin{equation}
    \begin{aligned}
   0.0 < \mathrm{F410M}-\mathrm{F444W} < 0.2\\
    \mathrm{F444W}-\mathrm{F560W} > 0.8
\end{aligned}
\end{equation}

The boundaries of the Pop III, $Z=\mathrm{Z}_\odot$ region in the alternate F356W$-$F444W, F444W$-$F560W colour--colour plane are instead: 
\begin{equation}
    \begin{aligned}
   -0.15 < \mathrm{F356W}-\mathrm{F444W} < 0.1\\
    \mathrm{F444W}-\mathrm{F560W} > 0.8
\end{aligned}
\end{equation}

\subsubsection{[\ion{O}{III}], Ly$\alpha$ colour selection} \label{subsubsec:colour2}

\begin{figure*}
\centering
\includegraphics[width=.65\linewidth]{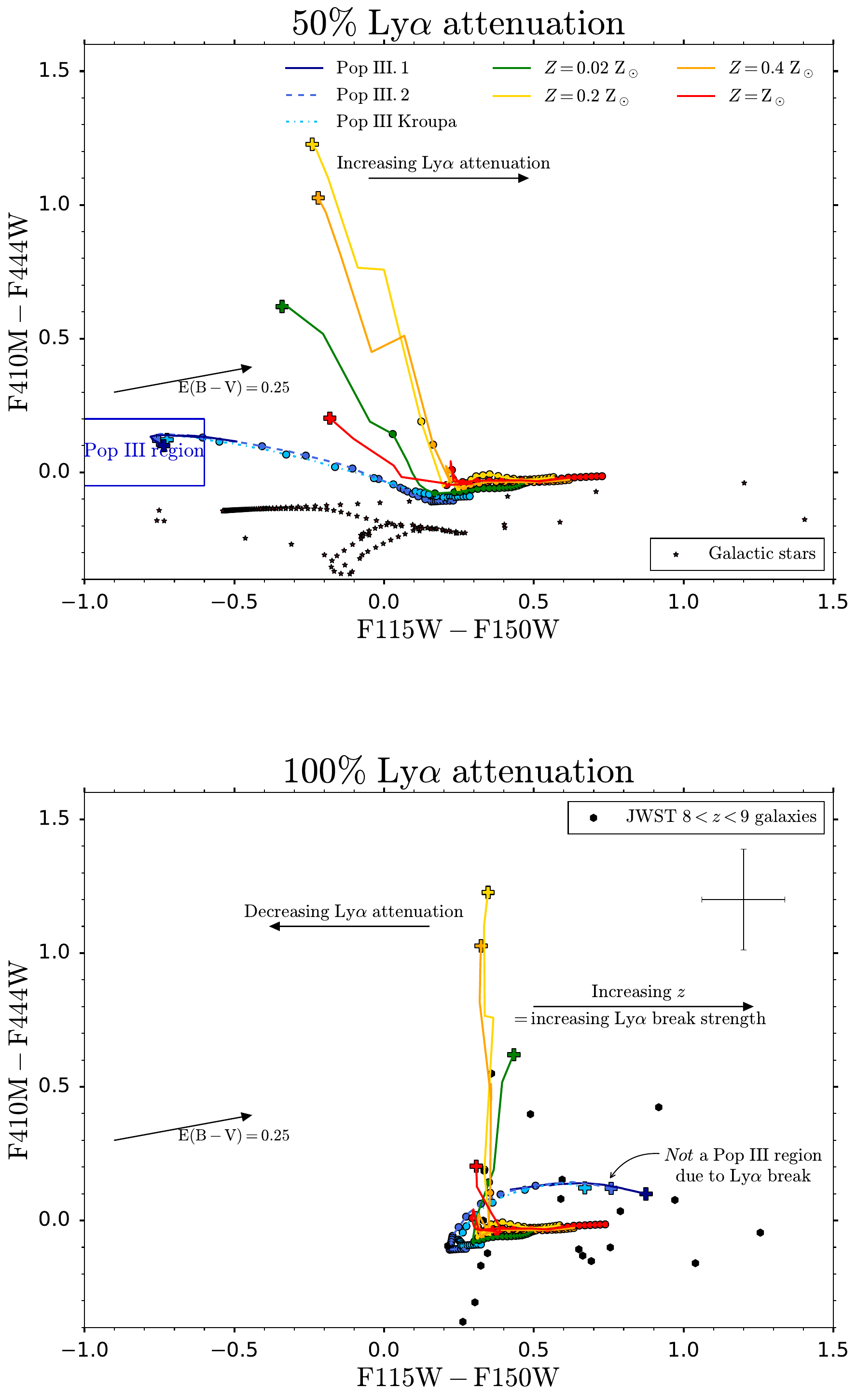}
\caption{Top panel: Similar to Fig.\@~\ref{fig:zackrisson_colours1}, but now showing the F410M$-$F444W, F115W$-$F150W colour--colour plane. The basis behind this selection is the substantially higher Ly$\alpha$ EW for Pop III galaxies in the F115W filter, together with the lack of [\ion{O}{III}] $\lambda 5007$ emission in the F444W filter. This colour selection is applicable over the redshift range $8.0 < z < 9.0$. Note that the \citet{Inoue2014} IGM attenuation has been applied, which effectively removes all flux blueward of the Ly$\alpha$ central wavelength. However, no attenuation/scattering redward of the Ly$\alpha$ centre has been applied (i.e.\@ the Ly$\alpha$ attenuation is 50\%). Hence in practice the Ly$\alpha$ line may be much more heavily attenuated than in our models (except when Pop III galaxies form in the vicinity of galaxies/quasars that have already carved out ionised bubbles that enable Ly$\alpha$ to escape), which will erode this Pop III signature and thus may render this colour selection ineffective at identifying Pop III galaxies from observational data. Bottom panel: The F410M$-$F444W, F115W$-$F150W colours when 100\% of the Ly$\alpha$ emission is attenuated. The F115W$-$F150W colours become more red due to the lack of Ly$\alpha$ emission in the F115W filter. We stress that the relatively red F115W$-$F150W colours for $z=8$ Pop III galaxies with 100\% Ly$\alpha$ attenuation should \emph{not} be used as a Pop III indicator, as F115W$-$F150W is also sensitive to the strength of the Ly$\alpha$ break and thus increases with increasing redshift, causing Pop III and non-Pop III galaxies to overlap in F115W$-$F150W colour. We also show the colour of $8 < z < 9$ galaxy candidates (black hexagons) observed with \emph{JWST} in the CEERS field, and the median error bar on their colours. The colours of these candidates are consistent with strong attenuation of Ly$\alpha$, and thus the F410M$-$F444W, F115W$-$F150W colour selection cannot be used to distinguish between Pop III and non-Pop III galaxies.}
\label{fig:zackrisson_colours2}
\end{figure*}

\subsubsection*{Colour selection}

\begin{figure*}
\centering
\includegraphics[width=.65\linewidth]{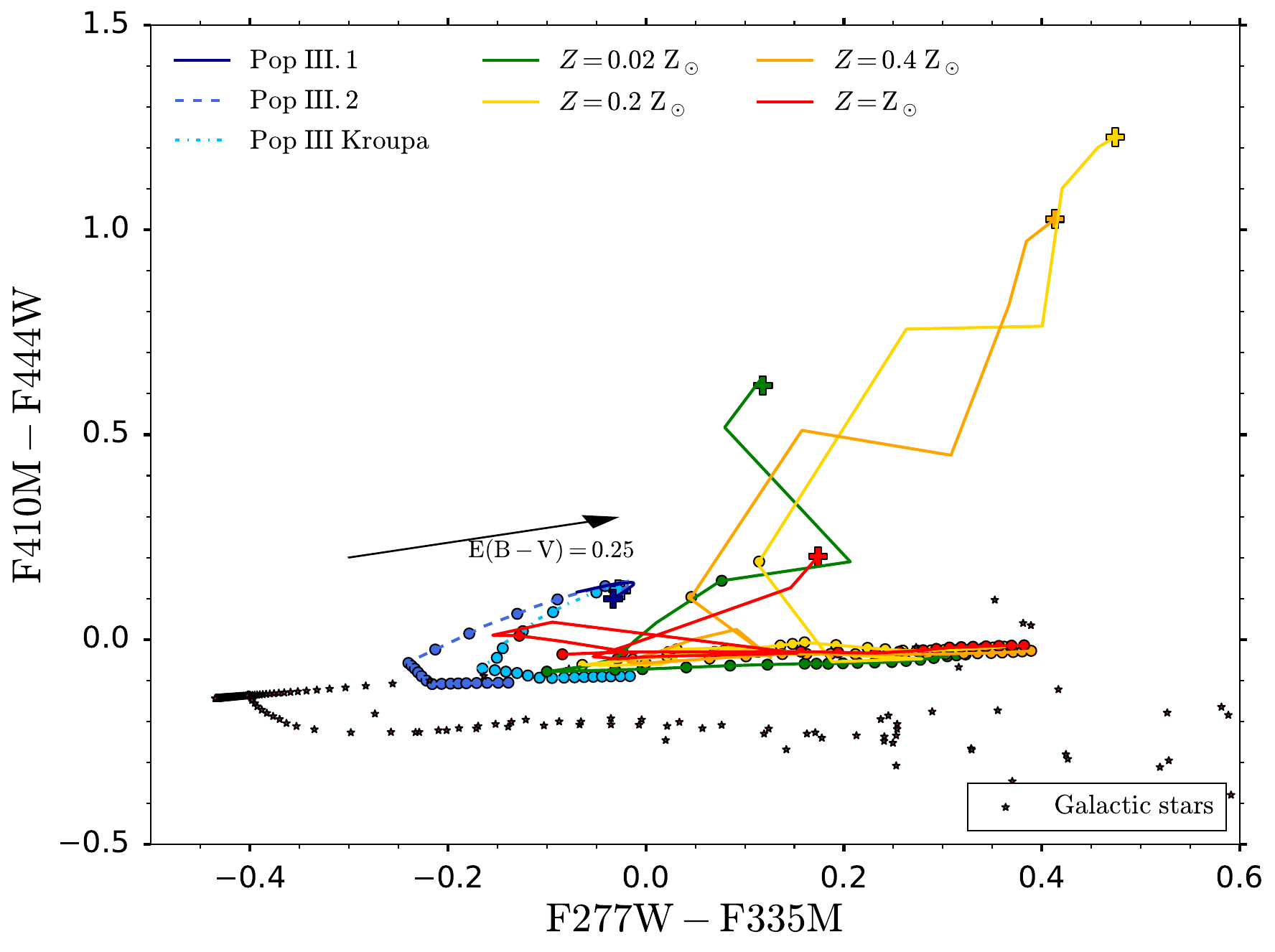}
\caption{Similar to Fig.\@~\ref{fig:zackrisson_colours1}, but now showing the F410M$-$F444W, F277W$-$F335M colour--colour plane. The basis behind this selection is the lack of [\ion{O}{III}] $\lambda 5007$ and [\ion{O}{II}] emission for Pop III galaxies in the F444W and F335M filters, respectively. This colour selection is applicable over the redshift range $7.95 < z < 8.35$, and is readily contaminated by other sources with flat colours. Owing to the limitations with this colour selection, we do not define a Pop III region to identify potential Pop III candidates in this colour--colour plane.}
\label{fig:zackrisson_colours3}
\end{figure*}

Given NIRCam's greater sensitivity and imaging footprint (9.7~arcmin$^2$ vs\@ 2.35~arcmin$^2$) compared to MIRI, we now discuss alternative colour selections that only require NIRCam photometry. We show our F410M$-$F444W, F115W$-$F150W colour selection in Fig.\@~\ref{fig:zackrisson_colours2}. The basis behind this colour selection is as follows.



At $z=8$, the Ly$\alpha$ line falls within the F115W filter. Unlike the Balmer recombination lines, and which we shall discuss more extensively in Section~\ref{sec:spectroscopy}, the equivalent width of Ly$\alpha$ is substantially higher for Pop III galaxies compared to non-Pop III. Hence the magnitude excess in the F115W filter caused by Ly$\alpha$  is much greater for Pop III galaxies. The F150W filter is chosen to measure the neighbouring continuum level, with the F115W$-$F150W colour thus being sensitive to the EW of Ly$\alpha$. Hence Pop III galaxies will have bluer F115W$-$F150W colours than non-Pop III galaxies as they have a larger Ly$\alpha$ EW.

This colour selection can be applied across the redshift range $8 < z < 9$. The alternative F356W$-$F444W, F115W$-$F150W colour selection can be applied over the wider redshift range $7.50 < z < 9.00$. 

The boundaries of the Pop III region in the F410M$-$F444W, F115W$-$F150W colour--colour plane are given by: 
\begin{equation}
    \begin{aligned}
    -0.05 < \mathrm{F410M}-\mathrm{F444W} < 0.20\\
    \mathrm{F115W}-\mathrm{F150W} < -0.6
\end{aligned}
\end{equation}

The boundaries of the Pop III region in the alternative F356W$-$F444W, F115W$-$F150W colour--colour plane are instead: 
\begin{equation}
    \begin{aligned}
   -0.15 < \mathrm{F356W}-\mathrm{F444W} < 0.10\\
    \mathrm{F115W}-\mathrm{F150W} < -0.6 
\end{aligned}
\end{equation}

\subsubsection*{Contamination}

Galactic stars are not likely potential contaminants. Any stars that have comparable F410M$-$F444W, F115W$-$F444W colours to Pop III galaxies (such as those in the lower left region of Fig.\@~\ref{fig:zackrisson_colours2}) should be able to be identified as such from an inspection of their full SED. Additionally, starburst galaxies at $z\sim0.75$ with bright H$\alpha$ emission in the F115W filter can mimic the colours of $z\sim8$ Pop III galaxies. However, continuum detections in F090W and bluer, non-\emph{JWST} filters should enable one to distinguish between $z\sim8$ galaxies (which are non-detected in those bands) and these low-redshift interlopers.

\subsubsection*{Ly$\alpha$ attenuation}

It should be noted that the colours shown in the top panel of Fig.\@~\ref{fig:zackrisson_colours2} were obtained by applying the \citet{Inoue2014} prescription for IGM attenuation to the \citet{Zackrisson2011} spectra. This IGM attenuation essentially removes all flux blueward of the central wavelength of Ly$\alpha$. However, it does not account for any scattering/attenuation of the flux redward of the Ly$\alpha$ peak, which can substantially diminish the total line flux that will actually be seen in observations. Indeed, \citet{Castellano2022} showed that galaxies in the epoch of reionisation first need to carve out an ionising bubble of radius 1~Mpc before Ly$\alpha$ is able to effectively escape. Given the Pop III visibility window (following an instantaneous starburst) of ${\sim}3$~Myr, this is insufficient time for ionising photons to traverse such a distance, let alone completely ionise the gas enclosed in this volume. Hence the true Ly$\alpha$ flux observed will likely be substantially less than what was adopted to generate the F115W$-$F150W colours in Fig.\@~\ref{fig:zackrisson_colours2}. Hence in practice this colour selection may not be effective at identifying Pop III galaxies, as we will no longer be sensitive to the \emph{intrinsic} Ly$\alpha$ EW. 

We do note, however, that in most models, Pop III galaxies form because minihalos ($M_\mathrm{h} \sim 10^5$--$10^6~\mathrm{M}_\odot$, which are otherwise capable of H$_2$ cooling) get strongly irradiated by the Lyman--Werner radiation from nearby, metal-enriched galaxies \citep[e.g.\@][]{Stiavelli2010} or quasars \citep[e.g.][]{Johnson2019}, which prevents any star formation within these systems until they reach the \ion{H}{I} cooling mass ($M_\mathrm{h} \sim 10^7$--$10^8~\mathrm{M}_\odot$). Thus, these nearby galaxies/quasars may have contributed to the growth of an ionising bubble that would allow Ly$\alpha$ to escape from Pop III galaxies. In this case, the Ly$\alpha$ flux should not be as heavily attenuated, suggesting that our aforementioned colour selection should in principle still be effective at identifying such Pop III galaxies.

We show the effect of increasing the Ly$\alpha$ attenuation to 100\% (up from 50\%) in the bottom panel of Fig.\@~\ref{fig:zackrisson_colours2}. The F115W$-$F150W colours clearly become more red due to the lack of Ly$\alpha$ emission in the F115W filter. We stress that the relatively red F115W$-$F150W colours for $z=8$ Pop III galaxies with 100\% Ly$\alpha$ attenuation (compared to non-Pop III) should \emph{not} be used as a Pop III indicator. The reason for this is that at $7.5 < z < 9.5$, the F115W$-$F150W colour is sensitive to the strength of the Ly$\alpha$ break, and thus becomes increasingly more red with increasing redshift (up to $z=9.5$) as the Ly$\alpha$ break occupies progressively more of the F115W filter. Hence the red colours of $z=8$ Pop III galaxies with 100\% Ly$\alpha$ attenuation can overlap with those of non-Pop III galaxies at slightly higher redshift (than the $z=8$ galaxies shown in Fig.\@~\ref{fig:zackrisson_colours2}). Hence the F115W$-$F150W colour cannot be used to distinguish between Pop III and non-Pop III galaxies in the case of substantial Ly$\alpha$ attenuation. 

\subsubsection*{Comparison against early JWST galaxies}

We examine whether any potential Pop III candidates are present in the early \emph{JWST} data, by showing the colours for $8 < z < 9$ galaxy candidates observed with \emph{JWST} in Fig.\@~\ref{fig:zackrisson_colours2}. These galaxy candidates were imaged as part of the CEERS ERS program \citep{Bagley2023}, using both the June 2022 and December 2022 data. Briefly, the NIRCam data was reduced following the procedure in \citet{Adams2023a}, \cite{Adams2023b} and \citet{Ferreira2022a}, sources were identified using {\footnotesize SExtractor} \citep{Bertin1996} and photometric redshifts were derived using {\footnotesize LePhare} \citep{Arnouts1999, Ilbert2006} and {\footnotesize EAZY} \citep{Brammer2008}. Colours were measured from our calibrated photometry using the post-launch zero points, e.g., \citet{Adams2023a}.

The $8 < z < 9$ galaxy candidates shown in Fig.\@~\ref{fig:zackrisson_colours2} were selected by requiring a $\geq 5\sigma$ detection in the filters comprising the colour selection, non-detections (i.e.\@ $< 2\sigma$) in the \emph{HST} F606W and F814W bands \citep[using \emph{HST} data from the public CANDELS-EGS catalogs of][]{Stefanon2017}, as well as a best-fit photometric redshift $8 < z_\mathrm{phot} < 9$. 

We see that the F115W$-$F150W colours of these $8 < z < 9$ galaxy candidates are consistent with strong attenuation of Ly$\alpha$, and thus the F410M$-$F444W, F115W$-$F150W colour selection cannot be used to easily distinguish between Pop III and non-Pop III galaxies. Hence it is not possible to establish whether any Pop III candidates are present in the data. However these galaxies do generally fall in the region of where we find galaxies with normal stellar populations. This example shows how difficult it can be to use purely photometry to find Pop III galaxies, and likely several avenues will need to be investigated to verify any given system as Pop III.

\subsubsection{[\ion{O}{III}], [\ion{O}{II}] colour selection} \label{subsubsec:colour3}

In order to highlight the prospects (or lack thereof) of identifying Pop III candidates through colour selections that target weaker (i.e.\@ lower EW) metal emission lines, we show our final colour--colour selection in Fig.\@~\ref{fig:zackrisson_colours3}. Here the F410M$-$F444W, F277W$-$F335M colour pairs have been adopted. At $z=8$ the [\ion{O}{II}] $\lambda\lambda 3726,\ 3729$ doublet resides in the F335M filter, while the F277W filter contains no bright metal lines and just measures the continuum level. Hence the F277W$-$F335M colour is sensitive to the [\ion{O}{II}] EW and will be red for non-Pop III galaxies and relatively flat for Pop III galaxies. Here the less-used F335M filter was adopted, rather than the standardly used F356W, because the medium-band F335M filter will be more sensitive to the (rather weak) [\ion{O}{II}] doublet. 

\begin{figure*}
\centering
\includegraphics[width=\linewidth]{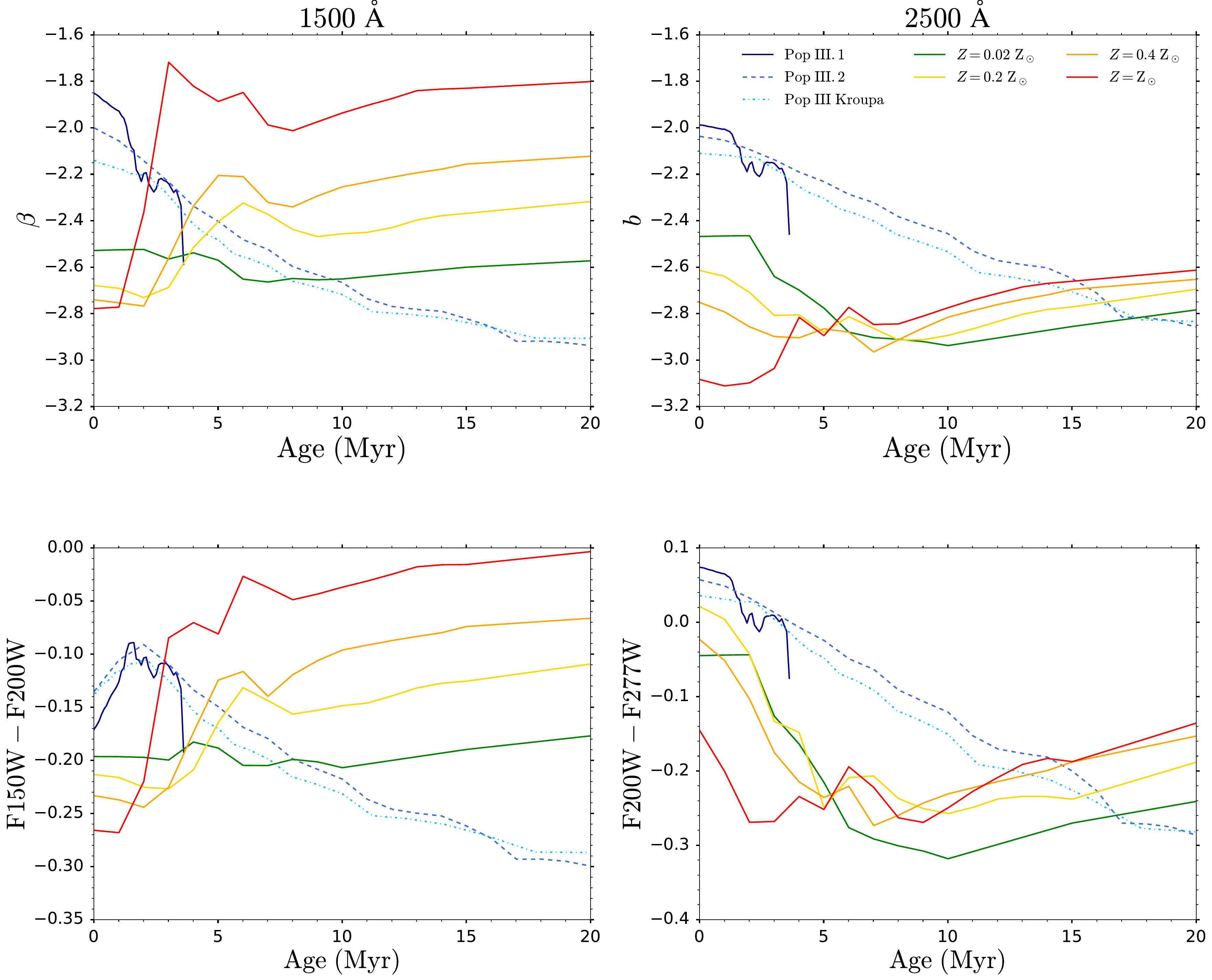}
\caption{The UV 1500~\AA\ continuum slope $\beta$ (top left panel) and 2500~\AA\ continuum slope $b$ (top right). Bottom panels: The imprint of the UV 1500~\AA\ continuum slope and 2500~\AA\ slope on NIRCam colours in the corresponding filters (F150W$-$F200W for 1500~\AA, F200W$-$F277W for 2500~\AA) for $z=8$ galaxies. The ages refer to the time elapsed after the instantaneous starburst. The colour coding and line styles follow the format of Fig.\@~\ref{fig:zackrisson_colours1}. Note that Pop III galaxies actually have redder continuum slopes than non-Pop III galaxies. The imprint of these redder continuum slopes on NIRCam colours is likely too small (immediately after a starburst) to use these colours as a Pop III diagnostic in practice, as even a minor amount of dust reddening and colour measurement error ($\sigma_\mathrm{C}=0.28$ for 5$\sigma$ flux density detections) will cause the Pop III and non-Pop III colours to overlap.}
\label{fig:continuum_slopes}
\end{figure*}

This colour selection has two main drawbacks. Firstly, it is only applicable over the narrow redshift range $7.95 < z < 8.35$. Secondly, the flat Pop III colours in both filter pairs means that this region of the colour--colour plane can readily be contaminated by other sources. At late times after the initial starburst, both the [\ion{O}{II}] and [\ion{O}{III}] lines will have faded and non-Pop III $z\sim8$ galaxies will have rather flat colours. Additionally, other sources at a diverse range of redshifts are likely to also have similar flat colours in these filter pairs. 

Owing to the limitations with this colour selection, we therefore do not define a Pop III region to identify potential Pop III candidates in this colour--colour plane.

\subsubsection{Continuum slope selection} \label{subsec:continuum_slope}

In this selection we briefly examine how effective measurements of the continuum slope, as inferred from photometry, will be at identifying Pop III candidates. We show the well-known UV power law index, i.e. the 1500~\AA\ UV slope $\beta$ in the top-left panel of Fig.\@~\ref{fig:continuum_slopes}. Given \emph{JWST}'s extensive coverage and sensitivity in the near- and mid-infrared, we will now be capable of probing the continuum slope at longer wavelengths than was possible before. Hence in the top-right panel we now introduce the new 2500~\AA\ slope $b$, with this wavelength range being selected for two reasons. Firstly, because it is relatively sensitive to Pop III stars. Secondly, because this region of the rest-frame spectrum is mostly devoid of high EW emission lines (with the possible exception of \ion{Mg}{II} at ${\sim}$2800~\AA), which would otherwise distort the inferred spectral slope.

As has already been pointed out in e.g.\@ \citet{Raiter2010b}, \citet{Zackrisson2011} and \citet{Dunlop2013}, the UV slope $\beta$ (and the 2500~\AA\ slope $b$) are in fact \emph{higher}, i.e. the spectrum is \emph{more red}, for Pop III galaxies. Despite the fact that the most luminous Pop III stars will be hotter and therefore bluer than their non-Pop III counterparts, the very bright but relatively red nebular continuum emission, which is very notable in Pop III spectra, is ultimately what drives the redder UV and 2500~\AA\ slopes in Pop III galaxies.

In the bottom panels we show the likely inferences that can be made on the 1500~\AA\ and 2500~\AA\ slopes from measurements with NIRCam photometry. At $z=8$, the 1500~\AA\ and 2500~\AA\ rest-frame wavelengths get redshifted to 1.35 \textmu m and 2.25 \textmu m, respectively. Thus we adopt the nearest accessible NIRCam filters available (avoiding F115W as it resides on Ly$\alpha$), namely the F150W$-$F200W and F200W$-$F277W filter pairs. The expected colours in these NIRCam filters can roughly be estimated by assuming the continuum $f_\nu$ follows a power law: $C = m_\mathrm{b} - m_\mathrm{r} = -2.5 \log (A\lambda_\mathrm{b}^n / A\lambda_\mathrm{r}^n) = -2.5n \log (\lambda_\mathrm{b}/\lambda_\mathrm{r})$. Here $n$ is the power law index, $\lambda_\mathrm{b}$ and $\lambda_\mathrm{r}$ are the effective wavelengths in the blue and red filters defining the colour $C$, with $A$ being the power law normalisation constant. In practice, if we wish to separate Pop III galaxies from non-Pop III, we are interested in the colour difference $\Delta C = -2.5\Delta n \log (\lambda_\mathrm{b}/\lambda_\mathrm{r})$. 

As can be seen from Fig.\@~\ref{fig:continuum_slopes}, in practice these colour shifts are rather small (immediately after a starburst), being only ${\sim}0.05$--$0.1$~mag. This is in part due to the \ion{He}{II} $\lambda 1640$ line, which falls in the F150W filter, thus driving bluer-than-otherwise Pop III F150W$-$F200W colours. Furthermore, the \ion{Mg}{II} ${\sim}$2800~\AA\ doublet, which falls in the F277W filter, drives redder-than-otherwise non-Pop III F200W$-$F277W colours. Given the marginally \emph{redder} Pop III colours, this small signal will likely be erased by even a minor amount of dust reddening. Additionally, assuming $5\sigma$ detections in the NIRCam filters, the colour uncertainty $\sigma _\mathrm{C} = 0.28$, which likely renders this Pop III colour diagnostic impractical. 

\subsubsection{\ion{He}{II} $\lambda 1640$ medium-band selection} \label{subsec:He_II_selection}

\begin{figure*}
\centering
\includegraphics[width=\linewidth]{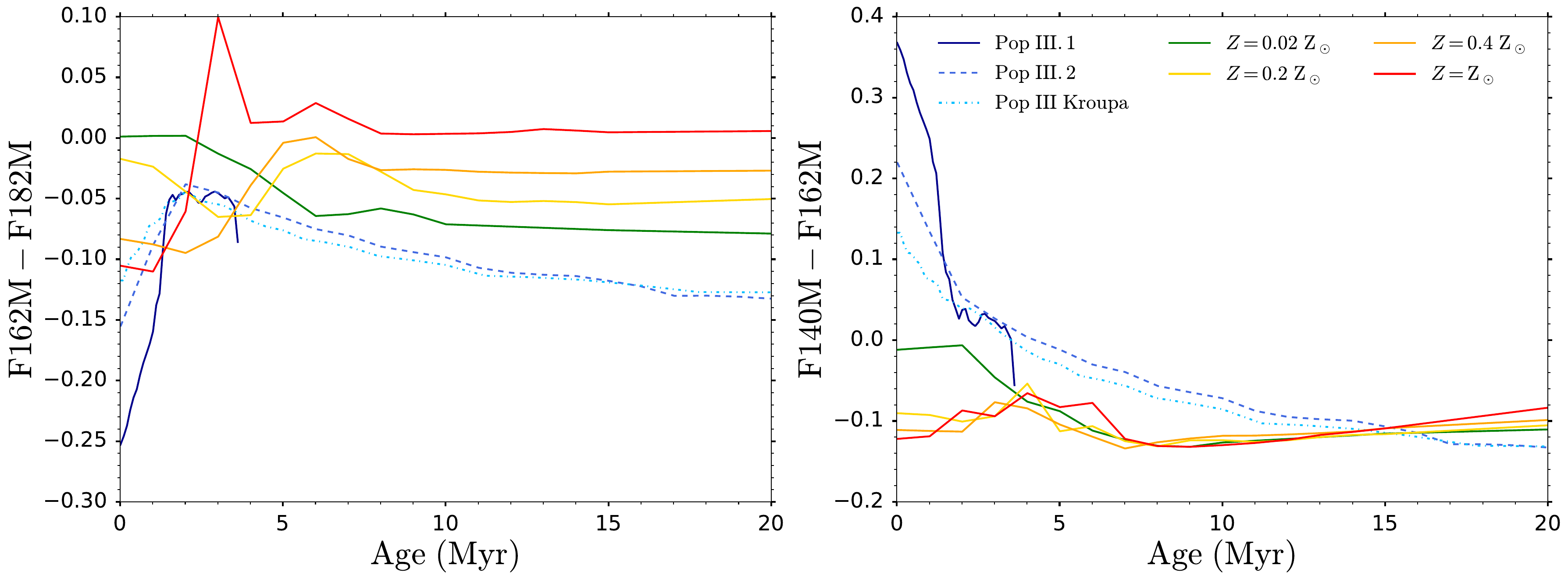}
\caption{The F162M$-$F182M (left panel) and F140M$-$F162M colours for $z=9$ (rather than our usual $z=8$) galaxies. At $z=9$, the \ion{He}{II} $\lambda 1640$ line resides in the F162M filter. Thus the F162M$-$F182M and F140M$-$F162M colours are in principle sensitive to the \ion{He}{II} $\lambda 1640$ EW at this redshift, which is characteristically high for Pop III galaxies. Despite the magnitude excess driven by \ion{He}{II} $\lambda 1640$ emission in the F162M filter for Pop III galaxies, the F162M$-$F182M colours for non-Pop III galaxies are comparably blue due to their intrinsically bluer UV continua. In contrast, these bluer non-Pop III UV slopes helps increase the separation in F140M$-$F162M colour between Pop III and non-Pop III galaxies, with this colour therefore being a potentially viable Pop III indicator for galaxies at $z\sim9$. Dust-reddened non-Pop III galaxies, \ion{C}{IV} $\lambda \lambda 1548, 1550$ emitters and \ion{O}{III}] $\lambda \lambda 1661, 1666$ emitters are potential contaminants, though in principle these can be removed through an inspection of their full SED. Indeed, the inclusion of the F140M filter (or F182M filter) is likely not necessary to probe the \ion{He}{II} $\lambda 1640$ EW (at $z=9$), as the presence of a F162M excess can be inferred from a consideration of the full SED, using the photometry in the other \emph{JWST} bands. Thus, medium-band NIRCam imaging surveys (see Table~\ref{tab:mediumband}) that search for high EW \ion{He}{II} $\lambda 1640$ emitters may provide a viable alternative for identifying Pop III candidates to deep and expensive F560W and F770W imaging campaigns with MIRI.}
\label{fig:he_II_mediumband}
\end{figure*}

\begin{table*}
\begin{center}
\resizebox{\linewidth}{!}{
\begin{tabular}{ |c|c|c|c|c|c|} 
\hline
Filter & Redshift range & IMF & Apparent magnitude (AB~mag) & $5\sigma$ exposure time (h) & Magnitude excess $\Delta m$ \\ 
\hline
& & Pop III.1 & 28.27 & 1.73 & 0.30 \\
F140M & $7.10 < z < 8.00$ & Pop III.2 & 29.49 & 16.36 & 0.15 \\
&  & Pop III Kroupa & 31.44 & 593.99 & 0.09 \\
\hline
& & Pop III.1 & 28.54 & 2.16 & 0.30 \\
F162M & $8.40 < z < 9.45$ & Pop III.2 & 29.75 & 20.03 & 0.15 \\
& & Pop III Kroupa & 31.71 & 740.93 & 0.09 \\
\hline
&  & Pop III.1 & 28.74 & 2.20 & 0.23 \\
F182M & $9.50 < z < 11.00$ & Pop III.2 & 29.94 & 20.03 & 0.12 \\
&  & Pop III Kroupa & 31.89 & 727.41 & 0.07 \\
\hline
& & Pop III.1 & 28.95 & 4.19 & 0.32 \\
F210M & $11.15 < z < 12.40$ & Pop III.2 & 30.17 & 39.61 & 0.16 \\
& & Pop III Kroupa & 32.13 & 1464.80 & 0.09 \\
\hline
\end{tabular}}
\caption{The NIRCam medium-band filters that can be used to search for Pop III candidates by targeting their characteristically strong \ion{He}{II} $\lambda 1640$ emission. Also shown are the redshift ranges over which the respective filters can be applied, the expected apparent magnitudes in those filters (assuming a Pop III galaxy with $\log (M_*/\mathrm{M}_\odot) = 6$, imaged immediately after an instantaneous starburst), the exposure times (in hours) required to reach $5\sigma$ depth, as well as the approximate magnitude excess $\Delta m = 2.5\log _{10}(1 + \mathrm{EW}_\mathrm{He\ II, rest}(1+z)/\Delta \lambda)$ in the adopted filter. Here $\Delta \lambda$ is the bandpass width of the filter, and the sources are assumed to be at $z = $ 7.5, 9, 10, 12 for the F140M, F162M, F182M and F210M filters, respectively. We assume \ion{He}{II} $\lambda 1640$ rest-frame equivalent widths $\mathrm{EW}_\mathrm{He\ II, rest}$ of 50~\AA, 26~\AA, 15~\AA\ (see Section~\ref{subsec:he_II}) for the Pop III.1, Pop III.2 and Pop III Kroupa IMFs, respectively.}
\label{tab:mediumband}
\end{center}
\end{table*}

As discussed in the literature \citep[e.g.\@][]{Schaerer2002, Schaerer2003, Raiter2010b}, and in more detail in Section~\ref{subsec:he_II}, high EW \ion{He}{II} $\lambda 1640$ emission likely serves as the definitive Pop III indicator. Thus, in this section we highlight the prospects of identifying Pop III candidates at $z \sim 9$ through NIRCam medium-band imaging campaigns that target the \ion{He}{II} $\lambda 1640$ emission from Pop III galaxies. Although the EW of \ion{He}{II} $\lambda 1640$ is \emph{relatively} high for Pop III galaxies, it is still small in an absolute sense, typically ranging from roughly 10--100~\AA\ across different Pop III models and IMFs. Hence the added emission line sensitivity from medium-band imaging (over a wide-band) will likely be crucial for identifying such emission lines through photometry, if possible. The expected magnitude excess driven by $z=9$ \ion{He}{II} $\lambda 1640$ emission in the F162M filter (which we discuss below) is 0.30~mag for Pop III.1 galaxies (which have rest-frame \ion{He}{II} $\lambda 1640$ EW of 50~\AA), and 0.48~mag for the highest EW \ion{He}{II} $\lambda 1640$ Pop III galaxies in the \citet{Nakajima2022} models (with rest-frame EW of 80~\AA).

At $z=9$ (rather than our usual $z=8$), \ion{He}{II} $\lambda 1640$ resides in the NIRCam F162M filter. As before, we seek neighbouring filters that probe the continuum level around \ion{He}{II} $\lambda 1640$, with the resulting colour thus being sensitive to the EW of the \ion{He}{II} line, and thus in principle enabling one to distinguish between Pop III and non-Pop III galaxies. To find a way to identify these galaxies we consider filters both redward and blueward of F162M, with Pop III galaxies having relatively blue and red colours in the corresponding F162M filter pairs, respectively.  Owing to the intrinsically \emph{bluer} UV slopes for non-Pop III galaxies (as discussed in the previous section), the filter used redward of F162M must be close in wavelength to F162M, otherwise Pop III and non-Pop III galaxies will exhibit similarly blue colours. We thus adopt the NIRCam F182M filter, rather than the F200W filter. Additionally, owing to the small wavelength gap between \ion{He}{II} $\lambda 1640$ and Ly$\alpha$, it is in fact rather challenging to select a filter blueward of F162M that does not include Ly$\alpha$ (otherwise complicating the interpretation of the colour) and also does not cover \ion{He}{II} $\lambda 1640$ (otherwise weakening the Pop III colour signature). We thus adopt the NIRCam F140M filter (rather than the F115W or F150W filters). 

We show the F162M$-$F182M and F140M$-$F162M colours of $z=9$ galaxies in the left and right panels of Fig.\@~\ref{fig:he_II_mediumband}, respectively. As mentioned above, although the \ion{He}{II} $\lambda 1640$ emission for Pop III galaxies drives a magnitude excess in the F162M filter, Pop III and non-Pop III galaxies have comparably blue F162M$-$F182M colours, due to the intrinsically bluer UV slopes of non-Pop III galaxies. In contrast, these bluer non-Pop III slopes help to increase the separation in F140M$-$F162M colour between Pop III and non-Pop III galaxies, with this colour therefore being a potentially viable Pop III indicator for galaxies at $z\sim9$.

However, the \emph{redder} F140M$-$F162M colours exhibited by Pop III galaxies can in principle be replicated by non-Pop III galaxies with a moderate amount of dust reddening. Indeed, although the F140M and F162M filters probe (at $z=9$) the rather narrowly separated rest-frame wavelengths of 1400~\AA\ and 1620~\AA, respectively, most dust attenuation laws $k(\lambda)$ rise sharply with decreasing wavelength in this wavelength regime. Hence the amount of dust reddening, i.e.\@ colour excess $\mathrm{E(B}-\mathrm{V)}$, needed to shift $\sim 0.3~$mag $ = \mathrm{E(B}-\mathrm{V)} (k(1400~$\AA$) - k(1620~$\AA$))$ the non-Pop III galaxy F140M$-$F162M colours onto the Pop III is $\mathrm{E(B}-\mathrm{V)} = 0.35$, assuming the \citet{Calzetti2000} dust attenuation law.

Still, such dust-reddened non-Pop III galaxies should in principle be able to be distinguished from Pop III galaxies, from an inspection of their full SED, which should be relatively red. In the same vein, other potential contaminants such as \ion{C}{IV} $\lambda\lambda 1548, 1550$ emitters \citep[i.e.\@ AGN or metal-poor galaxies, see e.g.\@][]{Stark2015} or \ion{O}{III}] $\lambda\lambda 1661,1666$ emitters at comparable redshifts, can in principle also be removed (as these should have bright [\ion{O}{III}] $\lambda 5007$ emission). Indeed, by considering the entire SED (such as through SED fitting, which we will investigate in a future work), the inclusion of the F140M filter (or F182M filter) likely is not necessary to probe the \ion{He}{II} $\lambda 1640$ EW (at $z=9$), as the presence (or absence) of a magnitude excess in the F162M filter should be able to be inferred from a comparison against the photometry in the other \emph{JWST} bands. 

Thus, owing to the enhanced sensitivity and footprint of NIRCam relative to MIRI, medium-band NIRCam imaging surveys that search for high EW \ion{He}{II} $\lambda 1640$ emitters may provide a viable alternative for identifying Pop III candidates to the deep F560W and F770W imaging campaigns with MIRI. However, it is likely that only Pop III galaxies with the most top-heavy IMFs (such as Pop III.1), imaged immediately after the starburst ($\Delta t < 1~$Myr) can be identified in this way. 

In Table~\ref{tab:mediumband}, we show the various NIRCam medium-bands that can be used to target \ion{He}{II} $\lambda 1640$ emitters at different redshifts, together with the expected apparent magnitudes, exposure times needed for 5$\sigma$ detections, and the likely magnitude excess $\Delta m$ in the medium-band, for the \citet{Zackrisson2011} Pop III models considered in this work. We note that the NIRCam F140M, F162M, F182M and F210M bands can be used to target \ion{He}{II} $\lambda 1640$ \citep[and thus Pop III, AGN or DCBH candidates, but also galaxies containing Wolf--Rayet stars and X-ray binaries, see e.g.\@][]{Katz2023} at $z = $ 7.5, 9, 10 and 12, respectively. The exposure times required to reach $5\sigma$ depth in the aforementioned filters are relatively short (compared to the MIRI requirements), at $< 4~$h and $< 40~$h for the Pop III.1 and Pop III.2 IMFs, respectively. However, deeper imaging (going beyond $5\sigma$ depth, i.e.\@ $\sigma_\mathrm{m} = 0.2$) will likely be required for the small photometric \ion{He}{II} $\lambda 1640$ signature we are searching for ($\Delta m \sim 0.15$--$0.30$~mag) to be convincing. 

\subsection{Slitless emission-line-selection} \label{subsec:line_selection}

\begin{table*}
\begin{center}
\resizebox{\linewidth}{!}{
\begin{tabular}{ |c|c|c|c|c|c|c|} 
\hline
IMF & H$\beta$ flux (cgs) & H$\beta$ 5$\sigma$ exposure time (h) & Ly$\alpha$ flux (cgs) & Ly$\alpha$ 5$\sigma$ exposure time (h) & \ion{He}{II} $\lambda$1640 flux (cgs) & \ion{He}{II} $\lambda$1640 5$\sigma$ exposure time (h) \\ 
\hline
Pop III.1 & $6.76\times10^{-19}$ & 17.12 & $3.31\times10^{-17}$ & 0.07 & $8.71\times10^{-19}$ & 55.25 \\
\hline
Pop III.2 & $2.34\times10^{-19}$ & 142.92 & $1.10\times10^{-17}$ & 0.62 & $1.58\times10^{-19}$ & 1678.90 \\ 
\hline
Pop III Kroupa & $3.63\times10^{-20}$ & 5938.84 & $1.78\times10^{-18}$ & 23.72 & $1.58\times10^{-20}$ & 167889.76\\
\hline
\end{tabular}}
\caption{The expected H$\beta$, \emph{unattenuated} Ly$\alpha$ and \ion{He}{II} $\lambda 1640$ fluxes (in cgs units, i.e.\@ erg~s$^{-1}$~cm$^{-2}$), as well as the integration times (in hours) needed to achieve a $5\sigma$ detection with NIRISS/NIRCam slitless spectroscopy, for the three Zackrisson Pop III models at $z=8$ with $\log (M_*/\mathrm{M}_\odot) = 6$ observed \emph{immediately after} ($0.01$~Myr) an instantaneous starburst. The expected line fluxes and required integration times will be different (see the time evolution of the emission line luminosities in Section~\ref{sec:spectroscopy}) if the Pop III galaxy is observed at a later time after the starburst. The integration times were estimated by extrapolating from the NIRISS and NIRCam sensitivities reported in \citet{Bagley2023b} and \citet{Oesch2023}, respectively.}
\label{tab:slitless_observations}
\end{center}
\end{table*}

In this subsection we discuss the prospects for identifying Pop III candidates from blind emission-line surveys carried out through slitless spectroscopy with NIRISS ($0.8 \leq \lambda$ (\textmu m) $\leq 2.0$) and NIRCam ($2.0 \leq \lambda$ (\textmu m) $\leq 5.0$). We focus on the brightest lines that yield the greatest constraints on the potential Pop III nature of the galaxy. These lines are H$\beta$, Ly$\alpha$ and \ion{He}{II} $\lambda$1640. The expected line fluxes for $z=8$ Pop III galaxies with $\log (M_*/\mathrm{M}_\odot) = 6$ detected immediately after an instantaneous starburst are shown in Table~\ref{tab:slitless_observations}. We also include the expected integration times needed to achieve a $5\sigma$ line detection, estimated by extrapolating the exposure times and sensitivities reported for NIRISS and NIRCam in the NGDEEP survey \citep{Bagley2023b} and the FRESCO survey \citep{Oesch2023}, respectively. We discuss the properties of these individual lines below and how well we can use these as a tracer of Pop III galaxies. 

\subsubsection{H$\beta$ detection}

At $z=8$, the H$\beta$ line is redshifted to $\lambda = 4.37$~\textmu m, placing it within the F444W band for slitless spectroscopic observations with NIRCam. As can be seen from Table~\ref{tab:slitless_observations}, the integration times needed for a $5\sigma$ detection are demanding (17.12~h, 142.92~h), even for the Pop III.1 and Pop III.2 IMFs. Of course, the integration times can be reduced substantially with moderate lensing $\mu$ and/or a mass boost $\mathcal{M}$ above the assumed nominal stellar mass. 

As will be discussed more extensively in Section~\ref{sec:spectroscopy}, the merits to detecting H$\beta$ for Pop III identification are twofold. Firstly, a measurement of the H$\beta$ equivalent width can place constraints on the potential Pop III nature of the source, though in practice this will require a line detection at much more than just $5\sigma$ significance. Secondly, a non-detection of the neighbouring bright [\ion{O}{III}] $\lambda 5007$ line (which will also lie within the F444W band), can be used to place constraints on the upper limit of the metallicity of the galaxy \citep[see also][]{Nakajima2022}.

\subsubsection{Ly$\alpha$ detection}

At $z=8$, the Ly$\alpha$ line is redshifted to $\lambda = 1.09$~\textmu m, placing it within the F115W band for slitless spectroscopic observations with NIRISS. The Ly$\alpha$ line fluxes in Table~\ref{tab:slitless_observations} correspond to the intrinsic line fluxes, i.e.\@ assuming no IGM attenuation. In this case the integration times needed are exceptionally short, owing to the great intrinsic brightness of this line for Pop III galaxies. In reality, Ly$\alpha$ will be heavily attenuated and scattered by the IGM. Thus this line will actually be much fainter in observations, and may therefore be an unreliable Pop III indicator. 

\subsubsection{\ion{He}{II} $\lambda1640$ detection}

At $z=8$, the \ion{He}{II} $\lambda$1640 line is redshifted to $\lambda = 1.48$~ \textmu m, placing it within the F150W band for slitless spectroscopic observations with NIRISS. For Pop III.1, the \ion{He}{II} $\lambda$1640 flux is comparable to that of H$\beta$. Furthermore, NIRISS F150W and NIRCam F444W are also comparable in sensitivity. Hence the integration time of 55.25~h is roughly similar to that of H$\beta$ (17.12~h). However, with Pop III.2 and Pop III Kroupa, the \ion{He}{II} line begins to drop off with respect to H$\beta$ and the integration times become much longer, at ${\sim} 1000$~h and ${\sim} 100\,000$~h, respectively. Thus a \ion{He}{II} $\lambda$1640 line detection with NIRISS will only be possible for Pop III.2 sources that are much more massive than the nominal stellar mass ($M_* \sim 5\times 10^{6}~\mathrm{M}_\odot$) and/or have been strongly gravitationally lensed ($\mu \sim 5$).

As has been discussed in the literature \citep[see e.g.\@][]{Schaerer2002, Schaerer2003, Raiter2010b, Grisdale2021, Nakajima2022} and will also be discussed more thoroughly in Section~\ref{sec:spectroscopy}, the main merit for detecting \ion{He}{II} $\lambda 1640$ is that it is a clear Pop III signature. Although AGN and/or DCBH can also produce bright \ion{He}{II} emission \citep[see e.g.\@][]{Nakajima2022}, these can be readily ruled out from photometry due to their much redder colours \citep[see e.g.\@][or Fig.\@~\ref{fig:nakajima_colours1}]{Inayoshi2022}. Furthermore, a measurement of the \ion{He}{II} equivalent width can also distinguish between different Pop III IMFs, and thus (in principle, though in practice can be difficult due to model uncertainties) is able to firmly separate Pop III.1, Pop III.2 and Pop III Kroupa galaxies.

\section{Pop III constraints from follow-up spectroscopy} \label{sec:spectroscopy}

Having identified potential Pop III candidates either from colour selection, emission-line selection and/or SED fitting (not covered in this work), deep follow-up spectroscopy will be essential to place tighter constraints on the Pop III nature of these sources. In this section we outline the NIRSpec spectroscopic observations that could be undertaken, as well as the spectroscopic diagnostics that need to be applied to achieve this. Our emphasis will be on emission line equivalent widths, emission line mass-to-light ratios and line ratios. We will focus on the brightest lines that will be accessible by NIRSpec that are most sensitive to Pop III star formation. As discussed earlier, at $z=8$ these lines are H$\beta$, Ly$\alpha$ and \ion{He}{II} $\lambda 1640$. 

We show the expected emission line fluxes for $z=8$ Pop III galaxies with $\log (M_*/\mathrm{M}_\odot) = 6$ immediately after an instantaneous starburst in Table~\ref{tab:nirspec_observations}. We also show the expected integration times needed to achieve a $5\sigma$ detection of these lines, assuming that the $R=1000$ NIRSpec gratings (which are the most sensitive gratings for emission line detection) have been used. For H$\beta$, Ly$\alpha$ and \ion{He}{II} $\lambda 1640$, this corresponds to the G395M, G140M and G140M gratings, respectively. We use the JWST ETC to perform these integration time estimations, assuming a point source, with a continuum-subtracted spectrum with a 5 spectral pixel, and 4 spatial pixel extraction window. 

\begin{table*}
\begin{center}
\resizebox{\linewidth}{!}{
\begin{tabular}{ |c|c|c|c|c|c|c|} 
\hline
IMF & H$\beta$ flux (cgs) & H$\beta$ 5$\sigma$ exposure time (h) & Ly$\alpha$ flux (cgs) & Ly$\alpha$ 5$\sigma$ exposure time (h) & \ion{He}{II} $\lambda$1640 flux (cgs) & \ion{He}{II} $\lambda$1640 5$\sigma$ exposure time (h) \\ 
\hline
Pop III.1 & $6.76\times10^{-19}$ & 1.02 & $3.31\times10^{-17}$ & 0.0059 & $8.71\times10^{-19}$ & 4.76 \\
\hline
Pop III.2 & $2.34\times10^{-19}$ & 8.50 & $1.10\times10^{-17}$ & 0.0537 & $1.58\times10^{-19}$ & 144.51 \\ 
\hline
Pop III Kroupa & $3.63\times10^{-20}$ & 353.06 & $1.78\times10^{-18}$ & 2.05 & $1.58\times10^{-20}$ & 14451.46\\
\hline
\end{tabular}}
\caption{Similar to Table\@~\ref{tab:slitless_observations}, but now showing the integration times needed to achieve a $5\sigma$ line detection with NIRSpec $R{\sim}1000$ observations. We estimated the integration times using the JWST ETC, assuming a point source, with a continuum-subtracted spectrum with a 5 spectral pixel, and 4 spatial pixel extraction window.}
\label{tab:nirspec_observations}
\end{center}
\end{table*}

\subsection{H$\beta$ diagnostics}

\subsubsection{H$\beta$ equivalent width}

\begin{figure*}
\centering
\includegraphics[width=\linewidth]{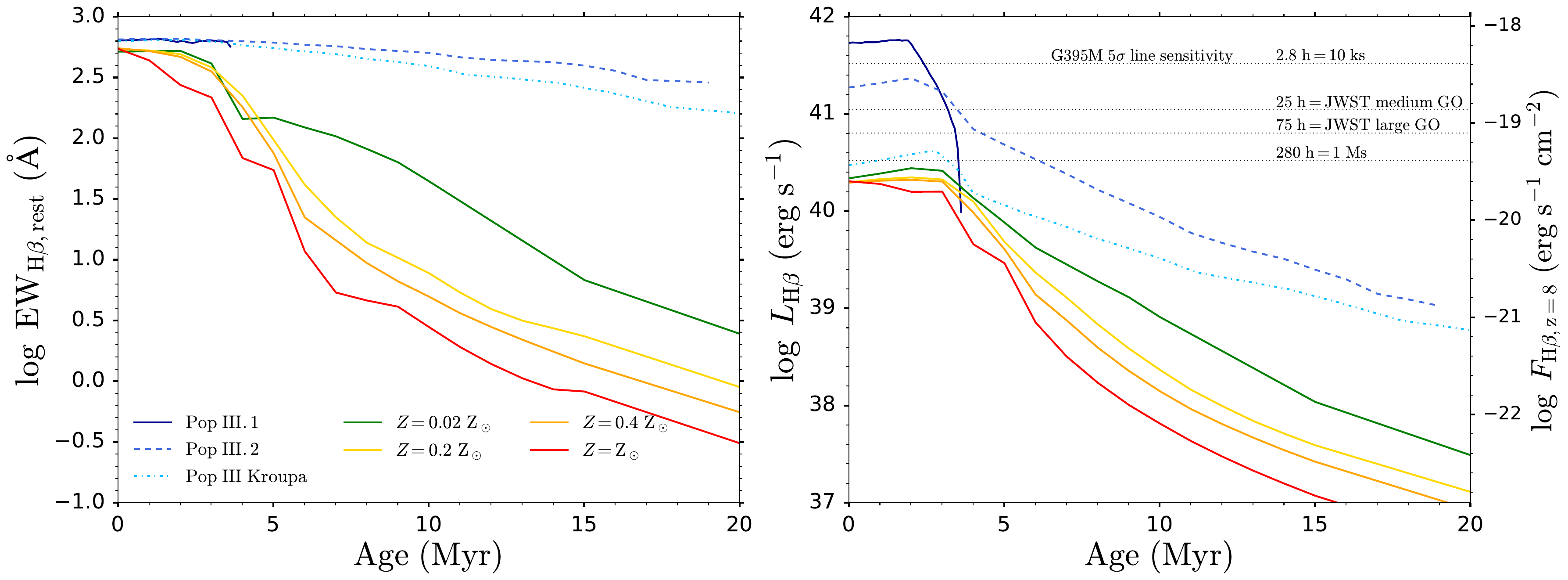}
\caption{The H$\beta$ rest-frame equivalent width (left panel) and H$\beta$ luminosity (as well as $z=8$ H$\beta$ flux, right panel) for galaxies at a nominal stellar mass of  $\log (M_*/\mathrm{M}_\odot) = 6$. The ages refer to the time elapsed after the instantaneous starburst. The dotted horizontal lines in the right panel are the expected $5\sigma$ line sensitivities achieved with NIRSpec/G395M (which covers H$\beta$ at $z=8$) in integration times of 2.8~h = 10~ks, 25~h = \emph{JWST} medium GO program, 75~h = \emph{JWST} large GO program and 280~h = 1~Ms. The colour coding and line styles follow the format of Fig.\@~\ref{fig:zackrisson_colours1}. The H$\beta$ equivalent widths of Pop III galaxies are only marginally higher (${\sim}0.1$~dex) than for non-Pop III galaxies. The substantially larger (${\sim}$1~dex) H$\beta$ luminosities per unit stellar mass for Pop III galaxies likely cannot be used as a Pop III indicator, as that would require an accurate measurement of the stellar mass, which in turn would require one to already know if the source is Pop III.}
\label{fig:zackrisson_lines_Hb}
\end{figure*}

In Fig.\@~\ref{fig:zackrisson_lines_Hb} we show how the H$\beta$ rest-frame equivalent widths (left panel) and the H$\beta$ line luminosity (right) vary with time after an instantaneous starburst. We again assume a nominal stellar mass of $\log (M_*/\mathrm{M}_\odot) = 6$. Hence the H$\beta$ line luminosity in the right panel is essentially a mass-to-light ratio, representing how much line luminosity is expected per $10^6~\mathrm{M}_\odot$ formed. As pointed out earlier in this paper, Pop III galaxies have much greater H$\beta$ line luminosities per unit stellar mass formed. However, their continuum normalisation per unit stellar mass formed (not shown) is also substantially higher. As a result, their H$\beta$ equivalent widths are only marginally higher (${\sim}0.1$~dex) than for non-Pop III galaxies.

In principle, a measurement of the H$\beta$ equivalent width can therefore be used to distinguish between Pop III and non-Pop III galaxies. With a $5\sigma$ H$\beta$ detection, the uncertainty on the measured flux will be 20\%. However, the separation between Pop III and non-Pop III is ${\sim}0.1$~dex $\approx 25$\%. Hence in practice, barring any model uncertainties, a much deeper integration (e.g.\@ $4\times$ the integration time) with H$\beta$ detected at $10\sigma$ will likely be required to more definitively identify a source as likely being a Pop III galaxy. Given the relatively short exposure times needed to detect H$\beta$ at $5\sigma$ with NIRSpec/G395M, at $1.02$~h and $8.50$~h for Pop III.1 and Pop III.2 galaxies, respectively, such deep integrations would certainly be possible to achieve even within a small-to-medium \emph{JWST} GO observing program.

\begin{figure*}
\centering
\includegraphics[width=\linewidth]{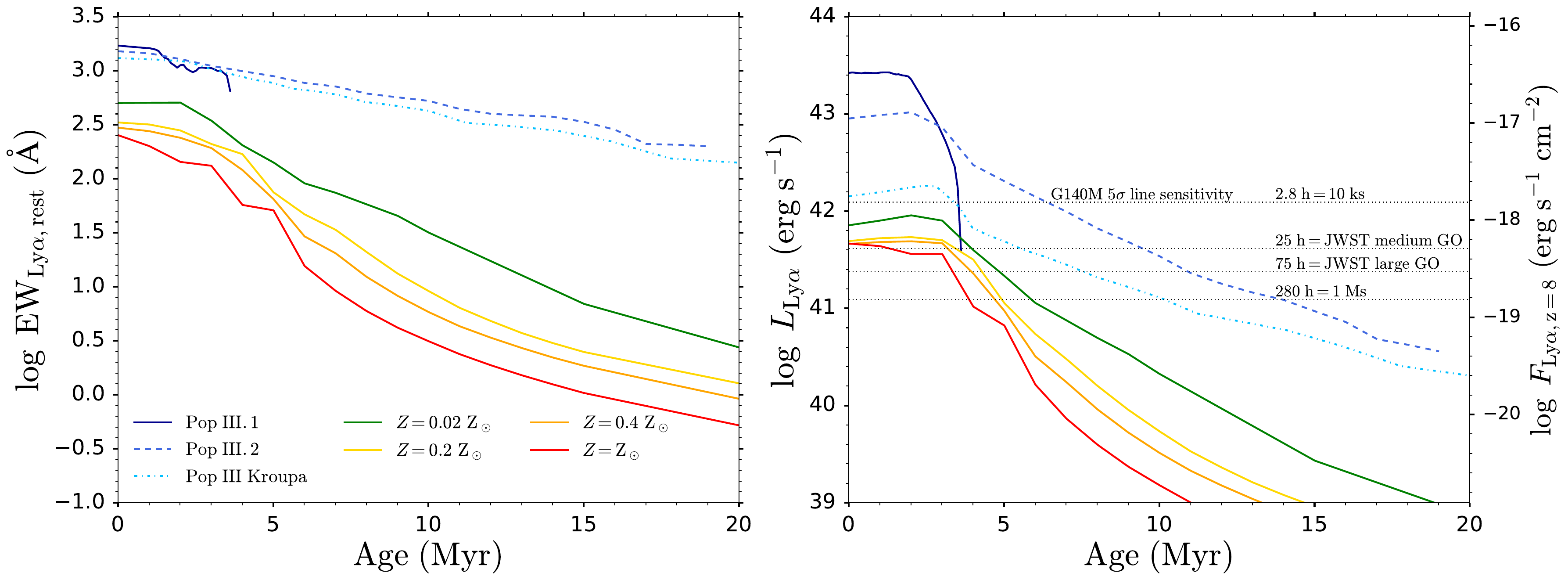}
\caption{Similar to Fig.\@~\ref{fig:zackrisson_lines_Hb}, but now showing Ly$\alpha$. Note that we show the \emph{intrinsic} (i.e.\@ not attenuated by the IGM) Ly$\alpha$ EW and luminosity. In practice the Ly$\alpha$ line will likely be heavily attenuated/scattered by the IGM, both blueward and redward of the line centre. The dotted horizontal lines in the right panel are now the expected $5\sigma$ line sensitivities achieved with NIRSpec/G140M (which covers Ly$\alpha$ at $z=8$).}
\label{fig:zackrisson_lines_Lya}
\end{figure*}

However, there are other complications with using the measured H$\beta$ equivalent width as a Pop III indicator. Firstly, the exact EW associated with Pop III will likely be model dependent, with different models likely predicting different H$\beta$ line luminosities and continuum levels. Indeed, in the \citet{Nakajima2022} models (not shown) the expected H$\beta$ EWs for Pop III galaxies are $\log\,  \mathrm{EW} = 2.95$--$3.00$, whereas for non-Pop III the range is $\log \, \mathrm{EW} \leq 2.85$, thus overlapping with the predicted Pop III EWs in the \citet{Zackrisson2011} models ($\log \, \mathrm{EW} \approx 2.85$).

Secondly, the H$\beta$ EW also depends on the covering fraction $f_\mathrm{cov}$ of the ISM gas \citep[see e.g.\@][]{Zackrisson2013}. The lower the covering fraction, the lower the H$\beta$ line luminosity but also the lower the nebular continuum. Hence in principle one would also need to carefully take into account the $f_\mathrm{cov}$ (or $f_\mathrm{esc}$) dependence of the H$\beta$ EW \citep[for more details see][]{Zackrisson2013} before conclusive inferences on the Pop III nature of the source can be made.

It should be noted that the EWs shown in Fig.\@~\ref{fig:zackrisson_lines_Hb} correspond to a single stellar population all formed after an instantaneous starburst. For non-Pop III galaxies, there may be an underlying older stellar population, in addition to the newly formed starburst. Provided that this older stellar population provides a non-negligible flux density to the continuum, the observed EWs for non-Pop III galaxies will actually be lower than those shown in Fig.\@~\ref{fig:zackrisson_lines_Hb}.

Furthermore, determining the H$\beta$ EW requires both the H$\beta$ line flux and the continuum level to be measured. As Pop III galaxies are typically very faint (see Fig.\@~\ref{fig:popIII_spectrum} and Table~\ref{tab:imaging_observations}), detecting their continuum through e.g.\@ $R$$\sim$$100$ NIRSpec/PRISM spectroscopy will be a challenging endeavour (requiring $\sim$28.5--30.5~AB~mag depth, barring any flux boost from gravitational lensing or elevated Pop III stellar masses). Therefore, the continuum flux density of Pop III galaxies will likely have to be estimated from broadband photometry. However, the difficulty with this is that the bandpass-averaged flux densities measured via photometry can be strongly boosted (by e.g.\@ $>0.1$~dex $=0.25$~mag) above the continuum level by the high EW emission lines in Pop III galaxies (as shown in Fig.\@~\ref{fig:popIII_spectrum}). This makes estimating the true continuum level around H$\beta$ particularly difficult as it itself has a high EW and it is also bracketed by a series of bright emission lines at shorter (e.g.\@, H$\gamma$) and longer wavelengths (e.g.\@ H$\alpha$). In principle, this flux boost effect can be corrected for by taking into account the measured H$\beta$ flux (and all other high EW lines that reside within the filter of interest), as the observed bandpass-averaged flux density $f_{\lambda, \mathrm{obs}} = f_{\lambda, \mathrm{cont}} + f_\mathrm{lines}/\Delta \lambda$, where  $f_{\lambda, \mathrm{cont}}$ is the true continuum level, $f_\mathrm{lines}$ is the total flux of all the emission lines that reside within the filter, and $\Delta \lambda$ is the bandpass width of the filter. Thus, the potential difficulty in accurately estimating the continuum flux density (and the measurement error on the H$\beta$ flux), together with the small difference in H$\beta$ EW between Pop III and non-Pop III (and systematics and secondary dependences therein), likely renders the H$\beta$ EW impractical as a diagnostic for actually distinguishing between Pop III and non-Pop III galaxies.

\subsubsection{H$\beta$ luminosity}

As can be seen from Fig.\@~\ref{fig:zackrisson_lines_Hb}, the H$\beta$ luminosity starts to drop off substantially $\sim$3~Myr after an instantaneous starburst. Thus the visibility window over which the H$\beta$ emission from Pop III galaxies can likely be detected with \emph{JWST} ($\sim$3~Myr) is comparable to the timescale over which these galaxies can be detected in broadband photometry (2.8--5.6~Myr, see Table~\ref{tab:imaging_observations}). Note that the H$\beta$ luminosities (and fluxes) shown in Fig.\@~\ref{fig:zackrisson_lines_Hb} are for galaxies at a fixed stellar mass $M_* = 10^{6}~\mathrm{M}_\odot$, and hence can be taken as an indication of the H$\beta$ luminosity per unit stellar mass ($=L_{\mathrm{H}\beta}/10^6$).

Now, given the large separation in H$\beta$ luminosity per unit stellar mass between Pop III and non-Pop III (${\sim}$1~dex), as well as the various Pop III IMFs (${\sim}$0.5~dex between Pop III.1 and Pop III.2), an accurate measurement of the line luminosity per unit stellar mass would enable one to readily confirm the Pop III nature of a source. Whilst the determination of the line luminosity is relatively straightforward, the difficulty lies with establishing the stellar mass. Indeed, the continuum flux density per unit stellar mass is vastly different for Pop III and non-Pop III galaxies (not shown), i.e.\@ Pop III galaxies have much lower mass-to-light ratios. This can be inferred from considering the fact that the H$\beta$ line luminosities are much greater for Pop III, while their H$\beta$ EWs are almost the same as non-Pop III. As a result, an accurate determination of the stellar mass therefore requires an accurate assessment of the mass-to-light ratio. However, accurately knowing the mass-to-light ratio is akin to knowing whether the source is Pop III or not (as the M/L ratio is much lower for Pop III). Thus, in order to determine whether a source is Pop III from its line luminosity per unit stellar mass, we would have to already know if it was Pop III or not, making this diagnostic unusable in practice. 

\subsection{Ly$\alpha$ diagnostics}

We show the \emph{intrinsic} Ly$\alpha$ rest-frame equivalent widths and line luminosities in Fig.\@~\ref{fig:zackrisson_lines_Lya}. As alluded to earlier, unlike the Balmer recombination lines, the Ly$\alpha$ EW is much larger for Pop III galaxies relative to non-Pop III. Thus a measurement of the \emph{intrinsic} Ly$\alpha$ EW would not only enable one to readily separate Pop III from non-Pop III, but in principle it would also enable one to distinguish between different Pop III IMFs. 

However, in practice, the Ly$\alpha$ line will likely be heavily attenuated and scattered by the IGM. Hence the \emph{observed} Ly$\alpha$ line luminosities and EWs will likely be substantially lower than those shown in Fig.\@~\ref{fig:zackrisson_lines_Lya}, which would erode this otherwise strong Pop III signature. 

\subsection{\ion{He}{II} $\lambda 1640$ diagnostics} \label{subsec:he_II}

\subsubsection{\ion{He}{II} $\lambda 1640$ equivalent width}

\begin{figure*}
\centering
\includegraphics[width=\linewidth]{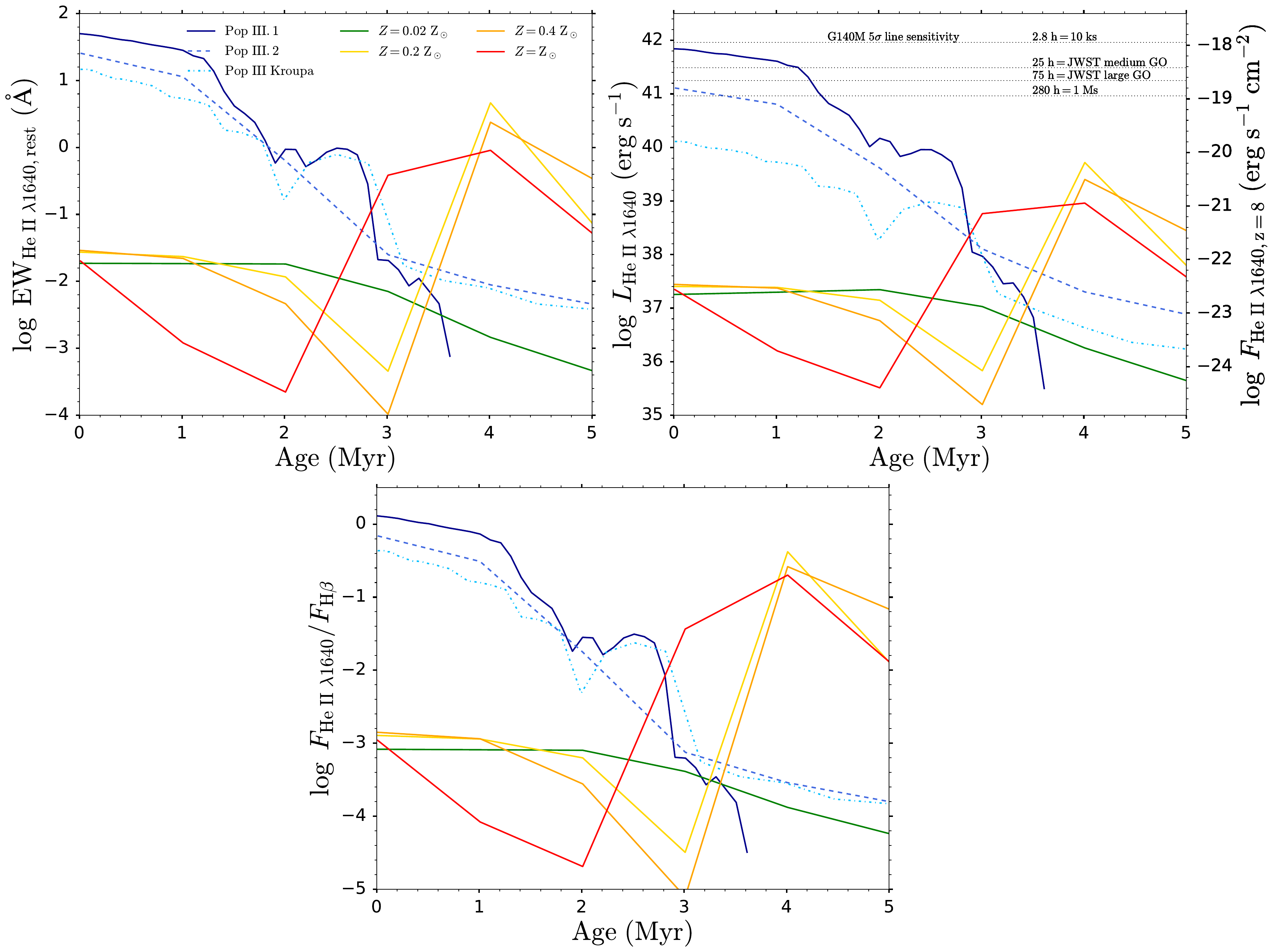}
\caption{Similar to Fig.\@~\ref{fig:zackrisson_lines_Hb}, but now showing \ion{He}{II} $\lambda 1640$. The dotted horizontal lines in the top right panel are now the expected $5\sigma$ line sensitivities achieved with NIRSpec/G140M (which covers \ion{He}{II} $\lambda 1640$ at $z=8$). We also show the \ion{He}{II} $\lambda 1640$/H$\beta$ line ratio (bottom panel). Note that the ages are only shown up to 5~Myr, rather than 20~Myr as in Fig.\@~\ref{fig:zackrisson_lines_Hb}. High EW \ion{He}{II} $\lambda 1640$ emission will serve as the definitive indicator of Pop III galaxies. }
\label{fig:zackrisson_lines_He_II}
\end{figure*}

We show the \ion{He}{II} $\lambda 1640$ rest-frame equivalent width, line luminosity and flux ratio with H$\beta$ in Fig.\@~\ref{fig:zackrisson_lines_He_II}. We see that bright \ion{He}{II} $\lambda 1640$ emission is a clear spectroscopic signature for Pop III galaxies \citep[as has been previously suggested in e.g.\@][]{Schaerer2002, Schaerer2003, Raiter2010b, Grisdale2021, Nakajima2022}, as they have substantially higher EWs than non-Pop III. Hence a measurement of the EW will establish the Pop III nature of a source. Furthermore, the various Pop III IMFs also yield distinctly different \ion{He}{II} $\lambda 1640$ EWs (with a ${\sim}0.5$~dex spread). 

Using the \citet{Nakajima2022} Pop III models (not shown here but see their Fig.\@~6), we have investigated the dependence of the \ion{He}{II} $\lambda 1640$ EW on ionisation parameter $U$, finding only a very weakly decreasing trend (with $\Delta \log \, \mathrm{EW} \approx 0.07$~dex) with increasing $U$ (from $U=-2.0$ to $U=-0.5$). Thus the EW ranges spanned by each IMF is distinct, even within the range of possible $U$ values. However, the \ion{He}{II} $\lambda 1640$ EWs do depend somewhat on the covering fraction $f_\mathrm{cov}$, with $\Delta \log \, \mathrm{EW} \approx 0.1$~dex between $f_\mathrm{cov} = 0.5, 1$. We note that while the dependence of the \ion{He}{II} $\lambda 1640$ EW on ionisation parameter $U$ is relatively weak in the \citet{Nakajima2022} models (over the range $-2.0 < U < -0.5$, it is much stronger ($\geq 0.2$~dex) in the models of \citet{Raiter2010b} (over the range $-4.0 < U < -1.0$ with $n_\mathrm{H} = 10^{3}$~cm$^{-3}$). However, most of this ionisation parameter dependence stems from the evolution between $-4 < U < -2$, with only weak trends in the range $-2 < U < -1$, similar to the \citet{Nakajima2022} models. 

Furthermore, we note that while models generally predict Pop III galaxies to have characteristically high \ion{He}{II} $\lambda 1640$ EWs (compared to non-Pop III galaxies), with the EW tending to increase as the IMF becomes increasingly top-heavy, there is some variation in the actual EW predicted by the different models, which could render a determination of the Pop III IMF difficult. In the Zackrisson models shown in Fig.\@~\ref{fig:zackrisson_lines_He_II}, the predicted Pop III \ion{He}{II} $\lambda 1640$ rest-frame EWs are in the range 16--50~\AA, while for the \citet{Nakajima2022} models the range is instead 25--80~\AA, with \citet{Raiter2010b} and \citet{Inoue2011} predicting 20--90~\AA \ and 15~\AA, respectively.    Despite the aforementioned secondary dependences and systematics, a measurement of the \ion{He}{II} $\lambda 1640$ EW should help to place valuable constraints on the Pop III IMF.

Now, determining the \ion{He}{II} $\lambda 1640$ EW also requires the continuum level to be estimated (which is likely more viable via broadband photometry, rather than spectroscopy, as 28.5--29.5~AB~mag depth is required). In contrast to H$\beta$, the EW of \ion{He}{II} $\lambda 1640$ is relatively small and the spectrum around $1640$~\AA\ is relatively devoid of bright emission lines (with the exception of Ly$\alpha$, see Fig.\@~\ref{fig:popIII_spectrum}). Hence the true continuum level around \ion{He}{II} $\lambda 1640$ can be relatively well estimated from broadband photometry (using e.g.\@ the F150W or F200W filters at $z=8$). However, given the narrow width of the NIRSpec slits (0.2~arcsec), the physical scale associated with the \ion{He}{II} $\lambda 1640$ emission (probed with NIRSpec) may be different to the physical scale associated with the broadband continuum emission (probed by NIRCam). Thus it may in principle be difficult to accurately determine the \ion{He}{II} $\lambda 1640$ EW from a combination of NIRSpec and NIRCam data. While this potential issue should not matter for distinguishing between Pop III and non-Pop III galaxies (due to their very large separation in \ion{He}{II} $\lambda 1640$ EW), it may complicate the determination of the Pop III IMF. Nevertheless, provided that Pop III galaxies are very compact/point-sources (which seems reasonable to assume), then the \ion{He}{II} $\lambda 1640$ EW and thus the Pop III IMF should be able to be determined relatively accurately.

\subsubsection{\ion{He}{II} $\lambda 1640/$H$\beta$ ratio}

We also show the \ion{He}{II} $\lambda 1640$ / H$\beta$ ratio in Fig.\@~\ref{fig:zackrisson_lines_He_II}, where there is a ${\sim}0.5$~dex spread between the different Pop III IMFs. The benefits for considering this ratio in Pop III searches are twofold. 

Firstly, in the case of both a \ion{He}{II} $\lambda 1640$ and H$\beta$ detection, the line ratio can be computed. This ratio is less sensitive to the ionisation parameter $U$ than the \ion{He}{II} $\lambda 1640$ EW, with $\Delta \log \, \mathrm{EW} \approx 0.02$~dex from $U=-2.0$ to $U=-0.5$. The dependence on $f_\mathrm{cov}$ is similar, with $\Delta \log \, \mathrm{EW} \approx 0.1$~dex between $f_\mathrm{cov} = 0.5, 1$. Thus a measurement of the \ion{He}{II} $\lambda 1640$/H$\beta$ ratio would therefore in principle enable tighter constraints on the Pop III IMF. 

Secondly, although the \ion{He}{II} $\lambda 1640$ line is comparable in brightness to H$\beta$ for the Pop III.1 and Pop III.2 IMFs (see Table~\ref{tab:nirspec_observations}), it resides in the G140M NIRSpec grating at $z=8$. This grating is roughly $3\times$ less sensitive than the G395M grating which covers H$\beta$. Hence the integration times needed to detect \ion{He}{II} $\lambda 1640$ at $5\sigma$ are roughly $5$--$15\times$ longer than those needed to detect H$\beta$ at $5\sigma$. Indeed the integration times needed to detect \ion{He}{II} $\lambda 1640$ at $5\sigma$ will require medium-deep to ultra-deep integrations, with 4.76~h to 144.51~h needed for Pop III.1 and Pop III.2, respectively (assuming $M_* = 10^6~\mathrm{M}_\odot$). Thus, in many cases, only H$\beta$ will be detected with \emph{JWST}. Nevertheless, a \ion{He}{II} $\lambda 1640$ non-detection can still be used to place valuable upper limits on the Pop III IMF, as from the non-detection one can begin to rule out the more top heavy IMFs. Of course, if the source is reasonably magnified ($\mu \approx 2$--$3$), or is some small multiple of the nominal stellar mass, then even for the Pop III.2 IMF a line detection can be made within a medium \emph{JWST} GO program (25--75 hours).

\subsubsection{\ion{He}{II} $\lambda 1640$ luminosity}

We note that the \ion{He}{II} $\lambda 1640$ luminosity for Pop III galaxies begins decreasing immediately after an instantaneous starburst, with an even steeper decline after $1$--$1.5$~Myr. Thus the visibility window over which the \ion{He}{II} $\lambda 1640$ emission from Pop III galaxies can be detected with JWST ($\sim$$1$--$1.5$~Myr) is even shorter than the visibility windows for H$\beta$ and Ly$\alpha$ detections ($3$~Myr) and detections in broadband photometry (2.8--5.6~Myr). Thus, even if a true Pop III galaxy has been identified as a Pop III candidate from photometry, there is no guarantee that its Pop III nature can be definitively established from deep follow-up spectroscopy. As if the galaxy is observed too late after the initial starburst, its \ion{He}{II} $\lambda 1640$ emission will likely be too faint to detect, with the inferred \ion{He}{II} $\lambda 1640$ EW therefore being too low for the galaxy to be classified as Pop III. 

\subsection{\ion{He}{II} $\lambda 4686$}

The \ion{He}{II} $\lambda 4686$ recombination line is also accessible with NIRSpec. In principle this line could also be used as a spectroscopic Pop III indicator. Indeed, \ion{He}{II} $\lambda 4686$ resides in the $3\times$ more sensitive G395M grating, together with H$\beta$. However the \ion{He}{II} $\lambda 4686$ line is roughly an order of magnitude fainter than \ion{He}{II} $\lambda 1640$ (in the \citealt{Zackrisson2011} Pop III models, but see also \citealt{Osterbrock2006}). Hence a $5\sigma$ line detection would take approximately $10\times$ longer than for \ion{He}{II} $\lambda 1640$ and roughly $100\times$ longer than for H$\beta$, thereby making it impractical to use this line as a Pop III indicator.

\subsection{Metallicity constraints from [\ion{O}{III}] non-detections}

\begin{figure*}
\centering
\includegraphics[width=\linewidth]{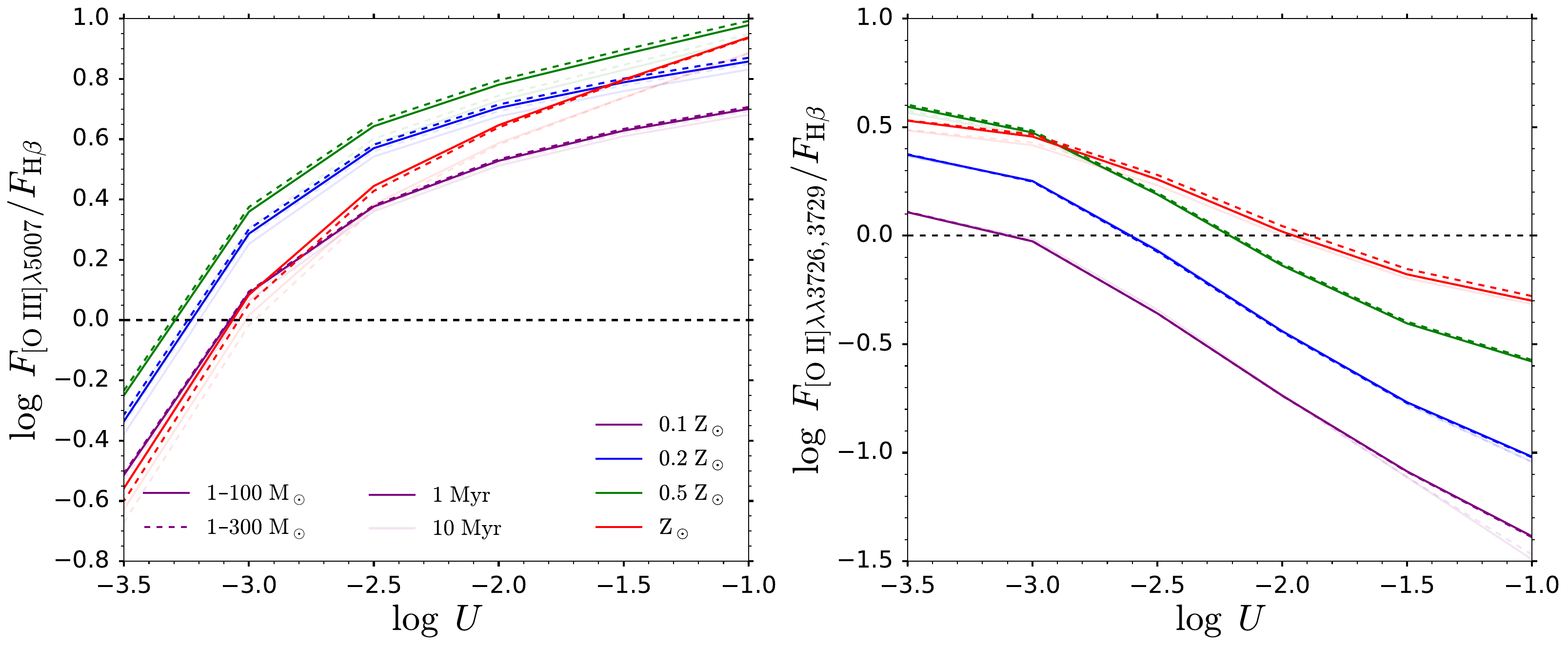}
\caption{The [\ion{O}{III}] $\lambda 5007$/H$\beta$ line ratio (left panel) and [\ion{O}{II}] $\lambda\lambda 3726,3729$/H$\beta$ ratio (right), as a function of ionisation parameter $U$ using the \citet{Nakajima2022} models. We show the line ratios for metal-rich Pop I galaxies at metallicities $Z = 0.1$~$\mathrm{Z}_\odot$ (purple), $Z = 0.2$~$\mathrm{Z}_\odot$ (blue), $Z = 0.5$~$\mathrm{Z}_\odot$ (green), $Z = $~$\mathrm{Z}_\odot$ (red), adopting a \citet{Kroupa2001} IMF with upper mass limits of 100~$\mathrm{M}_\odot$ (solid) and 300~$\mathrm{M}_\odot$ (dashed), at 1~Myr (unfaded) and 10~Myr (faded) after the starburst. At these high metallicities, always one of [\ion{O}{III}] $\lambda 5007$ or [\ion{O}{II}] is brighter than H$\beta$ (assuming no dust attenuation dimming the [\ion{O}{II}] doublet). Hence if both [\ion{O}{III}] $\lambda 5007$ and [\ion{O}{II}] are non-detected, but H$\beta$ is detected, then the metallicity of the galaxy is likely below $Z = 0.1~\mathrm{Z}_\odot$.}
\label{fig:nakajima_OIII_OII}
\end{figure*}

At $z=8$ the [\ion{O}{III}] $\lambda 5007$ line is also redshifted into the G395M grating, similar to H$\beta$. Conveniently, the [\ion{O}{II}] $\lambda\lambda 3726,\ 3729$ doublet also falls within the G395M grating at these redshifts. At intermediate-to-high metallicities and ionisation parameters, the [\ion{O}{III}] $\lambda 5007$ line is brighter than H$\beta$. Hence a H$\beta$ detection (at e.g.\@ $5\sigma$), together with a [\ion{O}{III}] $\lambda 5007$ non-detection, can place valuable upper limits on the metallicity of a galaxy.

In Fig.\@~\ref{fig:nakajima_OIII_OII} we show the [\ion{O}{III}] $\lambda 5007$/H$\beta$ line ratio (left panel) and the [\ion{O}{II}]/H$\beta$ line ratio (right panel) for the \citet{Nakajima2022} models. We have chosen to use the \citet{Nakajima2022} models for this analysis because they enable one to explore the dependence of the line ratios on ionisation parameter $U$, and also because they extend down to lower metallicities. We find that at most metallicity and ionisation parameter combinations the [\ion{O}{III}] $\lambda 5007$ line is brighter than H$\beta$. Hence in this regime the [\ion{O}{III}] $\lambda 5007$ line should be detected if H$\beta$ has been detected. It is only at low ionisation parameter that [\ion{O}{III}] $\lambda 5007$ begins to become fainter than H$\beta$, and we may expect a [\ion{O}{III}] non-detection. However, it is precisely at these low ionisation parameters that the [\ion{O}{II}] line becomes brighter than H$\beta$. Hence there is a sort of ``see-saw effect'', where one of [\ion{O}{III}] or [\ion{O}{II}] is always brighter than H$\beta$ (assuming no dust attenuation dimming the [\ion{O}{II}] doublet) across the range of possible ionisation parameters. Hence if both [\ion{O}{III}] $\lambda 5007$ and [\ion{O}{II}] are not detected, but H$\beta$ is detected, then the metallicity of the galaxy is likely below $Z = 0.1~\mathrm{Z}_\odot$. 

\begin{figure*}
\centering
\includegraphics[width=\linewidth]{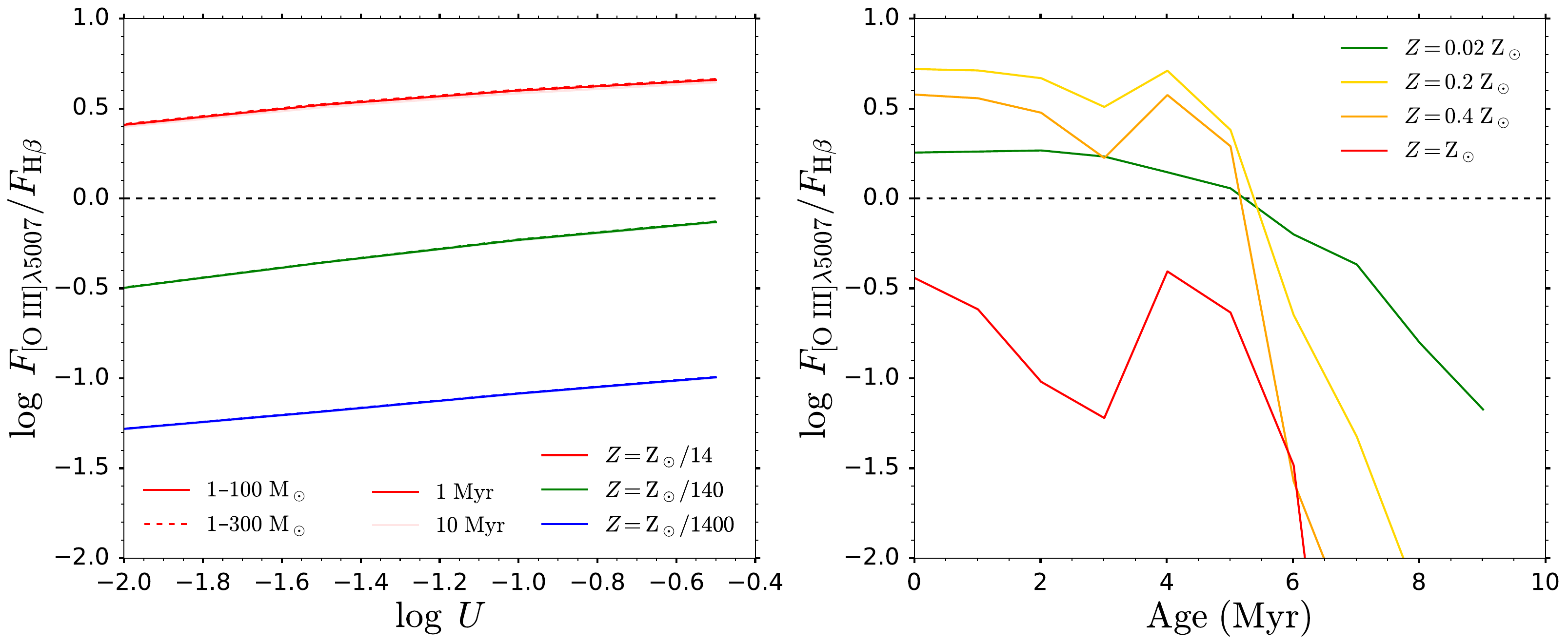}
\caption{The [\ion{O}{III}] $\lambda 5007$ /H$\beta$ line ratio for metal-poor Pop II galaxies using the \citet{Nakajima2022} models (left panel) at metallicities $Z = \mathrm{Z}_\odot / 14$ (red), $Z = \mathrm{Z}_\odot / 140$ (green) and $Z = \mathrm{Z}_\odot / 1400$ (blue). The line ratios for the non-Pop III \citet{Zackrisson2011} models (right), following the colour coding of Fig.\@~\ref{fig:zackrisson_colours1}. In the \citet{Zackrisson2011} models, [\ion{O}{III}] $\lambda 5007$ is brighter than H$\beta$ even at $Z = 0.02~\mathrm{Z}_\odot$. In the \citet{Nakajima2022} models, [\ion{O}{III}] $\lambda 5007$ begins to become fainter than H$\beta$ at $Z = \mathrm{Z}_\odot / 140$. Hence a H$\beta$ detection, together with an [\ion{O}{III}] $\lambda 5007$ non-detection, implies that the metallicity is likely below $Z = 0.02~\mathrm{Z}_\odot$. A $10\times$ longer integration time than was needed to detect H$\beta$ is needed to establish that the metallicity is $Z \leq \mathrm{Z}_\odot /140$.}
\label{fig:OIII_Hb_nakajima_zackrisson}
\end{figure*}

Pushing to even lower metallicities, we now turn to Fig.\@~\ref{fig:OIII_Hb_nakajima_zackrisson}. We show the [\ion{O}{III}] $\lambda 5007$/H$\beta$ ratios for the \citet{Nakajima2022} and \citet{Zackrisson2011} models in the left and right panels, respectively. We see that for the \citet{Zackrisson2011} models (which adopt a spherical geometry for the \ion{H}{II} region and therefore a radially varying ionisation parameter for each metallicity), the [\ion{O}{III}] $\lambda 5007$ line is brighter than H$\beta$ even at $Z = 0.02~\mathrm{Z}_\odot$, provided that we see the galaxy within ${\sim}5.5$~Myr after the starburst. We note that it is unlikely for non-Pop III galaxies to be mistakenly identified as Pop III candidates $\geq 5$~Myr after a starburst, as demonstrated by our colour selections in Section~\ref{subsec:colour_selection}, hence the weaker [\ion{O}{III}] $\lambda 5007$ emission in this regime is likely not a concern. If the radially-varying ionisation parameters in the \citet{Zackrisson2011} models are representative of the ionisation parameters we expect to find in galaxies at $z\sim8$, then a H$\beta$ detection, together with an [\ion{O}{III}] $\lambda 5007$ non-detection, implies that the metallicity is likely below $Z = 0.02~\mathrm{Z}_\odot$.

Now if we consider the \citet{Nakajima2022} models in the left panel, we find that the [\ion{O}{III}] $\lambda 5007$ line begins to dip below H$\beta$ in brightness at $Z = \mathrm{Z}_\odot /140$. Indeed, at $\log \ U = -2.0$, we find that the logarithm of the [\ion{O}{III}] $\lambda 5007$/H$\beta$ ratio is $-0.5$. Therefore if H$\beta$ is detected at $5\sigma$ in this scenario, a $10\times$ longer integration time would be needed to detect [\ion{O}{III}] $\lambda 5007$ (at 5$\sigma$). In other words, once H$\beta$ is detected, a $10\times$ longer integration time is needed to establish that the metallicity is $Z \leq \mathrm{Z}_\odot /140$. 

If such a deep integration in the G395M grating is considered, it is likely better placed to instead opt for an equally deep integration in the G140M grating instead. As discussed in Section~\ref{subsec:he_II}, and shown in Table~\ref{tab:nirspec_observations}, one requires a roughly $5$--$15\times$ longer integration to detect \ion{He}{II} $\lambda 1640$ compared to H$\beta$. A \ion{He}{II} $\lambda 1640$ line detection would yield much more conclusive constraints on both the Pop III nature of a source and the Pop III IMF, than can possibly be inferred from upper limits on the metallicity from non-detections of [\ion{O}{III}] $\lambda 5007$.

\subsection{H$\alpha$ vs.\@ H$\beta$ at $z=8$}

Our discussion of Balmer recombination lines has thus far centred around H$\beta$ for it lies within the NIRSpec spectral range at $z=8$. In contrast, the intrinsically brighter H$\alpha$ line is redshifted into the MIRI/MRS spectral range at this redshift. Although H$\alpha$ is roughly $3\times$ brighter than H$\beta$ \citep[barring any dust, which seems reasonable to ignore for pristine Pop III galaxies, see e.g.\@][]{Zackrisson2011}, MIRI is roughly $6\times$ less sensitive (at the H$\alpha$ wavelength) than NIRSpec (at the H$\beta$ wavelength). Hence the integration time needed for a $5\sigma$ detection of H$\alpha$ at $z=8$ will be roughly $4\times$ longer than that needed for H$\beta$. Furthermore, while there are additional merits to observing H$\beta$ due to added metallicity constraints from [\ion{O}{III}] non-detections (a line that is typically \emph{brighter} than H$\beta$), non-detections of the neighbouring [\ion{N}{II}] and [\ion{S}{II}] doublets around H$\alpha$ yield no such constraints as these lines are typically \emph{fainter} than H$\alpha$ \citep[see e.g.\@][]{Baldwin1981, Kewley2001, Steidel2014, Shapley2015a, Curti2022a}.

\subsection{Metal line (non-)detections with ALMA}

In the epoch of reionisation, [\ion{O}{III}] 88~\textmu m is likely the brightest emission line in the rest-frame FIR, usually being brighter than [\ion{C}{II}] 158~\textmu m \citep[see e.g.\@][]{Laporte2019, Harikane2020, Katz2022}. Here we briefly discuss the merits for following-up potential Pop III candidates with ALMA to rule out metal line emission and hence place constraints on the metallicity. Depending on the ISM temperature, the [\ion{O}{III}] $\lambda 5007$ line is between $1$--$10\times$ more luminous than [\ion{O}{III}] 88 \textmu m \citep[see e.g.\@][]{Moriwaki2018, Yang2020}. Furthermore, NIRSpec/G395M is roughly $2\times$ more sensitive than ALMA at emission line detection. Hence the integration times needed to detect [\ion{O}{III}] 88~\textmu m with ALMA are $4$--$400\times$ longer than what is needed to detect [\ion{O}{III}] $\lambda 5007$ with JWST. Thus NIRSpec is currently the optimal instrument for spectroscopic follow-up of potential Pop III candidates.

\subsection{Spectroscopic follow-up with 25+ m telescopes}

The next generation of 25+~m class, extremely large telescopes will likely play a complementary, synergistic role with \emph{JWST}. Indeed, with their ${\sim}5\times$ wider mirror diameters, they will have both a vastly larger light collecting area, as well as significantly less light smearing (by diffraction, provided that adaptive optics is used). This makes such telescopes ideal for deep follow-up spectroscopic observations of potential Pop III candidates identified by \emph{JWST}. Indeed, provided that the emission line of interest does not sit on top of a skyline, these 25+~m telescopes will be substantially more sensitive at emission line detection than \emph{JWST}. 

At $8 \leq z \leq 10$, the \ion{He}{II} $\lambda 1640$ line is redshifted into the H-band of TMT/IRMS, the planned infrared multi-object spectrograph for the Thirty Meter Telescope. Assuming a point source (which seems reasonable for a Pop III galaxy), the expected emission-line sensitivity in the H-band is $5.6\times10^{-19}$~erg~s$^{-1}$~cm$^{-2}$ at $10\sigma$ within a 1~h integration \citep{Skidmore2015}. Similarly, ELT/MOSAIC, the infrared multi-object spectrograph for the upcoming Extremely Large Telescope, is expected to have an emission-line sensitivity of $6.3\times10^{-19}$~erg~s$^{-1}$~cm$^{-2}$ \citep{Evans2015}. Thus these next-generation ground-based spectrographs will be roughly an order of magnitude more sensitive than NIRSpec/G140M, which has a sensitivity of $3.8\times10^{-18}$~erg~s$^{-1}$~cm$^{-2}$. Hence the integration times needed to achieve a line detection of \ion{He}{II} $\lambda 1640$ with 25+~m telescopes will essentially be ${\approx}100\times$ shorter than the integration times needed with \emph{JWST}. Thus such telescopes should readily be able to spectroscopically follow-up potential Pop III candidates (identified through \emph{JWST} NIRCam+MIRI photometry and/or slitless spectroscopy) and confirm their Pop III nature through a detection of the \ion{He}{II} $\lambda 1640$ line \citep[for more details, see][]{Grisdale2021}, requiring roughly only 0.05~h and 1.45~h of integration for the Pop III.1 and Pop III.2 IMFs, assuming a $M_* = 10^6~\mathrm{M}_\odot$ Pop III galaxy. For lower mass Pop III galaxies, such as those residing in minihalos with $M_* = 10^{2\text{--}3}~\mathrm{M}_\odot$, even more sensitive instrumentation will be needed, such as the potential Moon-based 100~m Ultimately Large Telescope discussed in \citet{Schauer2020}. 

Ultimately to find Pop III galaxies we will need to use a combination of photometry to find good candidates and then spectroscopy to follow them up and ensure that no metal lines are present.  However, with the large number of distant galaxies that \emph{JWST} is finding, and will continue to find, a photometric method for selecting candidates, such as what we present here, is necessary to remove systems that have properties that are not consistent with being Pop III galaxies.

\section{On the stellar masses, redshifts and likelihood of encountering Pop III galaxies} \label{sec:caveats}

In this section we briefly discuss our assumptions regarding the stellar masses of Pop III galaxies, their redshifts and the general likelihood of encountering Pop III galaxies with \emph{JWST}.

Throughout this paper we have assumed a nominal Pop III stellar mass of $M_* = 10^6~\mathrm{M}_\odot$. We have adopted this stellar mass not because it is realistic (as theory/simulations generally expect lower Pop III masses), but rather because it is practical, likely representing the least massive Pop III galaxies that can be detected with \emph{JWST} (without gravitational lensing). Indeed, as has already been discussed in the introduction, based off of simple virial arguments for an atomic cooling halo, one might expect a typical Pop III stellar mass of $M_* = 10^5~\mathrm{M}_\odot$. Depending on the assumed star formation efficiency, and as shown in e.g.\@ semi-analytic models of Pop III star formation \citep[see e.g.\@][]{Hartwig2018, Hartwig2022}, the typical total mass of Pop III stars formed may be $M_* = 10^4~\mathrm{M}_\odot$ or lower. Moreover, this Pop III star formation likely proceeds over some small (but finite) timescale, rather than the instantaneous Pop III starburst we assume in our analysis. This non-instantaneous Pop III star formation timescale can be important when it comes to the observability of Pop III galaxies, due to the short window over which the continuum emission (2.8--5.6~Myr) and line emission (1--1.5~Myr for \ion{He}{II} $\lambda 1640$) can be detected, even for an instantaneous starburst. Thus the massive ($M_* = 10^6~\mathrm{M}_\odot$), instantaneous starburst Pop III galaxies we assume in this work may be very rare \citep[though see also][]{Inayoshi2018}. Now, for each order of magnitude decrease in Pop III stellar mass, the required integration times will increase by 100$\times$ the tabulated values in Tables 1--4. Hence, as has been discussed earlier, strong gravitational lensing will be essential to detect these lower mass Pop III galaxies, likely requiring larger magnifications than the $\mu=$ 2--3 we typically assume in this work. Indeed, the magnification factors for compact objects (like Pop III galaxies) can indeed be very high \citep[though rare, see e.g.\@][]{Zackrisson2012, Zackrisson2015, Windhorst2018}, with for example magnifications of $\mu =$ 30--70 having been observed for young massive star clusters of $M_* \sim 10^{6\text{--}7}~\mathrm{M}_\odot$ and with effective radii $R_\mathrm{e}\sim$ 1--20~pc in the $z \sim 6$ Sunrise arc \citep{Vanzella2023}. Thus although such lower stellar mass Pop III galaxies may be more realistic (and hence more abundant), the steep gravitational lensing requirement (and rarity thereof) will likely still make it challenging to encounter (and detect) such systems with \emph{JWST}. 

With regards to the redshifts of these Pop III galaxies, we have based our colour selections (in the main body of the paper) around a nominal redshift of $z \sim 8$ (this is extended to $z \sim 10$ in Appendix~\ref{app:z10}). This redshift was adopted, again, not because it is the most realistic redshift to find Pop III galaxies (as the Universe is more chemically pristine at even higher redshifts), but rather because it is practical, being roughly the highest redshift at which some key Pop III tracers are still accessible by NIRCam/NIRSpec, which allows for a much more efficient identification and spectroscopic follow-up of Pop III candidates. Indeed, above this redshift, the lack of metal line emission (traced by [\ion{O}{III}] $\lambda$5007), as well as the (marginally) stronger H$\beta$ emission, are redshifted into the MIRI regime, with both the MIRI imager and spectrograph being $>$7$\times$ less sensitive than their NIRCam/NIRSpec counterparts (thus requiring $>$50$\times$ the integration time) and having a smaller imaging footprint ($4\times$ smaller, 2.35~arcmin$^2$ vs.\@ 9.7~arcmin$^2$) and lack of spectroscopic multiplexing (1 object vs.\@ dozens of galaxies). Of course, the greater the redshift of the galaxy, roughly the greater the likelihood of it being Pop III, simply due to the Universe still being relatively chemically pristine due to a lack of enrichment by prior star formation. Indeed, according to the simulations in \citet{Pallottini2014}, the mean baryon metallicity is $Z = 10^{-3.75}~\mathrm{Z}_\odot$ at $z=8$, while only being $Z = 10^{-4.75}~\mathrm{Z}_\odot$ at $z=10$, with $Z_\mathrm{crit} = 10^{-4}~\mathrm{Z}_\odot$ commonly being assumed to be the critical metallicity marking the transition point between Pop III and Pop II star formation.

Finally, we also assume that our Pop III galaxies are purely composed of Pop III stars. Of course, galaxies with a mixture of Pop III and chemically-enriched Pop II stars are also possible \citep[see e.g.\@][]{Sarmento2017, Riaz2022, Venditti2023} Provided that these two stellar populations are spatially separate (as seen through the resolution of \emph{JWST}), it should in principle be possible to disentangle the Pop III component from the metal-enriched components, provided that a spatially resolved analysis is undertaken (and the necessary signal-to-noise is available), enabling our pure Pop III selections and diagnostics to still be applied. However, if these two populations are mixed (from the observer's perspective), then the Pop II stars will likely erode the characteristic Pop III signatures that we have discussed in this article, with the extent of the erosion depending on the relative brightness \citep[for more details, see e.g.\@][]{Riaz2022} of the Pop II stars/gas to the Pop III (and also on the metallicity of the Pop II stars/gas). For example, [\ion{O}{III}] and [\ion{S}{III}] emission from the Pop II gas will contaminate the pure Pop III emission, causing the galaxy to (begin to) drift out of the Pop III region of the F444W$-$F560W, F560W$-$F770W colour--colour plane, causing such systems to be potentially missed by our pure Pop III colour selections, depending on the brightness and metallicity of the Pop II stars/gas. Additionally,  the additional stellar+nebular continuum, plus the lack of \ion{He}{II} $\lambda$1640 emission originating from the Pop II stars/gas, can result in a net (Pop III + Pop II) \ion{He}{II} $\lambda$1640 equivalent width that falls below the expected range for pure Pop III galaxies, again, causing such Pop III + Pop II mixed systems to be missed in our pure Pop III classifications.

With the aforementioned caveats in mind, we now briefly discuss the likelihood of encountering Pop III galaxies with \emph{JWST}. We will adopt three different approaches to estimate the expected number of bright/massive Pop III galaxies on the sky. 

Firstly, we use the expected Pop III cosmic star formation rate density, assuming a nominal value of 10$^{-4}$~$\mathrm{M}_\odot$~yr$^{-1}$~Mpc$^{-3}$ at $z=8$ and $z=10$ \citep[see e.g.\@][]{Sarmento2018}. Similar to the rest of our analysis, we make the simplifying assumption that all Pop III galaxies are massive, i.e.\@ with $M_* = 10^6~\mathrm{M}_\odot$, and form their stars over a very short/almost instantaneous timescale, i.e.\@ $\Delta t = 0.1$~Myr (which is comparable to the free-fall time for a gas cloud of the same mass and a radius of 5~pc). Following the methodology in \citet{Windhorst2018}, the 10$^{-4}$~$\mathrm{M}_\odot$~yr$^{-1}$~Mpc$^{-3}$ Pop III cosmic star formation rate density thus translates into a Pop III comoving number density of $n = 10^{-5}$~Mpc$^{-3}$. This in turn corresponds to an expected number of 0.20 and 0.16 massive Pop III galaxies per NIRCam field of view and per unit redshift interval, at $z=8$ and $z=10$, respectively, i.e.\@ we expect to detect one massive Pop III galaxy every 5 or 6 NIRCam pointings. Of course, not all Pop III galaxies will be so massive (in fact, we likely expect such systems to be rare, though see \citealt{Inayoshi2018} for a formation mechanism for such massive Pop III systems), so the expected number of actual NIRCam pointings needed will likely be much higher (\citealt{Inayoshi2018} find this to be $\sim$30 NIRCam pointings at $z=15$).

Secondly, we use another result from \citet{Sarmento2018}, namely that the expected fraction of Pop III-bright galaxies (i.e.\@ galaxies with at least 75\% of their flux coming from Pop III stars, thus akin to the pure Pop III galaxies we assume in this work) is 1\% and 5\% of all galaxies brighter than $m_\mathrm{UV} = 31.4$~mag at $z=8$ and $z=10$, respectively. Using the \citet{Adams2023b} UV luminosity functions, this translates into an expected comoving density of $1.8\times10^{-4}$~Mpc$^{-3}$ and $1.6\times10^{-4}$~Mpc$^{-3}$ Pop III-bright galaxies at $z=8$ and $z=10$, respectively. In turn, this amounts to 3.6 and 2.7 Pop III-bright galaxies (with $m_\mathrm{UV} < 31.4$~mag) per NIRCam field of view and per unit redshift interval. As shown in Table~\ref{tab:imaging_observations}, ultra-deep ($\sim$100~h) integrations are required to reach such depths (31.4~AB mag) with \emph{JWST}/NIRCam.

Thirdly, we use the result from \citet{Stiavelli2010}, who find that the expected Pop III galaxy formation rate is ${\sim}2\times10^{-8}$ Mpc$^{-3}$~yr$^{-1}$ at $z=10$. Following their line of argument, assuming a Pop III visibility time of 2~Myr, this corresponds to an expected Pop III surface density of $\sim$800 galaxies per NIRCam FoV per unit redshift interval. However, the majority of these objects will likely be too faint to detect (without gravitational lensing), likely being less massive than the nominal $M_* = 10^6~\mathrm{M}_\odot$ we assume for detection with \emph{JWST}. Indeed, \citet{Stiavelli2010} note that at $z=10$, an atomic cooling halo (with $T = 10^4$~K) has a halo mass of ${\sim}3.7\times10^{7}~\mathrm{M}_\odot$ and a baryonic mass of ${\sim}6\times10^{6}~\mathrm{M}_\odot$, which would therefore require a star formation efficiency $\epsilon = 0.15$ to produce $10^6~\mathrm{M}_\odot$ in stars, which they deem unlikely. 

For a detailed investigation into the prospects of detecting Pop III galaxies in blind surveys, we refer the reader to  \citet{Vikaeus2022}. They find that photometric surveys will likely pick up lensed Pop III galaxies if the Pop III star formation efficiencies are sufficiently high ($\epsilon > 0.005$), though this is model-dependent. Moreover, the photometric prospects are better for the \emph{Roman Space Telescope} (\emph{RST}) than for \emph{JWST} or \emph{Euclid}. However, simulations predict that the Pop III number densities are likely too low for spectroscopic detection of \ion{He}{II} $\lambda$1640 for gravitationally lensed Pop III galaxies in wide-area surveys with \emph{JWST}, \emph{RST} and \emph{Euclid}. Instead, they argue that targeted cluster lensing surveys with \emph{JWST} likely offer the best prospects for spectroscopic detection of Pop III galaxies.

\section{Summary and conclusions} \label{sec:conclusions}

In this paper we investigate the prospects for observing and identifying Pop III galaxies with \emph{JWST}. We based our analysis on two different, but complementary Pop III models: \citet{Zackrisson2011} and \citet{Nakajima2022}. The basis for this approach was twofold. Firstly, to establish the robustness of our colour selections and spectroscopic diagnostics in identifying and confirming Pop III candidates. Secondly, to draw upon the synergy between these two different models. With the \citet{Zackrisson2011} models we are able to determine the expected apparent magnitudes of $z=8$ Pop III sources in various NIRCam and MIRI filters, as well as the expected line fluxes of the brightest, most sensitive tracers of Pop III stars. On the other hand, with the \citet{Nakajima2022} models we explore the dependence of emission line diagnostics on the ionisation parameter $U$ and push our analysis of non-Pop III galaxies down to even lower (but still non-pristine) metallicities. In the main body of the paper we concentrated on $z\sim8$ galaxies, as at these redshifts the H$\beta$ and [\ion{O}{III}] $\lambda 5007$ lines are still accessible with NIRCam and NIRSpec. In the appendix, we undertook a similar analysis for $z\sim10$ Pop III galaxies. Our main findings at $z\sim8$ are summarised below.

Assuming a $z=8$ Pop III galaxy at the nominal stellar mass of $M_* = 10^6~\mathrm{M}_\odot$, we find that deep NIRCam imaging (${\sim}28.5$--$30.0$~AB~mag, $1$--$20$~h) and deep-to-ultra-deep MIRI F560W imaging (${\sim}27.5$--$29.0$~AB~mag, $10$--$100$~h) will be needed to achieve a $5\sigma$ detection, even for the more top-heavy Pop III.1 and Pop III.2 IMFs. Indeed, detections in the MIRI F770W band and beyond will likely be very challenging, with the flux boost from either strong gravitational lensing ($\mu = 10$) and/or fortuitous imaging of exceptionally massive ($M_* = 10^7~\mathrm{M}_\odot$) Pop III galaxies being essential. Hence deep imaging of lensing clusters and/or deep-and-wide imaging of many fields will likely be needed to find such bright Pop III sources.

We discuss a number of colour--colour selections to identify Pop III galaxy candidates, outlining the physical basis behind the colour selection, the redshift window of applicability and potential Galactic and low-$z$ contaminants. The main physical driver behind our Pop III colour selections are the (marginally) higher Balmer EWs for Pop III galaxies, together with their lack of (high EW) metal lines, which each contribute a significant ``magnitude excess'' in their respective filters. 

We find that a combination of NIRCam and MIRI photometry yields the most reliable Pop III colour selection, advocating for the use of the F444W$-$F560W, F560W$-$F770W colour--colour selection at $z\sim8$. Assuming $5\sigma$ flux density detections, there will likely be some contamination of the Pop III region of the colour--colour plane by both metal-poor ($Z = 0.02~\mathrm{Z}_\odot$) and metal-rich ($Z = \mathrm{Z}_\odot$) galaxies. These systems have comparable H$\alpha$ EWs (in F560W) to Pop III, while also having relatively low EW [\ion{O}{III}] $\lambda 5007$ (in F444W) and [\ion{S}{III}] $\lambda\lambda 9069,9531$ (in F770W) emission lines. This colour selection can be applied at $7.55 < z < 8.10$. However, if $Z = \mathrm{Z}_\odot$ galaxies are not considered as a contaminant concern at these redshifts, then the colour selection can be applied over the wider $7.00 < z < 8.35$ redshift interval. Even with perfect photometry (i.e.\@ zero observational errors) however, there will still be contamination by extremely metal-poor galaxies ($Z \leq 0.02~\mathrm{Z}_\odot$). However, with deep follow-up spectroscopy, one will be able to readily distinguish between Pop III galaxies and these non-pristine contaminants.

Given that detections in the MIRI F770W filter will be challenging, we also discuss an alternative NIRCam+MIRI colour selection that substitutes the MIRI F770W filter for the much more sensitive NIRCam F410M (or F356W) filter. This colour selection therefore does not require as substantial of a flux boost from gravitational lensing and/or elevated Pop III stellar masses to be applied, but does suffer from contamination by $Z = \mathrm{Z}_\odot$ galaxies. 

We also introduce two other colour--colour selections, which only require NIRCam photometry, thus needing even shorter integration times to actually apply. These are the [\ion{O}{III}] $\lambda 5007$ vs.\@ Ly$\alpha$ selection (i.e.\@ F410M$-$F444W, F115W$-$F150W), and the [\ion{O}{III}] $\lambda 5007$ vs.\@ [\ion{O}{II}] selection (i.e.\@ F410M$-$F444W, F277W$-$F335M). However, these colour selections have their own caveats which likely limit their practical use.

Additionally, we consider the prospects for identifying potential Pop III candidates with slitless spectroscopic emission-line surveys with \emph{JWST}. We focused on three bright, sensitive tracers of Pop III stars: H$\beta$, Ly$\alpha$ and \ion{He}{II} $\lambda 1640$. Ly$\alpha$ is intrinsically very bright and so in principle would be readily available with NIRISS, requiring only ${\sim}1$~h integrations. However, owing to IGM attenuation/scattering the observed Ly$\alpha$ flux will be substantially reduced, thereby likely making this line impractical to search for in slitless observations. Instead, H$\beta$ and \ion{He}{II} $\lambda 1640$ seem more promising, though would still require deep-to-ultra-deep observations, at 15--150~h and 50--1700~h, respectively. Hence the flux boost from moderate-to-high lensing ($\mu \sim 5$) and or from observing moderately massive Pop III galaxies ($M_* \sim 5\times10^6~\mathrm{M}_\odot$) will likely again be essential.

Finally, we focus on the additional Pop III constraints that can be derived from follow-up spectroscopy on potential Pop III candidates with NIRSpec. We find that with the NIRSpec/G395M grating, $5\sigma$ detections of H$\beta$ can readily be achieved (${\sim}$1--10~h) for the Pop III.1 and Pop III.2 IMFs. In principle, a measurement of the H$\beta$ rest-frame EW can be used to distinguish between Pop III and non-Pop III galaxies. However the difference in EW immediately after a starburst is in fact very small, being only ${\sim}0.1$~dex. Furthermore, owing to slight differences in the H$\beta$ EWs predicted by different models, as well as a secondary dependence on covering fraction $f_\mathrm{cov}$, it may be difficult in practice to make definitive inferences on the Pop III nature of a galaxy from a measurement of the H$\beta$ EW.

Echoing earlier works \citep[see e.g.\@][]{Schaerer2002, Schaerer2003, Raiter2010b, Grisdale2021, Nakajima2022}, we find that the \ion{He}{II} $\lambda 1640$ emission line serves as the optimal Pop III indicator. Indeed, a measurement of the \ion{He}{II} $\lambda 1640$ line enables one to both definitively identify Pop III galaxies, as well as place constraints on the Pop III IMF. If both \ion{He}{II} $\lambda 1640$ and H$\beta$ are detected then there are merits to computing their line ratio, as this is less sensitive to the (perhaps unknown) ionisation parameter $U$. Deep-to-ultra-deep integrations (5--150~h) with NIRSpec/G140M will be needed to detect \ion{He}{II} $\lambda 1640$ at $5\sigma$ at $z\sim8$. With moderate ($\mu=2$--$3$) lensing and/or moderately massive ($2$--$3\times10^6~\mathrm{M}_\odot$) Pop III galaxies, such line detections can be achieved in medium-sized \emph{JWST} GO programs.

Given the characteristically high \ion{He}{II} $\lambda 1640$ EWs for Pop III galaxies (relative to non-Pop III), we investigate the prospects for identifying Pop III candidates from NIRCam imaging surveys that target this line. As Pop III galaxies still have a relatively low (in an absolute sense) \ion{He}{II} $\lambda 1640$ EW ($\sim$10--100~\AA, rest-frame), we find that medium-band imaging will be essential, as even then the photometric signal we are looking for is rather small ($\Delta m =$ 0.30~mag, 0.15~mag for the Pop III.1 and Pop III.2 IMFs, respectively). Therefore, owing to NIRCam's greater sensitivity and imaging footprint compared to MIRI, NIRCam medium-band imaging campaigns that search for high EW \ion{He}{II} $\lambda 1640$ emitters may perhaps be a viable alternative to the demandingly deep imaging required with MIRI F560W and F770W.

Since [\ion{O}{III}] $\lambda 5007$ and H$\beta$ are both simultaneously observable with NIRSpec/G395M at $z=8$, we looked into the metallicity constraints that can be placed from a $5\sigma$ H$\beta$ detection and [\ion{O}{III}] $\lambda 5007$ non-detection. We find that this likely implies a metallicity $Z \leq 0.02~\mathrm{Z}_\odot$. Lowering the upper limit on the metallicity down to $Z \leq \mathrm{Z}_\odot/140$ would require an integration time that is $10\times$ longer than needed to detect H$\beta$ at $5\sigma$.

Thus, observing and identifying Pop III galaxies with certainty will be a challenging task for \emph{JWST}, likely requiring very deep imaging and spectroscopy to achieve the required S/N for robust detections. All of \emph{JWST}'s key scientific instruments, namely NIRCam, MIRI, NIRISS and NIRSpec, will likely play a crucial role in our search for the first stars. Indeed in this ambitious quest we likely will need to leverage on the flux boost from gravitational lensing and hope for fortuitous imaging of moderate-to-high mass Pop III galaxies. However, it is with the future synergy between space and ground, drawing on the enhanced sensitivity of the next-generation of extremely large telescopes, that we will be able to readily detect the bright \ion{He}{II} $\lambda 1640$ signature of Pop III galaxies.

\section*{Acknowledgements}

We thank the referee for their useful comments which helped to improve this article. JT thanks Rebecca Bowler, Andrew Bunker, Mark Dickinson, Jeyhan Kartaltepe, Harley Katz and Laura Pentericci for useful discussions which helped improve the quality of this manuscript. JT, CC and NA acknowledge support by the ERC Advanced Investigator Grant EPOCHS (788113) from the European Research Council (ERC) (PI Conselice). DA and TH acknowledge support from STFC in the form of PhD studentships. RM acknowledges support from the ERC Advanced Grant 695671 ‘QUENCH’. RM acknowledges support by the Science and Technology Facilities Council (STFC). EZ acknowledges funding from the Swedish National Space Board. LF acknowledges financial support from Coordenação de Aperfeiçoamento de Pessoal de Nível Superior - Brazil (CAPES) in the form of a PhD studentship.

This work is based on observations made with the NASA/ESA \textit{Hubble Space Telescope} (\emph{HST}) and NASA/ESA/CSA \textit{James Webb Space Telescope} (\emph{JWST}) obtained from the \texttt{Mikulski Archive for Space Telescopes} (\texttt{MAST}) at the \textit{Space Telescope Science Institute} (STScI), which is operated by the Association of Universities for Research in Astronomy, Inc., under NASA contract NAS 5-03127 for \emph{JWST}, and NAS 5–26555 for \emph{HST}. These observations are associated with \emph{JWST} program 1345.

This research made use of Astropy,\footnote{http://www.astropy.org} a community-developed core Python package for Astronomy \citep{astropy2013, astropy2018}. 

\section*{Data Availability}

The \citet{Zackrisson2011} models used in this article are available at: https://www.astro.uu.se/~ez/yggdrasil/yggdrasil.html. The \citet{Nakajima2022} models used in this article will be shared on reasonable request to Kimihiko Nakajima. Any remaining data underlying the analysis in this article will be shared on reasonable request to the first author.



\bibliographystyle{mnras}
\bibliography{main.bib} 




\appendix

\section{[\ion{O}{III}]$-$H$\alpha$, H$\alpha -$[\ion{S}{III}] colour selection with Nakajima models} \label{app:nakajima_colours}

\begin{figure*}
\centering
\includegraphics[width=.65\linewidth]{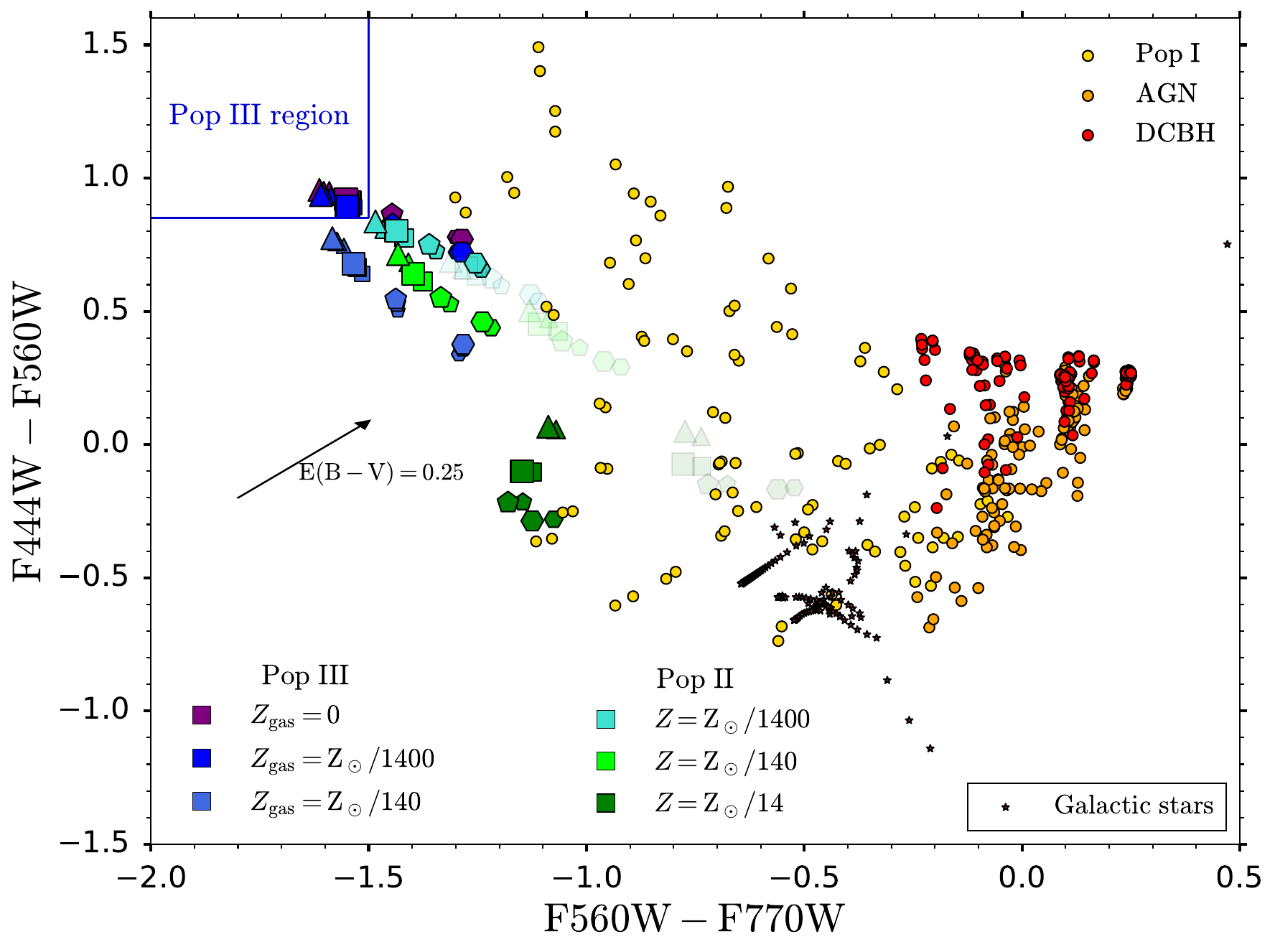}
\caption{Similar to Fig.\@~\ref{fig:zackrisson_colours1}, but now using the \citet{Nakajima2022} models. We show the colours for Pop III galaxies (shades of blue) with gas-phase metallicities $Z_\mathrm{gas} = 0$ (purple), $\mathrm{Z}_\odot/1400$ (dark blue) and $\mathrm{Z}_\odot/14$ (light blue). The colours for metal-poor Pop II galaxies are also shown (shades of green) with metallicities $Z = \mathrm{Z}_\odot/1400$ (cyan), $\mathrm{Z}_\odot/140$ (green) and $Z=\mathrm{Z}_\odot/14$ (dark green). Finally, we also show the colours of metal-rich Pop I galaxies (yellow), AGN (orange) and direct collapse black holes (DCBH, red). Pop III and Pop II galaxies are plotted with different symbols based off of their ionisation parameter: $U = -2.0$ (triangle), $U=-1.5$ (square), $U=-1.0$ (pentagon) and $U=-0.5$ (hexagon). Larger symbol sizes for Pop III and Pop II galaxies represent a higher upper mass limit for the IMF. These are 1--100~$\mathrm{M}_\odot$, 1--300~$\mathrm{M}_\odot$ and 50--500~$\mathrm{M}_\odot$ for Pop III galaxies, and 1--100~$\mathrm{M}_\odot$ and 1--300~$\mathrm{M}_\odot$ for Pop II galaxies. We show galaxy colours both 1~Myr (unfaded) and 10~Myr (faded) after an instantaneous starburst. As was seen with the \citet{Zackrisson2011} models, Pop III galaxies have distinct colours within this colour--colour plane. Extremely metal-poor galaxies with $Z = \mathrm{Z}_\odot/1400, \mathrm{Z}_\odot/140$ occupy similar regions of the colour--colour plane to the chemically pristine Pop III galaxies. Such galaxies will therefore likely be picked up, in addition to true Pop III galaxies, in colour--colour selections. For reference, the boundaries of the Pop III region (solid blue) from Fig.\@~\ref{fig:zackrisson_colours1}, which was based on the \citet{Zackrisson2011} models, are also shown.} 
\label{fig:nakajima_colours1}
\end{figure*}

In Fig.\@~\ref{fig:nakajima_colours1} we again show the F444W$-$F560W, F560W$-$F770W colour--colour selection for $z=8$ Pop III galaxies, but now using the \citet{Nakajima2022} models. Pop III galaxies are shown in various shades of blue, with the purple, dark blue and lighter blue symbols corresponding to $Z_\mathrm{gas} = 0,\ \mathrm{Z}_\odot/1400,\ \mathrm{Z}_\odot/14$, respectively. The metal-poor Pop II galaxies are shown in various shades of green, with the cyan, green and dark green symbols corresponding to $Z = \mathrm{Z}_\odot/1400,\ \mathrm{Z}_\odot/140,\ \mathrm{Z}_\odot/14$, respectively. Metal-rich Pop I galaxies are shown in yellow, and AGN and DCBH are shown in orange and red, respectively. For the Pop III and Pop II models, the triangles, squares, pentagons and hexagons correspond to ionisation parameters $U = -2.0, -1.5, -1.0, -0.5$, respectively. Larger symbols represent a higher upper mass limit for the IMF. The unfaded and faded symbols correspond to the galaxy colours 1~Myr and 10~Myr after the starburst, respectively. Unlike the Pop III and Pop II models, we do not further separate the Pop I or AGN/DCBH models in terms of metallicity, ionisation parameter, IMF or age.  

As was seen with the \citet{Zackrisson2011} models, Pop III galaxies have distinct colours within this colour--colour plane. The colours for Pop III galaxies with $Z_\mathrm{gas}=0$ are very similar to those in the \citet{Zackrisson2011} models. Owing to the wider range of ionisation parameters $U$ available in the \citet{Nakajima2022} models, together with the fact that models for Pop III stars with both pristine and very metal poor gas are available, the locus of points occupied by Pop III galaxies in this colour--colour plane immediately after the starburst is larger. Furthermore, as the \citet{Nakajima2022} models also provide spectra for non-pristine Pop II galaxies with very low metallicities, we see that these extremely metal-poor galaxies with $Z = \mathrm{Z}_\odot/1400, \mathrm{Z}_\odot/140$ occupy similar regions of the colour--colour plane to the chemically pristine Pop III galaxies. With the \citet{Zackrisson2011} models we found that Pop III galaxies and $Z = 0.02~\mathrm{Z}_\odot$ galaxies still occupied distinct regions of the colour--colour plane. Hence extremely metal-poor galaxies ($Z \leq 0.02~\mathrm{Z}_\odot$) will likely be picked up, in addition to true Pop III galaxies, in such colour--colour selections. However, with follow-up spectroscopy, one can still readily distinguish between Pop III galaxies and these non-pristine contaminants, through measurements of the \ion{He}{II} $\lambda 1640$ rest-frame EW and line ratio with H$\beta$. 

\section{Observing and identifying \lowercase{$z\sim10$} Pop III galaxies} \label{app:z10}

In the main body of the paper we focused on the observability and identification of Pop III galaxies at $z\sim8$, because this is a relatively high redshift window where [\ion{O}{III}] $\lambda 5007$ and H$\beta$ are still accessible by NIRCam and NIRSpec. Indeed, [\ion{O}{III}] $\lambda 5007$ gets redshifted out of the NIRCam and NIRSpec ranges at $z=9$ \& $9.6$, respectively. H$\beta$ is redshifted out of the NIRCam and NIRSpec coverage at $z=9.3$ \& $9.9$. In this section we instead consider the prospects for observing and identifying Pop III galaxies at $z\sim10$. The $z\sim10$ colour selections that we will introduce are analogous to the $z\sim8$ colour selections discussed in the main body of this paper, but now use redder filters as the high EW emission lines underpinning the colour selection are redshifted to longer wavelengths.

\subsection{The observability of $z\sim10$ Pop III galaxies}

In Tables~\ref{tab:imaging_observations2}, \ref{tab:slitless_observations2} and \ref{tab:nirspec_observations2} we show the expected apparent magnitudes and line fluxes for $z=10$ Pop III galaxies at the nominal stellar mass $\log (M_*/\mathrm{M}_\odot) = 6$, as well as the integration times needed to achieve $5\sigma$ detections. 

For the expected apparent magnitudes in Table~\ref{tab:imaging_observations2}, we see that $5\sigma$ detections NIRCam will still be possible in relatively short exposure times (1--30~h). However, in this case the H$\beta$ and [\ion{O}{III}] $\lambda 5007$ lines have been redshifted into the MIRI F560W band, which is substantially less sensitive than NIRCam F444W. Furthermore, the [S~III] $\lambda 9069, 9531$ doublet, which formed the basis for our most robust colour--colour selection, has now been redshifted into the MIRI F1000W band, which is again much less sensitive than MIRI F770W. Obtaining $5\sigma$ detections in the MIRI F560W and F770W bands will likely require moderate ($\mu > 3$) gravitational lensing and/or imaging of a moderately massive ($M_* \geq 3\times10^{6}$~$\mathrm{M}_\odot$) Pop III galaxy. $5\sigma$ detections in the F1000W filter are only possible with an extreme flux boost ($\geq 10$) from lensing/larger stellar mass.

\begin{table*}
\begin{center}
\resizebox{\linewidth}{!}{
\begin{tabular}{ |c|c|c|c|c|c|c|c|c|c|c|c|c|c| } 
\hline
IMF & F090W & F115W & F150W & F182M & F200W & F277W & F335M & F356W & F410M & F444W & F560W & F770W & F1000W \\
\hline
Pop III.1 & 38.63 & 37.45 & 27.94 & 28.74 & 28.83 & 28.99 & 28.93 & 28.89 & 28.95 & 28.98 & 29.07 & 28.34 & 29.61 \\
& N/A & N/A & 0.34 & 2.20 & 1.26 & 1.58 & 2.84 & 1.15 & 4.03 & 2.74 & 134.28 & 131.83 & 4699.15 \\
\hline
Pop III.2 & 39.78 & 38.62 & 29.08 & 29.94 & 30.00 & 30.12 & 30.07 & 30.03 & 30.09 & 30.12 & 30.25 & 29.50 & 30.79 \\
& N/A & N/A & 2.84 & 20.03 & 10.90 & 12.87 & 23.20 & 9.41 & 32.33 & 22.78 & 1180.37 & 1116.91 & 41306.55 \\ 
\hline
Pop III Kroupa & 41.73 & 40.62 & 31.04 & 31.89 & 31.94 & 32.07 & 32.03 & 31.99 & 32.05 & 32.09 & 32.24 & 31.49 & 32.80 \\
& N/A & N/A & 103.17 & 740.93 & 388.66 & 458.75 & 858.17 & 289.45 & 1196.56 & 842.51 & 46133.76 & 43653.48 & 1644443.18 \\ 
\hline
\end{tabular}}
\caption{Similar to Table~\ref{tab:imaging_observations}, but now for Pop III galaxies at $z=10$.}
\label{tab:imaging_observations2}
\end{center}
\end{table*}

In terms of slitless spectroscopy, H$\beta$ and [\ion{O}{III}] $\lambda 5007$ will no longer be accessible, as these lines get redshifted out of the NIRCam F444W filter at $z=9$ and $z=9.3$, respectively. Ly$\alpha$ and \ion{He}{II} $\lambda 1640$ now fall in the NIRISS F150W and F200W filters, respectively. 

\begin{table*}
\begin{center}
\resizebox{\linewidth}{!}{
\begin{tabular}{|c|c|c|c|c|} 
\hline
IMF & Ly$\alpha$ flux (cgs) & Ly$\alpha$ 5$\sigma$ exposure time (h) & \ion{He}{II} $\lambda$1640 flux (cgs) & \ion{He}{II} $\lambda$1640 5$\sigma$ exposure time (h) \\ 
\hline
Pop III.1 & $2.00\times10^{-17}$ & 0.10 & $5.13\times10^{-19}$ & 148.76 \\
\hline
Pop III.2 & $6.61\times10^{-18}$ & 0.96 & $9.55\times10^{-20}$ & 4292.65 \\ 
\hline
Pop III Kroupa & $1.05\times10^{-18}$ & 38.02 & $9.55\times10^{-21}$ & 429264.55 \\
\hline
\end{tabular}}
\caption{Similar to Table~\ref{tab:slitless_observations}, but now for Pop III galaxies at $z=10$. Note that H$\beta$ is no longer accessible for slitless spectroscopy with NIRCam.}
\label{tab:slitless_observations2}
\end{center}
\end{table*}

For follow-up spectroscopy with NIRSpec, H$\beta$ is only accessible up to $z=9.9$, though we still include this line in Table~\ref{tab:nirspec_observations2}, assuming a source at $z=10$. We note that, at $z=10$, the integration times needed to detect \ion{He}{II} $\lambda 1640$ (with NIRSpec/G235M), H$\gamma$ (with NIRSpec/G395M) and H$\alpha$ (with MIRI/MRS) are all roughly comparable. Given the only moderately larger H$\alpha$ and H$\gamma$ rest-frame EWs for Pop III galaxies, the optimal line to target for deep follow-up spectroscopy would undoubtedly be \ion{He}{II} $\lambda 1640$ with NIRSpec/G235M. 

\begin{table*}
\begin{center}
\resizebox{\linewidth}{!}{
\begin{tabular}{|c|c|c|c|c|c|c|} 
\hline
IMF & H$\beta$ flux (cgs) & H$\beta$ 5$\sigma$ exposure time (h) & Ly$\alpha$ flux (cgs) & Ly$\alpha$ 5$\sigma$ exposure time (h) & \ion{He}{II} $\lambda$1640 flux (cgs) & \ion{He}{II} $\lambda$1640 5$\sigma$ exposure time (h) \\ 
\hline
Pop III.1 & $3.98\times10^{-19}$ & 8.47 & $2.00\times10^{-17}$ & 0.01 & $5.13\times10^{-19}$ & 13.85 \\
\hline
Pop III.2 & $1.38\times10^{-19}$ & 70.45 & $6.61\times10^{-18}$ & 0.07 & $9.55\times10^{-20}$ & 399.76 \\ 
\hline
Pop III Kroupa & $2.24\times10^{-20}$ & 2673.95 & $1.05\times10^{-18}$ & 2.76 & $9.55\times10^{-21}$ & 39976.49 \\
\hline
\end{tabular}}
\caption{Similar to Table~\ref{tab:nirspec_observations}, but now for Pop III galaxies at $z=10$. Note that H$\beta$ is only observable with NIRSpec up to $z=9.9$, though we assume the source is at $z=10$.}
\label{tab:nirspec_observations2}
\end{center}
\end{table*}

\subsection{Pop III galaxy $z\sim10$ colour selection}

\subsubsection{[\ion{O}{III}]$-$H$\alpha$, H$\alpha -$[\ion{S}{III}] colour selection}

\begin{figure*}
\centering
\includegraphics[width=.65\linewidth]{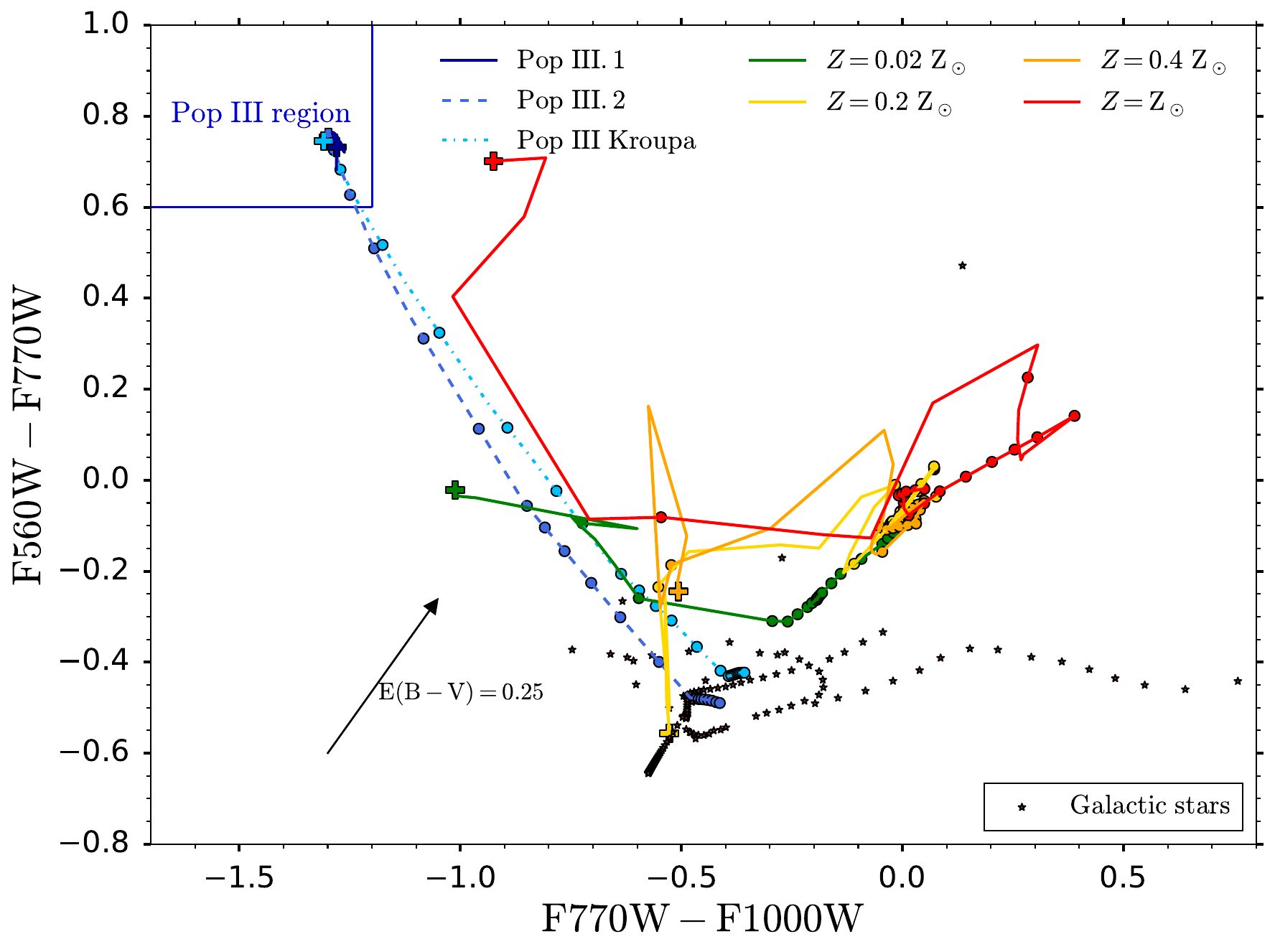}
\caption{Similar to the [\ion{O}{III}]$-$H$\alpha$, H$\alpha -$[\ion{S}{III}] colour selection in Fig.\@~\ref{fig:zackrisson_colours1}, but now using the analogous redder filters (as the high EW emission lines underpinning the colour selection are redshifted to longer wavelengths) to select Pop III galaxies at $z\sim10$. This colour selection can be used over the redshift range $9.8 < z < 10.3$. If $Z = \mathrm{Z}_\odot$ galaxy contamination is not a concern, then this colour selection can be applied over the wider redshift range $9.1 < z < 10.4$.}
\label{fig:zackrisson_colours4}
\end{figure*}

We show the $z\sim10$ [\ion{O}{III}]$-$H$\alpha$, H$\alpha-$[\ion{S}{III}] colour selection in Fig.~\ref{fig:zackrisson_colours4}, which uses the F560W$-$F770W, F770W$-$F1000W filter pairs. The colour selection can be applied over the redshift range $9.80 < z < 10.30$, at which point contamination by $Z = \mathrm{Z}_\odot$ galaxies begins. If such contamination is not a concern, this colour selection can be applied over the wider redshift range $9.10 < z < 10.40$. 

The boundaries of the Pop III region in the F560W$-$F770W, F770W$-$F1000W colour--colour plane are given by: 
\begin{equation}
    \begin{aligned}
    &\mathrm{F560W}-\mathrm{F770W} > 0.60\\
    &\mathrm{F770W}-\mathrm{F1000W} < -1.2 
\end{aligned}
\end{equation}

If contamination by $Z=\mathrm{Z}_\odot$ galaxies is not deemed a concern, these boundaries can be extended to:

\begin{equation}
    \begin{aligned}
    &\mathrm{F560W}-\mathrm{F770W} > 0.50\\
    &\mathrm{F770W}-\mathrm{F1000W} < -1.0
\end{aligned}
\end{equation}

\subsubsection{[\ion{O}{III}], [\ion{O}{III}]$-$H$\alpha$ colour selection}

\begin{figure*}
\centering
\includegraphics[width=.65\linewidth]{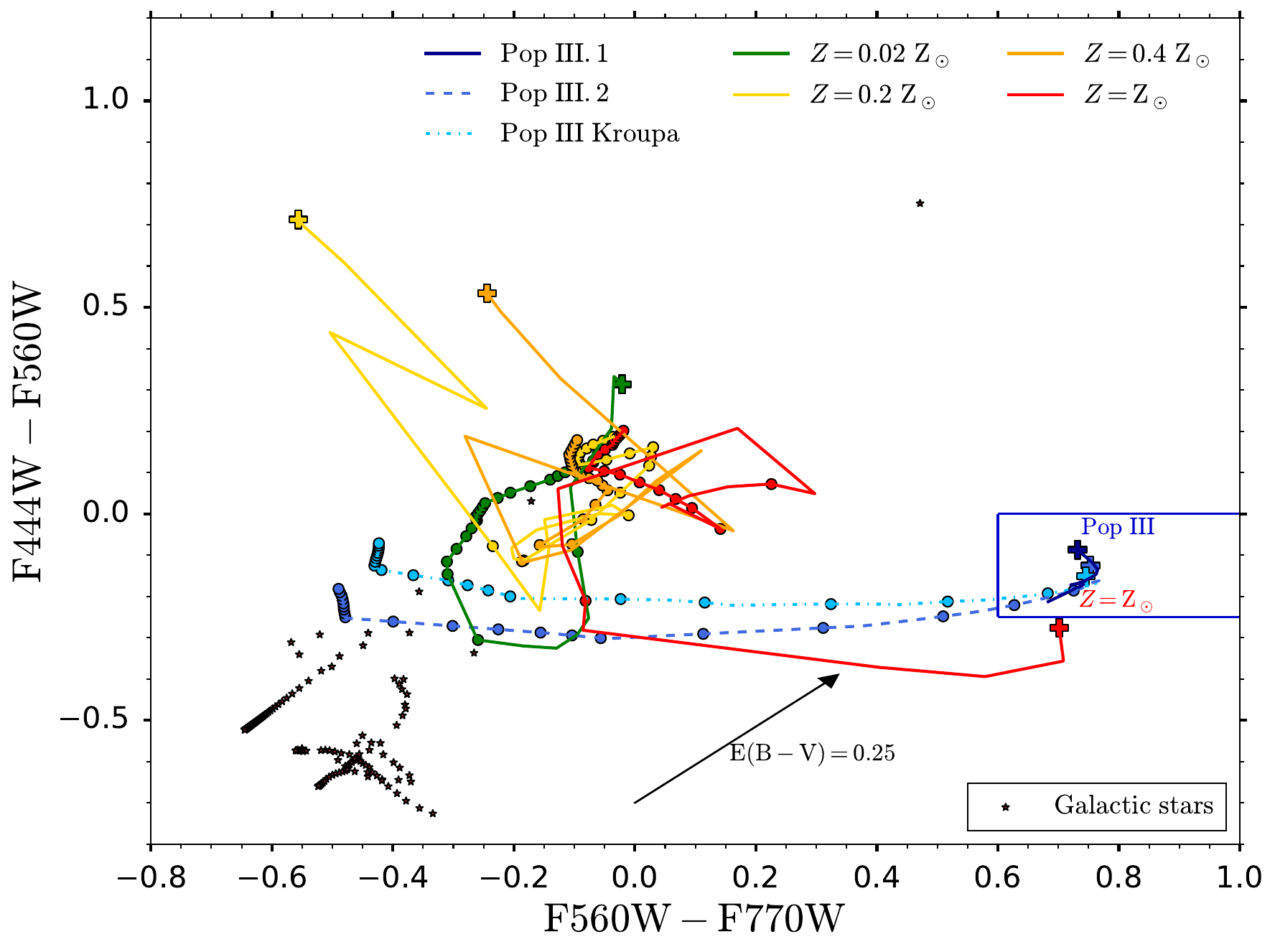}
\caption{Owing to the challenge in actually detecting Pop III galaxies in the MIRI F1000W filter, we advocate for this alternative [\ion{O}{III}], [\ion{O}{III}]$-$ H$\alpha$ colour selection (similar to Fig.\@~\ref{fig:zackrisson_colours_extra}, but now applied at $z\sim10$), which leverages on the deep F560W and F770W imaging that will need to be taken to identify $z\sim8$ Pop III candidates, as well as the much greater NIRCam F444W sensitivity (relative to MIRI F1000W). This colour selection can be used over the redshift range $9.50 < z < 10.50$, though suffers from contamination by $Z = \mathrm{Z}_\odot$ galaxies throughout.}  
\label{fig:zackrisson_colours7}
\end{figure*}

Given the great challenge in actually detecting Pop III galaxies in the MIRI F1000W filter, we introduce a substitute colour selection in Fig.\@~\ref{fig:zackrisson_colours7} that is still rather robust, but is much more practical to apply in practice. Here we use the F444W$-$F560W, F560W$-$F770W filter pairs for our colour--colour selection. The basis behind this Pop III colour selection is the same as Fig.\@~\ref{fig:zackrisson_colours_extra} but now applied at $z\sim10$, namely a lack of [\ion{O}{III}] emission in the F560W filter, together with moderately higher EW H$\alpha$ emission in the F770W filter. Thus the F444W$-$F560W colour is relatively flat for Pop III galaxies, and the F560W$-$F770W colour is relatively red. This group of filters leverages on the deep F560W and F770W imaging that will need to be taken to identify $z\sim8$ Pop III candidates, as well as the much greater NIRCam F444W sensitivity (relative to MIRI F1000W). 

This colour selection can be applied over the redshift interval $9.50 < z < 10.50$. Owing to the fact that Pop III galaxies and $Z = \mathrm{Z}_\odot$ galaxies have similar F444W$-$F560W and F560W$-$F770W colours, contamination may be a concern over this entire redshift range. 
The boundaries of the Pop III, $Z = \mathrm{Z}_\odot$ region in the F444W$-$F560W, F560W$-$F770W colour--colour plane are given by: 
\begin{equation}
    \begin{aligned}
    -0.25 < \mathrm{F444W}-\mathrm{F560W} < 0.00\\
    \mathrm{F560W}-\mathrm{F770W} > 0.60
\end{aligned}
\end{equation}

\subsubsection{[\ion{O}{III}], Ly$\alpha$ colour selection}

\begin{figure*}
\centering
\includegraphics[width=.65\linewidth]{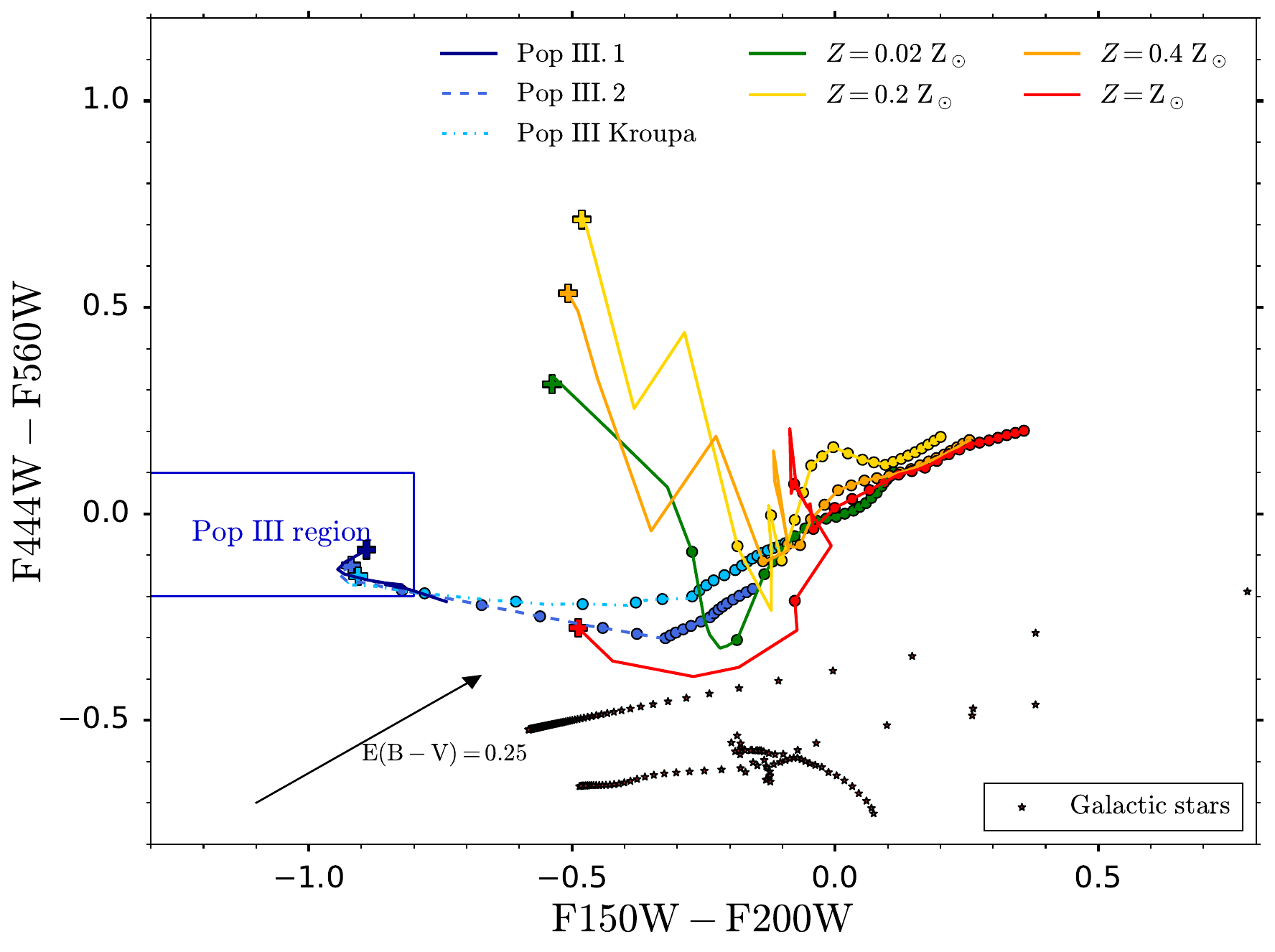}
\caption{Similar to the [\ion{O}{III}], Ly$\alpha$ colour selection in Fig.\@~\ref{fig:zackrisson_colours2}, but now using the analogous redder filters to select Pop III galaxies at $z\sim10$. This colour selection can be used over the redshift range $10.0 < z < 11.0$. Though we once again stress that no attenuation/scattering redward of the Ly$\alpha$ centre has been applied. In practice the Ly$\alpha$ line will likely be much more heavily attenuated than in our models, which will erode this Pop III signature and thus may render this colour selection ineffective at actually identifying Pop III galaxies from observational data.}
\label{fig:zackrisson_colours5}
\end{figure*}

\begin{figure*}
\centering
\includegraphics[width=.65\linewidth]{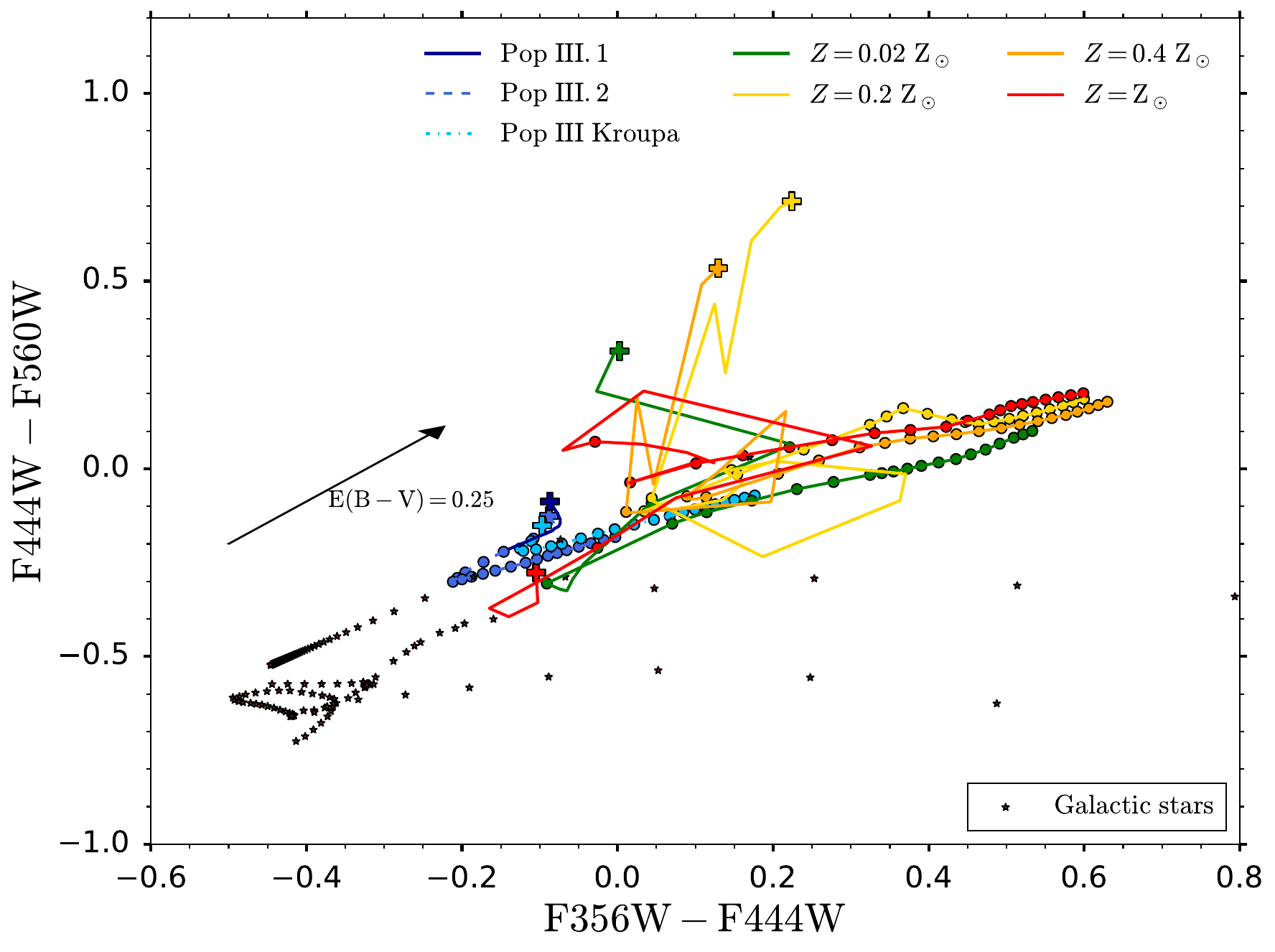}
\caption{Similar to the [\ion{O}{III}], [\ion{O}{II}] colour selection in Fig.\@~\ref{fig:zackrisson_colours2}, but now using the analogous redder filters to select Pop III galaxies at $z\sim10$. This colour selection can only be applied over the narrow redshift range $10 < z < 10.25$. If contamination from $Z = \mathrm{Z}_\odot$ galaxies is not a concern, the colour selection can be applied over the wider redshift range $9.75 < z < 10.25$. Owing to the limitations with this colour selection, we do not define a Pop III region to identify potential Pop III candidates in this colour--colour plane.}
\label{fig:zackrisson_colours6}
\end{figure*}
We show the $z\sim10$ [\ion{O}{III}], Ly$\alpha$ colour selection in Fig.\@~\ref{fig:zackrisson_colours5}, which uses the F444W$-$F560W, F150W$-$F200W filter pairs. This colour selection can be applied over the redshift range $10.0 < z < 11.0$. Starburst galaxies at $z=1.3$ with bright H$\alpha$ emission in the F150W filter can mimic the colours of $z\sim10$ Pop III galaxies and hence are potential contaminants. However, continuum detections in F090W and F115W, as well as bluer, non-\emph{JWST} filters should enable one to distinguish between $z\sim10$ galaxies and these low-redshift interlopers.

The boundaries of the Pop III region in the F444W$-$F560W, F150W$-$F2000W colour--colour plane are given by: 
\begin{equation}
    \begin{aligned}
    -0.20 < \mathrm{F444W}-\mathrm{F560W} < 0.10\\
    \mathrm{F150W}-\mathrm{F200W} < -0.8
\end{aligned}
\end{equation}

\subsubsection{[\ion{O}{III}], [\ion{O}{II}] colour selection}

We show the $z\sim10$ [\ion{O}{III}], [\ion{O}{II}] colour selection in Fig.\@~\ref{fig:zackrisson_colours6}, which uses the F444W$-$F560W, F356W$-$444W filter pairs. This colour selection can only be applied over the narrow redshift range $10 < z < 10.25$. If contamination from $Z = \mathrm{Z}_\odot$ galaxies is not a concern, the colour selection can be applied over the wider redshift range $9.75 < z < 10.25$. Owing to the limitations with this colour selection, we do not define a Pop III region to identify potential Pop III candidates in this colour--colour plane.


\bsp	
\label{lastpage}
\end{document}